\documentclass[a4paper,11pt,headings=big,DIV=12]{scrartcl}
\pdfoutput=1
\usepackage[utf8]{inputenc}
\usepackage{amsmath,amssymb}
\usepackage{slashed}
\usepackage{color,xcolor}
\definecolor{blue}{rgb}{0,0,0.5}
\usepackage[colorlinks,linkcolor=blue,urlcolor=blue,citecolor=blue]{hyperref}
\usepackage{graphicx}
\addtokomafont{disposition}{\rmfamily\boldmath}
\usepackage{cite}
\usepackage{multirow}
\usepackage{booktabs}
\definecolor{maroon}{cmyk}{0, 0.87, 0.68, 0.32}
\definecolor{halfgray}{gray}{0.55}
\definecolor{ipython_frame}{RGB}{207, 207, 207}
\definecolor{ipython_bg}{RGB}{247, 247, 247}
\definecolor{ipython_red}{RGB}{186, 33, 33}
\definecolor{ipython_green}{RGB}{0, 128, 0}
\definecolor{ipython_cyan}{RGB}{64, 128, 128}
\definecolor{ipython_purple}{RGB}{170, 34, 255}

\usepackage{listings}
\lstdefinelanguage{iPython}{
    morekeywords={access,and,break,class,continue,def,del,elif,else,except,exec,finally,for,from,global,if,import,in,is,lambda,not,or,pass,print,raise,return,try,while},%
    morekeywords=[2]{abs,all,any,basestring,bin,bool,bytearray,callable,chr,classmethod,cmp,compile,complex,delattr,dict,dir,divmod,enumerate,eval,execfile,file,filter,float,format,frozenset,getattr,globals,hasattr,hash,help,hex,id,input,int,isinstance,issubclass,iter,len,list,locals,long,map,max,memoryview,min,next,object,oct,open,ord,pow,property,range,raw_input,reduce,reload,repr,reversed,round,set,setattr,slice,sorted,staticmethod,str,sum,super,tuple,type,unichr,unicode,vars,xrange,zip,apply,buffer,coerce,intern},%
    sensitive=true,%
    morecomment=[l]\#,%
    morestring=[b]',%
    morestring=[b]",%
    morestring=[s]{'''}{'''},% used for documentation text (mulitiline strings)
    morestring=[s]{"""}{"""},% added by Philipp Matthias Hahn
    morestring=[s]{r'}{'},% `raw' strings
    morestring=[s]{r"}{"},%
    morestring=[s]{r'''}{'''},%
    morestring=[s]{r"""}{"""},%
    morestring=[s]{u'}{'},% unicode strings
    morestring=[s]{u"}{"},%
    morestring=[s]{u'''}{'''},%
    morestring=[s]{u"""}{"""},%
    literate=
    {á}{{\'a}}1 {é}{{\'e}}1 {í}{{\'i}}1 {ó}{{\'o}}1 {ú}{{\'u}}1
    {Á}{{\'A}}1 {É}{{\'E}}1 {Í}{{\'I}}1 {Ó}{{\'O}}1 {Ú}{{\'U}}1
    {à}{{\`a}}1 {è}{{\`e}}1 {ì}{{\`i}}1 {ò}{{\`o}}1 {ù}{{\`u}}1
    {À}{{\`A}}1 {È}{{\'E}}1 {Ì}{{\`I}}1 {Ò}{{\`O}}1 {Ù}{{\`U}}1
    {ä}{{\"a}}1 {ë}{{\"e}}1 {ï}{{\"i}}1 {ö}{{\"o}}1 {ü}{{\"u}}1
    {Ä}{{\"A}}1 {Ë}{{\"E}}1 {Ï}{{\"I}}1 {Ö}{{\"O}}1 {Ü}{{\"U}}1
    {â}{{\^a}}1 {ê}{{\^e}}1 {î}{{\^i}}1 {ô}{{\^o}}1 {û}{{\^u}}1
    {Â}{{\^A}}1 {Ê}{{\^E}}1 {Î}{{\^I}}1 {Ô}{{\^O}}1 {Û}{{\^U}}1
    {œ}{{\oe}}1 {Œ}{{\OE}}1 {æ}{{\ae}}1 {Æ}{{\AE}}1 {ß}{{\ss}}1
    {ç}{{\c c}}1 {Ç}{{\c C}}1 {ø}{{\o}}1 {å}{{\r a}}1 {Å}{{\r A}}1
    {€}{{\EUR}}1 {£}{{\pounds}}1
    {^}{{{\color{ipython_purple}\^{}}}}1
    {=}{{{\color{ipython_purple}=}}}1
    {+}{{{\color{ipython_purple}+}}}1
    {*}{{{\color{ipython_purple}$^\ast$}}}1
    {/}{{{\color{ipython_purple}/}}}1
    {+=}{{{+=}}}1
    {-=}{{{-=}}}1
    {*=}{{{$^\ast$=}}}1
    {/=}{{{/=}}}1,
    literate=
    *{-}{{{\color{ipython_purple}-}}}1
     {?}{{{\color{ipython_purple}?}}}1,
    identifierstyle=\color{black}\ttfamily,
    commentstyle=\color{ipython_cyan}\ttfamily,
    stringstyle=\color{ipython_red}\ttfamily,
    keepspaces=true,
    showspaces=false,
    showstringspaces=false,
    rulecolor=\color{ipython_frame},
    frame=single,
    frameround={t}{t}{t}{t},
    framexleftmargin=6mm,
    numbers=left,
    numberstyle=\tiny\color{halfgray},
    backgroundcolor=\color{ipython_bg},
    basicstyle=\footnotesize\ttfamily,
    keywordstyle=\color{ipython_green}\ttfamily,
    aboveskip=1.2em,
    belowskip=1.2em,
}

\newcommand\YAMLcolonstyle{\color{black}\ttfamily}
\newcommand\YAMLkeystyle{\color{red!70!black}\ttfamily}
\newcommand\YAMLvaluestyle{\color{green!40!black}\ttfamily}
\newcommand\YAMLliststyle{\color{green!40!black}\ttfamily}

\makeatletter

\newcommand\language@yaml{yaml}

\expandafter\expandafter\expandafter\lstdefinelanguage
\expandafter{\language@yaml}
{
  keywords={true,false,null,y,n},
  keywordstyle=\YAMLvaluestyle,
  basicstyle=\YAMLkeystyle\footnotesize,                                 % assuming a key comes first
  sensitive=false,
  showstringspaces=false,
  comment=[l]{\#},
  morecomment=[s]{/*}{*/},
  commentstyle=\color{purple}\ttfamily,
  stringstyle=\YAMLvaluestyle\ttfamily,
  moredelim=[l][\color{black!50}]{[...},
  moredelim=[l][\color{orange}]{\&},
  moredelim=**[il][\YAMLcolonstyle{:}\YAMLvaluestyle]{:},   % switch to value style at :
  morestring=[b]',
  morestring=[b]",
  literate =    {---}{{\ProcessThreeDashes}}3
                {>}{{\textcolor{red}\textgreater}}1     
                {|}{{\textcolor{red}\textbar}}1 
                {\ -\ }{{\YAMLliststyle\ {\YAMLcolonstyle-}\ }}3,
  frame=single,
    rulecolor=\color{ipython_frame},
    frame=single,
    frameround={t}{t}{t}{t},
    framexleftmargin=6mm,
    numbers=left,
    numberstyle=\tiny\color{halfgray},
    backgroundcolor=\color{ipython_bg},
    aboveskip=1.5em,
    belowskip=1.5em,
}

\lst@AddToHook{EveryLine}{\ifx\lst@language\language@yaml\YAMLkeystyle\fi}
\makeatother

\usepackage[normalem]{ulem}
\usepackage{placeins}
\usepackage{tabularx}

\usepackage{upquote}

\newcommand{\flavio}{\texttt{flavio}}
\newcommand{\wcxf}{WCxf}
\newcommand{\wilson}{\texttt{wilson}}
\newcommand{\smelli}{\texttt{smelli}}

\renewcommand{\Re}{\operatorname{Re}}
\renewcommand{\Im}{\operatorname{Im}}

\allowdisplaybreaks

\newcommand{\bea}{\begin{eqnarray}}
\newcommand{\eea}{\end{eqnarray}}

\begin{document}

\begin{center}
\vspace*{1cm}
{\LARGE\bfseries
A Global Likelihood\\[0.5em] for Precision Constraints and Flavour Anomalies
}\\[0.8 cm]
{\textsc{
Jason Aebischer$^a$, Jacky Kumar$^b$, Peter Stangl$^c$, David M.\ Straub$^a$
}\\[1 cm]
\small
$^a$ Excellence Cluster Universe, Boltzmannstr.~2, 85748~Garching, Germany \\
$^b$ Physique des Particules, Universite de Montreal,  C.P. 6128, succ.
centre-ville,\\ Montreal, QC, Canada H3C 3J7 \\
$^c$ Laboratoire d’Annecy-le-Vieux de Physique Th\'eorique, UMR5108, Universit\'e de Savoie Mont-Blanc et CNRS, 9 Chemin de Bellevue, B.P. 110, F-74941, Annecy-le-Vieux Cedex, France
}
\\[1 cm]
\footnotesize
E-Mail:
\texttt{jason.aebischer@tum.de},
\texttt{jacky.kumar@umontreal.ca},
\texttt{peter.stangl@lapth.cnrs.fr},
\texttt{david.straub@tum.de}
\\[1 cm]
\end{center}

\begin{abstract}\noindent
We present a global likelihood function in the space of dimension-six Wilson
coefficients in the Standard Model Effective Field Theory (SMEFT).
The likelihood includes contributions from flavour-changing neutral current
$B$ decays, lepton flavour universality tests in charged- and neutral-current
$B$ and $K$
decays, meson-antimeson mixing observables in the $K$, $B$, and $D$ systems,
direct CP violation in $K\to\pi\pi$,
charged lepton flavour violating $B$, tau, and muon decays, electroweak precision
tests on the $Z$ and $W$ poles, the anomalous magnetic moments of the electron, muon, and tau,
and several other precision observables, 265 in total.
The Wilson coefficients can be specified at any scale, with the one-loop running
above and below the electroweak scale automatically taken care of.
The implementation of the likelihood function is based on the open source tools \flavio\
and \wilson\ as well as the open Wilson coefficient exchange format
(\wcxf) and can be installed as a Python
package.
It can serve as a basis either  for model-independent fits or
for testing dynamical models, in particular models built to address
the anomalies in $B$ physics.
We discuss a number of example applications, reproducing
results from the EFT and model building literature.
\end{abstract}

\newpage
\tableofcontents
\newpage

\section{Introduction}

Precision tests at low energies, such as flavour physics in the quark
and lepton sectors, as well as precision tests at the electroweak (EW)
scale, such as $Z$ pole observables, are important probes of physics
beyond the Standard Model (SM). The absence of a direct
discovery of any particle beyond the SM spectrum at the LHC
makes these indirect tests all the more important.
Effective field theories (EFTs) are a standard tool to describe
new physics (NP) effects in these precision observables. For
low-energy quark flavour physics, their use is mandatory to separate
the long-distance QCD dynamics from the short-distance NP of interest.
But also for precision tests at electroweak-scale energies, EFTs have
become increasingly popular, given the apparent scale separation between
the EW scale and the scale of the NP.
With mild assumptions, namely the absence of non-SM states below or
around the EW scale as well as a linear realization of EW symmetry
breaking, NP effects in precision observables can be described in the
context of the Standard Model effective field theory (SMEFT), that extends the SM by the full set of
dimension-6 operators allowed by the SM gauge symmetry
\cite{Buchmuller:1985jz,Grzadkowski:2010es}
(see \cite{David:2015waa,deFlorian:2016spz,Brivio:2017vri} for reviews).
While this description facilitates model-independent investigations
of NP effects in precision observables, a perhaps even more important
virtue is that SMEFT can serve as an intermediate step between dynamical
models in the UV and the low-energy precision phenomenology. Computing all the relevant precision observables in a given UV model
and comparing the predictions to experiment is a formidable task.
Employing SMEFT, this task can be separated in two: computing the
SMEFT Wilson coefficients at the UV scale is model-dependent
but straightforward, while computing all the precision observables
in terms of these Wilson coefficients and comparing them to experiment
is challenging but, importantly, model-independent.

Eventually, to test a UV model given the  plethora of existing
precision measurements, we require a likelihood function
that quantifies the agreement of all existing precision observable
measurements to the model's predictions.
This likelihood function $L$
is a function of the model's Lagrangian parameters $\vec{\lambda}$
and certain model-independent phenomenological parameters $\vec{\theta}$ (form factors, decay constants, etc.),
$L=L(\vec{\lambda}, \vec{\theta})$.
Using SMEFT to describe NP effects in precision observables
model-independently in terms of the Wilson coefficients $\vec{C}$,
the likelihood can be reexpressed as
\begin{equation}
L(\vec{\lambda}, \vec{\theta}) = L_\text{SMEFT}(\vec C(\vec{\lambda}), \vec{\theta})\,,
\end{equation}
where $L_\text{SMEFT}(\vec C, \vec{\theta})$
is the {\em global SMEFT likelihood} in the space of Wilson coefficients
and phenomenological parameters. Having this function
at hand, the problem of testing any UV model is reduced to
computing the SMEFT Wilson coefficients $\vec{C}(\vec{\lambda})$
(and suitably accounting for the uncertainties in the parameters
$\vec{\theta}$).

A major challenge in obtaining this global likelihood function is that the SMEFT renormalization group evolution from the NP scale down to the EW scale does not preserve flavour, such that the likelihood in the space of SMEFT Wilson coefficients does not factorize into sectors with definite flavour quantum numbers. This is in contrast to the weak effective theory (WET) below the EW scale, that is frequently employed in low-energy flavour physics, where QCD and QED renormalization is flavour-blind.
Thanks to the calculation of the complete one-loop SMEFT RGEs \cite{Jenkins:2013zja, Jenkins:2013wua, Alonso:2013hga, Celis:2017hod},
the complete matching from SMEFT onto WET \cite{Aebischer:2015fzz, Jenkins:2017jig} and the complete one-loop QCD and QED RGEs within WET \cite{Aebischer:2017gaw, Jenkins:2017dyc} that have been incorporated in the public code \texttt{wilson} \cite{Aebischer:2018bkb} leveraging the open Wilson coefficient exchange format (WCxf) \cite{Aebischer:2017ugx}, the
relation between high-scale SMEFT Wilson coefficients and the coefficients in the appropriate low-energy EFT can now be automatized.

Having obtained the Wilson coefficients at the appropriate scales, the precision observables must be calculated and compared to the experimental measurements to obtain the likelihood function. This programme has been carried out in the literature for various subsets of observables or Wilson coefficients, e.g.
\begin{itemize}
\item simultaneous fits to Higgs and EW precision data have been performed by many groups, see \cite{deFlorian:2016spz} and references therein,
\item a fit to $Z$ pole observables not assuming lepton flavour universality (LFU) \cite{Efrati:2015eaa},
\item a likelihood incorporating low-energy precision measurements (but not flavour-changing neutral currents) \cite{Falkowski:2017pss},
\item fits of semi-leptonic operators to beta decays
\cite{Alioli:2017ces,Gonzalez-Alonso:2018omy},
\item fits of triple gauge boson coupling operators \cite{Falkowski:2015jaa,Bobeth:2015zqa},
\item a fit of four-lepton operators \cite{Falkowski:2015krw}.
\end{itemize}

So far, no global likelihood has been constructed however that contains the observables relevant for the anomalies in $B$ physics or the numerous measurements of flavour-changing neutral current (FCNC) processes that are in principle sensitive to very high scales.
The main aim of the present work is thus to provide a likelihood function that also takes into
account a large number of observables in flavour physics, with a focus on the ones that are relevant in models
motivated by the anomalies recently observed in $B$ decays based on the
$b\to c\tau\nu$ and $b\to s\mu\mu$ transition.
Our results build on the open source code \texttt{flavio} \cite{flavio}, that computes a large number of observables in flavour physics as a function of dimension-6 Wilson coefficients beyond the SM and contains a database of relevant experimental measurements. To incorporate constraints beyond quark flavour physics, we have also implemented EW precision tests, lepton flavour violation, and various other precision observables in \texttt{flavio}. By using open source software throughout, we hope our results can serve as the basis for a more and more {\em global} SMEFT  likelihood emerging as a community effort.

The rest of this paper is organized as follows.
In section~\ref{sec:setup}, we describe the statistical formalism,
in section~\ref{sec:obs}, we list the observables included in our
likelihood function,
in section~\ref{sec:pheno}, we discuss several example applications
relevant for the $B$ physics anomalies,
in section~\ref{sec:py}, we describe the usage
of the Python package provided by us, and finally we summarize
in section~\ref{sec:concl}.

\section{Formalism}\label{sec:setup}

Given a set of independent precision
measurements $\vec{O}_\text{exp}$ and the corresponding theory
predictions $\vec{O}_\text{th}$ in the presence of NP described
model-independently by dimension-6
SMEFT Wilson coefficients,
the general form of the SMEFT likelihood reads
\begin{equation}
L_\text{SMEFT}(\vec C, \vec{\theta})
= \prod_i L_\text{exp}^i\left(\vec{O}^\text{exp}, \vec{O}^\text{th}\left(\vec C, \vec{\theta}\right)\right)
\times L_\theta(\vec{\theta})\,,
\label{eq:lsmeft}
\end{equation}
where $L_\text{exp}^i$ are the distribution functions of the experimental measurements and $L_\theta(\vec{\theta})$ are experimental or theoretical constraints on
the theory parameters $\theta$. Since we are interested in the likelihood
as a function of the Wilson coefficients, all parameters $\theta$ are
nuisance parameters that have to be removed by an appropriate procedure.

In a Bayesian approach, $L_\theta(\vec{\theta})$ would be a prior
probability distribution for the theory parameters and the appropriate
procedure would be to obtain the posterior probability by means of Bayes'
theorem, integrating over the $\theta$ directions.
In a frequentist approach\footnote{%
See \cite{Charles:2016qtt} for a comprehensive discussion of the treatment
of theory  uncertainties in a frequentist approach,
also discussing methods that are not captured by \eqref{eq:lsmeft}.
}, one would instead determine the profile likelihood,
i.e.\ for a given Wilson coefficient point $\vec{C}$
maximize the likelihood with respect to all the $\vec{\theta}$.

While both the Bayesian and the frequentist treatment are valid approaches,
they both have the drawback that they are computationally very expensive for a
large number of parameters. Even if one were to succeed in deriving the
Bayesian posterior distribution or the profile likelihood in the entire space
of interest, the procedure would have to be repeated anytime the experimental
data changes, which in practice happens frequently given the large number of
relevant constraints.

Due to these challenges, here we opt for a more approximate, but much faster
approach. We split all the observables of interest into two categories,
\begin{enumerate}
  \item Observables where the theoretical uncertainty can be neglected at present
  compared to the experimental uncertainty.
  \item Observables where both the theoretical and experimental uncertainty can be approximated
  as (possibly multivariate) Gaussian and where the theoretical uncertainty is
  expected to be weakly dependent on $\vec C$ and $\vec \theta$.
\end{enumerate}
We then write the nuisance-free likelihood
\begin{equation}
L_\text{SMEFT}(\vec C)
= \prod_{i\,\in\,1.} L_{\text{exp}}\left(\vec{O}^\text{exp}_i, \vec{O}^\text{th}_i\left(\vec C, \vec{\theta}_0\right)\right)
\prod_{i\,\in\,2.} \widetilde{L}_{\text{exp}}\left(\vec{O}^\text{exp}_i, \vec{O}^\text{th}_i\left(\vec C, \vec{\theta}_0\right)\right)
.
\label{eq:nfSL}
\end{equation}
The first product contains the full experimental likelihood for a fixed
value of the theory parameters $\theta_0$, effectively ignoring theoretical
uncertainties.
The second product contains a modified experimental likelihood.
Assuming the measurements of $\vec{O}_i^\text{exp}$ to be normally distributed with the
covariance matrix $C_\text{exp}$
and the theory predictions to be normally distributed as well with covariance
$C_\text{th}$,
$\widetilde{L}_\text{exp}$ has the form
\begin{equation}
-2\ln\widetilde{L}_{\text{exp}}
= \vec{x}^T (C_\text{exp} + C_\text{th})^{-1} \vec x \,,
\end{equation}
where
\begin{equation}
\vec x = \vec{O}^\text{exp}_i - \vec{O}^\text{th}_i \,.
\end{equation}
Effectively, the theoretical uncertainties stemming from the uncertainties
in the theory parameters $\theta$ are ``integrated out'' and
treated as additional experimental uncertainties.

These two different approaches of getting rid of nuisance parameters are
frequently used in phenomenological analyses.
Neglecting theory uncertainties is well-known to be a good approximation
in EFT fits to electroweak precision tests (see e.g.\ \cite{Efrati:2015eaa,Falkowski:2017pss}).
The procedure of ``integrating out'' nuisance parameters
has been applied to EFT fits of rare $B$ decays first in \cite{Altmannshofer:2014rta}
and subsequently also applied elsewhere (see e.g.~\cite{Descotes-Genon:2015uva}).

While the nuisance-free likelihood is a powerful tool for fast exploration of the parameter space of SMEFT or any UV theory matched to it, we stress that there are observables where none of the two above assumptions are satisfied and which thus cannot be taken into account in our approach, for instance:
\begin{itemize}
\item We treat the four parameters of the CKM matrix as nuisance parameters, but these parameters are determined from tree-level processes that can be affected by dimension-6 SMEFT contributions themselves,
e.g. $B$ decays based on the $b\to c\ell\nu$ \cite{Jung:2018lfu} or $b\to u\ell\nu$ transition, charged-current kaon decays \cite{Gonzalez-Alonso:2016etj}, or the CKM angle $\gamma$ \cite{Brod:2014bfa}. Thus to take these processes into account, one would have to treat the CKM parameters as floating nuisance parameters. We do however take into account tests of lepton flavour universality (LFU) in these processes where the CKM elements drop out.
\item The electric dipole moments (EDMs) of the neutron or of diamagnetic atoms\footnote{The uncertainties of EDMs of paramagnetic atoms are instead under control \cite{Dekens:2018bci} and could be treated within our framework. We thank Jordy de Vries for bringing this point to our attention.}
are afflicted by sizable hadronic uncertainties, but are negligibly small in the SM. Thus the uncertainty can neither be neglected nor assumed to be SM-like and the poorly known matrix elements would have to be treated as proper nuisance parameters.
\end{itemize}
We will comment on partial remedies for these limitations in section~\ref{sec:concl}.

\section{Observables}\label{sec:obs}

Having defined the general form of the global, nuisance-free SMEFT
likelihood \eqref{eq:nfSL} and the two different options for treating
theory uncertainties, we now discuss the precision observables that are
currently included in our likelihood.

Generally, the observables we consider can be separated into two classes:
\begin{itemize}
  \item Electroweak precision observables (EWPOs) on the $Z$ or $W$ pole. In this case
  we evolve the SMEFT Wilson coefficients from the input scale to the $Z$ mass
  and then compute the NP contributions directly in terms of them.
  \item Low-energy precision observables. In this case we match the SMEFT
  Wilson coefficients onto the weak effective theory (WET) where the
  electroweak gauge bosons, the Higgs boson and the top quark have been integrated out. We then run the WET Wilson coefficients down to the
  scale appropriate for the process. For decays of particles without $b$ flavour,
  we match to the appropriate 4- or 3-flavour effective theories.
\end{itemize}
The Python package to be described in section~\ref{sec:py}
also allows to access a pure WET likelihood. In this case the constraints in the first category are ignored.
The complete tree-level matching from SMEFT onto WET \cite{Aebischer:2015fzz,Jenkins:2017jig} as well as the one-loop running in SMEFT \cite{Alonso:2013hga,Jenkins:2013zja,Jenkins:2013wua} and WET \cite{Aebischer:2017gaw,Jenkins:2017dyc} is done with the \texttt{wilson} package
\cite{Aebischer:2018bkb}.

In appendix~\ref{app:obstable},
we list all the observables along
with their experimental measurements and SM predictions.

\subsection{Electroweak precision observables}\label{sec:ewpo}

To consistently include EWPOs, we follow \cite{Brivio:2017vri}
by parameterizing the shifts in SM parameters and couplings
as linear functions of SMEFT Wilson coefficients.
Terms quadratic in the dimension-6 Wilson coefficients
are of the same order in the EFT power counting as the interference
of the SM amplitude with dimension-8 operators and thus should be dropped.
We use the
$\lbrace \hat \alpha_e, \hat G_F, \hat m_Z \rbrace$ input parameter scheme.
We include the full set of $Z$ pole pseudo-observables measured at LEP-I
without assuming lepton flavour universality.
Following \cite{Efrati:2015eaa} we also include $W$ branching ratios,
the $W$ mass (cf.~\cite{Bjorn:2016zlr}), and the $W$ width.
As a non-trivial cross-check, we have confirmed that the electroweak part of
our likelihood exhibits the reparametrization invariance pointed out
in \cite{Brivio:2017bnu}.
Finally, we include LEP and LHC constraints on LFV $Z$ decays.
The total number of observables in this sector is 25.
For all these observables, we neglected the theoretical uncertainties,
which are in all cases much smaller than the experimental uncertainties.

\subsection{Rare $B$ decays}

Measurements of rare $B$ decays based on the $b\to s$ transition are of particular
interest as several deviations from SM expectations have been observed there,
most notably the anomalies in $\mu$/$e$ universality tests in
$B\to K^{(*)}\ell^+\ell^-$ \cite{Aaij:2014ora,Aaij:2017vbb}
and the anomalies in angular observables in $B\to K^*\mu^+\mu^-$ \cite{Aaij:2015oid}.
We include the following observables.
\begin{itemize}
\item All relevant CP-averaged observables in inclusive and exclusive
semi-leptonic $b\to s\mu\mu$ decays that have also been included
in the global fit \cite{Altmannshofer:2017yso}.
In this case the theoretical uncertainties are sizable and strongly
correlated and we use the second approach described in section~\ref{sec:setup}.
\item T-odd angular CP asymmetries in $B\to K^*\mu^+\mu^-$.
These are tiny in the SM and we neglect the theory uncertainty.
\item High-$q^2$ branching ratios and angular observables of
$\Lambda_b\to \Lambda\mu^+\mu^-$ \cite{Boer:2014kda,Detmold:2016pkz}.
\item The branching ratios of the leptonic decays $B^0\to\mu^+\mu^-$ and $B_s\to\mu^+\mu^-$ \cite{DeBruyn:2012wj,DeBruyn:2012wk}.
\item The $\mu$/$e$ universality tests $R_K$ and $R_{K^*}$ following \cite{Altmannshofer:2017fio}.
Here we neglect the tiny theory uncertainties \cite{Bordone:2016gaq}.
\item The branching ratio of the inclusive decay $B\to X_se^+e^-$ \cite{Huber:2015sra}.
\item All observables in inclusive and exclusive radiative $b\to s\gamma$ decays \cite{Misiak:2015xwa}
(including $B\to K^*e^+e^-$ at very low $q^2$) that have also been included
in the global fit in \cite{Paul:2016urs}.
\item Bounds on the exclusive decays $B\to K^{(*)}\nu\bar\nu$
\cite{Buras:2014fpa}.
Even though these have sizable uncertainties in the SM, they can be neglected
compared to the experimental precision (which in turn allows us to take into
account the non-Gaussian form of the likelihoods).
A sum over the unobserved neutrino flavours is performed, properly
accounting for models where wrong-flavour neutrino modes can contribute.
\item Bounds on tauonic $B$ decays: $B\to K\tau^+\tau^-$,  $B^0\to \tau^+\tau^-$,  $B_s\to \tau^+\tau^-$.
We neglect theoretical uncertainties.
\item Bounds on LFV $B$ decays: $B\to (\pi, K, K^*)\ell\ell'$ \cite{Becirevic:2016zri} for all cases where bounds exist.
We neglect theoretical uncertainties.
\end{itemize}

In contrast to EWPOs, in flavour physics there is no formal need to drop terms quadratic in the dimension-6 SMEFT Wilson coefficients. For processes that are forbidden in the SM, such as LFV decays, this is obvious since the leading contribution is the squared dimension-6 amplitude and the dimension-8 contribution is relatively suppressed by four powers of the NP scale. But also for processes that are not forbidden but suppressed by a mechanism that does not have to hold beyond the SM, the dimension-8 contributions are subleading. Schematically, the amplitude reads $\epsilon A_\text{SM} + v^2/\Lambda^2 A_6 + v^4/\Lambda^2 A_8+\ldots$, where $\epsilon$ is a SM suppression factor (e.g.\ GIM or CKM suppression) and $A_{6,8}$ the dimension-6 and 8 contributions without the dimensional suppression factors, respectively. Obviously, in the squared amplitude the $A_\text{SM} A_8^*$ interference term is suppressed by $\epsilon$ compared to the $|A_6|^2$ term, so it is consistent to only keep the latter.

\subsection{Semi-leptonic $B$ and $K$ decays}

As discussed at the end of section~\ref{sec:setup}, we cannot
use the semi-leptonic charged-current $B$ and $K$ decays with light leptons in our approach since we do not allow the CKM parameters to float. Nevertheless, we can include tests of LFU in $b\to q\ell\nu$ decays where the CKM elements drop out. We include:
\begin{itemize}
  \item The ratio of $K^+\to e^+\nu$ and $K^+\to \mu^+\nu$,
  \item The branching ratios\footnote{
  While these observables are strictly speaking not independent of the CKM element $V_{ub}$, the much larger experimental uncertainty compared to $B\to\pi\ell\nu$ means that they are
  only relevant as constraints on large violations of LFU or large scalar operators, which allows us to take them into account nevertheless. Alternatively, these observables could be normalized explicitly to $B\to\pi\ell\nu$, but we refrain from doing so for simplicity.
  } of $B\to\pi\tau\nu$, $B^+\to\tau^+\nu$, $B^+\to\mu^+\nu$, and $B^+\to e^+\nu$,
  \item The ratios $R_{D^{(*)}} = \text{BR}{(B\to D^{(*)}\tau\nu)}/{\text{BR}(B\to D^{(*)}\ell\nu)}$, where the deviations from SM expectations are observed,
  \item The $q^2$ distributions of $B\to D^{(*)}\tau\nu$ from Belle \cite{Huschle:2015rga} and BaBar \cite{Lees:2013uzd}.
\end{itemize}
For the latter, we use the results of\cite{Celis:2016azn}, where these are given for an arbitrary normalization. For our purpose we normalize these values in each bin by the integrated rate, in order to leave $R_{D^{(*)}}$  as independent observables.

For the form factors of the $B\to D$ and $B\to  D^*$ transition,
we use the results of \cite{Jung:2018lfu}, combining results
from lattice QCD, light-cone sum rules, and heavy quark effective theory but not using any experimental data on $b\to c\ell\nu$ decays to determine the form factors. This leads to a larger SM uncertainty (and also lower central values) for  $R_D$ and $R_{D^*}$. Even though we require $b\to c\ell\nu$ with $\ell=e,\mu$ to be mostly SM-like for  consistency as discussed in section~\ref{sec:setup}, we prefer to use the form factors  from pure theory predictions to facilitate a future treatment of the CKM elements as nuisance parameters (see section~\ref{sec:concl}).

\subsection{Meson-antimeson mixing}

We include the following observables related to meson-antimeson mixing in the
$K^0$, $B^0$, $B_s$, and $D^0$ systems:
\begin{itemize}
  \item The $B^0$ and $B_s$ mass differences $\Delta M_d$ and $\Delta M_s$,
  \item The mixing-induced CP asymmetries $S_{\psi K_S}$ and $S_{\psi\phi}$ (neglecting
  contributions to the penguin amplitude from four-quark operators),
  \item The CP-violating parameter $\epsilon_K$ in the $K^0$ system,
  \item The CP-violating parameter $x_{12}^\text{Im}$ in the $D^0$ system
  defined as in \cite{Aebischer:2018csl}.
\end{itemize}
We include the SM uncertainties as described in section~\ref{sec:setup}.

\subsection{FCNC $K$ decays}

We include the following observables in flavour-changing neutral current kaon decays.
\begin{itemize}
  \item The branching ratios of $K^+\to\pi^+\nu\bar\nu$ and  $K_L\to\pi^0\nu\bar\nu$.
  \item The branching ratios of $K_{L,S}\to\ell^+\ell^-$ \cite{Chobanova:2017rkj}.
  \item The bound on the LFV decay $K_L\to e^\pm\mu^\mp$.
  \item The parameter $\varepsilon'/\varepsilon$ measuring the ratio of direct to indirect CP violation in $K_L\to\pi\pi$ \cite{Aebischer:2018quc,Aebischer:2018csl,Blum:2015ywa,Bai:2015nea,Aebischer:2018rrz}.
\end{itemize}
For $\varepsilon'/\varepsilon$, using our approach described in section~\ref{sec:setup} to assume the uncertainties to be SM-like also beyond the SM is borderline since beyond the SM, other matrix elements become relevant, some of them not known from lattice QCD \cite{Aebischer:2018quc}. We stress however that we do not make use of the partial cancellations of matrix element uncertainties between the real and imaginary parts of the SM amplitudes \cite{Buras:2015yba}, so our SM uncertainty is conservative in this respect. Moreover, visible NP effects in $\varepsilon'/\varepsilon$ typically come from operators contributing to the $\Delta I=3/2$ amplitude, where the matrix elements are known to much higher precision from lattice QCD \cite{Blum:2015ywa}, such that also in these cases our approach can be considered conservative.

\subsection{Tau and muon decays}

We include the following LFV decays of taus and muons:
\begin{itemize}
  \item $\mu\to 3e$ \cite{Brignole:2004ah}, $\tau\to 3\mu$ \cite{Kuno:1999jp,Brignole:2004ah}, $\tau^-\to \mu^-e^+e^-$ \cite{Brignole:2004ah},
  \item $\tau^-\to e^-\mu^+e^-$, $\tau^-\to\mu^-e^+\mu^-$,
  \item $\mu\to e\gamma$, $\tau\to \ell\gamma$ \cite{Brignole:2004ah},
  \item $\tau\to \rho \ell$,  $\tau\to \phi \ell$,
\end{itemize}
where $\ell=e$ or $\mu$. Theoretical uncertainties can be neglected.

For $\tau\to \rho \ell$ and  $\tau\to \phi \ell$, we have calculated the full WET expressions of the decay widths including contributions from semi-leptonic vector and tensor operators as well as leptonic dipole operators.
In all expressions, we have kept the full dependence on the mass of the light lepton $\ell$.
The results, which to our knowledge have not been presented in this generality in the literature before, are given in appendix~\ref{app:tau_lV}.
As expected, considering only the dipole contributions, $\tau\to \rho \ell$ and $\tau\to \phi \ell$ are not competitive with $\tau\to \ell\gamma$.
Interestingly, the semi-leptonic tensor operators are generated in the tree-level SMEFT matching only for up-type quarks (semi-leptonic down-type tensor operators violate hypercharge).
This means that in a SMEFT scenario and neglecting loop effects, tensor operators do contribute to $\tau\to \rho \ell$ but do not contribute to $\tau\to \phi \ell$.

In addition we include the charged-current tau decays
\begin{itemize}
  \item $\tau\to \ell\nu\nu$ \cite{Pich:2013lsa},
\end{itemize}
which represent important tests of lepton flavour universality (LFU).
Since these are present in the SM and measured precisely, theory uncertainties
cannot be neglected and we include them as described in section~\ref{sec:setup}.
A sum over unobserved neutrino flavours is performed, properly
accounting for models where wrong-flavour neutrino modes can contribute.

Note that the branching ratio of $\mu\to e\nu\nu$ is not a constraint in our
likelihood as it is used to define the input parameter $G_F$ via the muon
lifetime. Potential NP contributions to this decay enter the EWPOs
of section~\ref{sec:ewpo} via
effective shifts of the SM input parameters.

\subsection{Low-energy precision observables}

Finally, we include the following flavour-blind low-energy observables:
\begin{itemize}
  \item the anomalous magnetic moments of the electron, muon, and tau, $a_\ell = (g_\ell-2)/2$,
  \item the neutrino trident production cross section \cite{Altmannshofer:2014pba}.
\end{itemize}

\section{Applications}\label{sec:pheno}

In this section, we demonstrate the usefulness of the global likelihood
with a few example applications motivated in particular by the
$B$ anomalies. While we restrict ourselves to simplistic two-parameter
scenarios for reasons of presentation,
we stress that the power of the {\em global} likelihood is that it can be used
to test models {\em beyond} such simplified scenarios.

\subsection{Electroweak precision analyses}

\begin{figure}
\centering
\includegraphics[width=4cm]{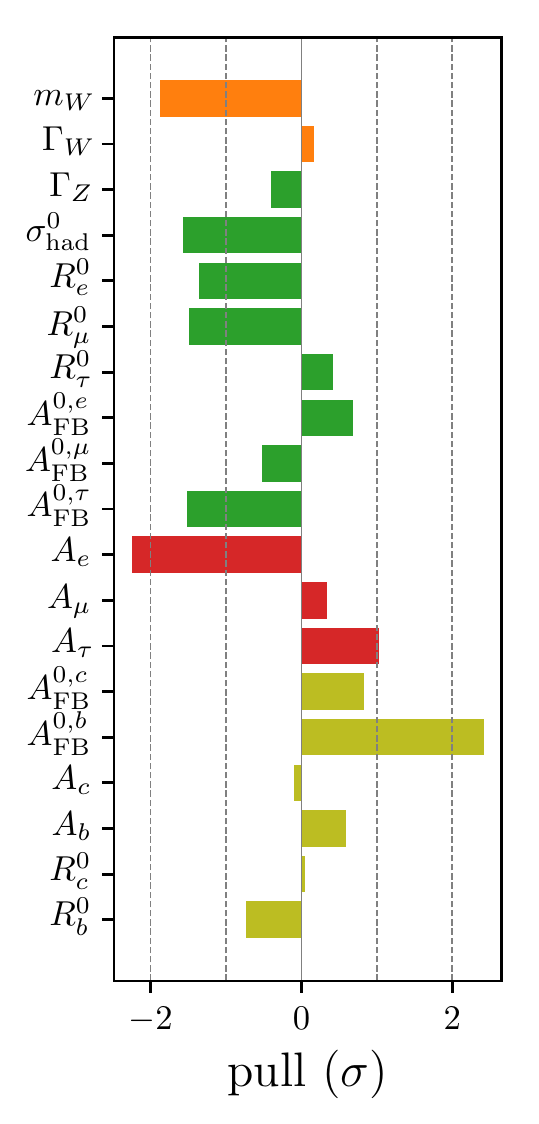}%
\includegraphics[width=0.5\textwidth]{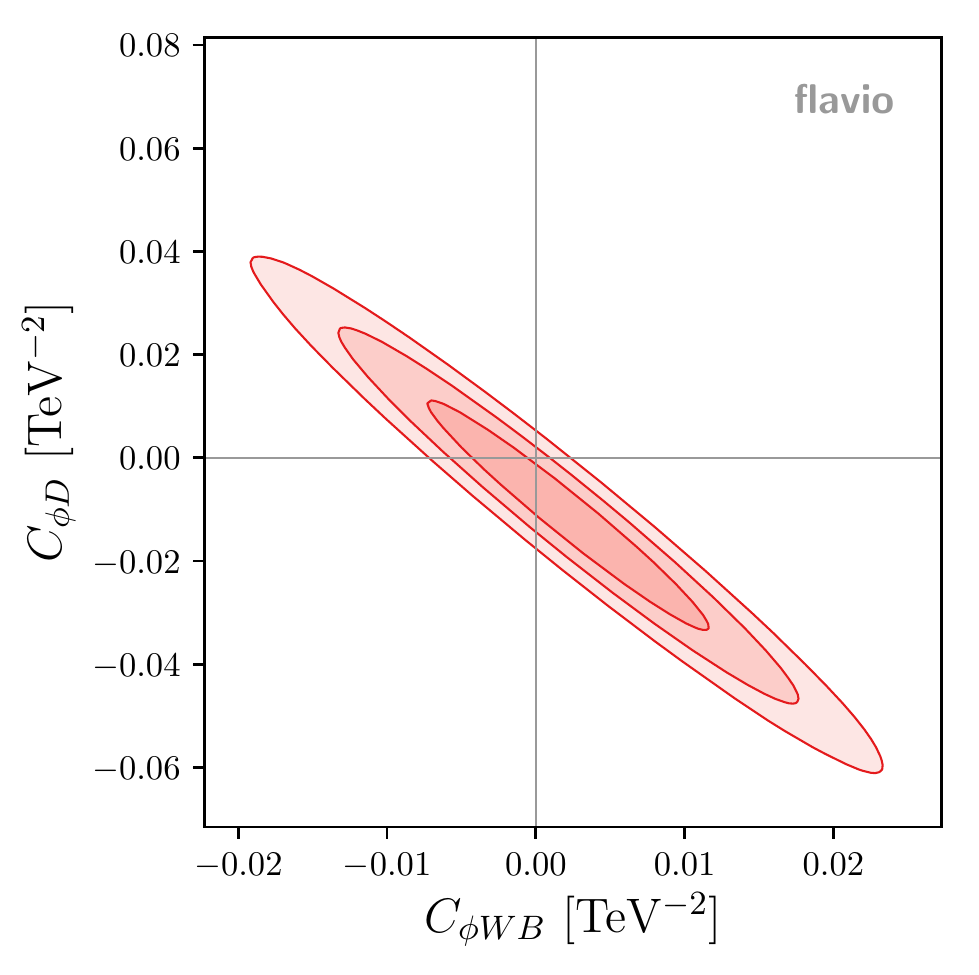}
\caption{Left: pulls for individual $Z$- and $W$-pole observables for the SM point.
Right: 1--3$\sigma$ likelihood contours in the plane of two Warsaw-basis Wilson coefficients that are proportional to the oblique parameters $S$ and $T$, assuming all other coefficients to vanish.}
\label{fig:ew}
\end{figure}

A non-trivial check of our implementation of EWPOs discussed in
sec.~\ref{sec:ewpo} is to compare the pulls between the SM prediction
and measurement for individual observables to sophisticated EW fits as
performed e.g.\ by the Gfitter collaboration \cite{Haller:2018nnx}.
We show theses pulls in fig.~\ref{fig:ew} left and observe good agreement
with the literature. The largest pull is in the forward-backward asymmetry
in $Z\to b\bar b$.

Another well-known plot is the EWPO constraint on the
oblique parameters $S$ and $T$, which are proportional to the SMEFT
Warsaw basis Wilson
coefficients $C_{\phi WB}$ and $C_{\phi D}$, respectively (see e.g.~\cite{Wells:2015uba}). Their corresponding operators read:
\begin{equation}
	O_{\phi WB}=\phi^\dagger \tau^I \phi W_{\mu \nu}^I B^{\mu \nu}\,,\quad O_{\phi D}= (\phi^\dagger D^\mu \phi)^* (\phi^\dagger D_\mu \phi)\,.
\end{equation}
In fig.~\ref{fig:ew} right, we show likelihood contours in the plane of these
coefficients at the scale $m_Z$, in good agreement with results in the
literature \cite{Haller:2018nnx,Ellis:2018gqa}.

\subsection{Model-independent analysis of $b\to s\ell\ell$ transitions}

\begin{figure}
\centering
\includegraphics[width=0.5\textwidth]{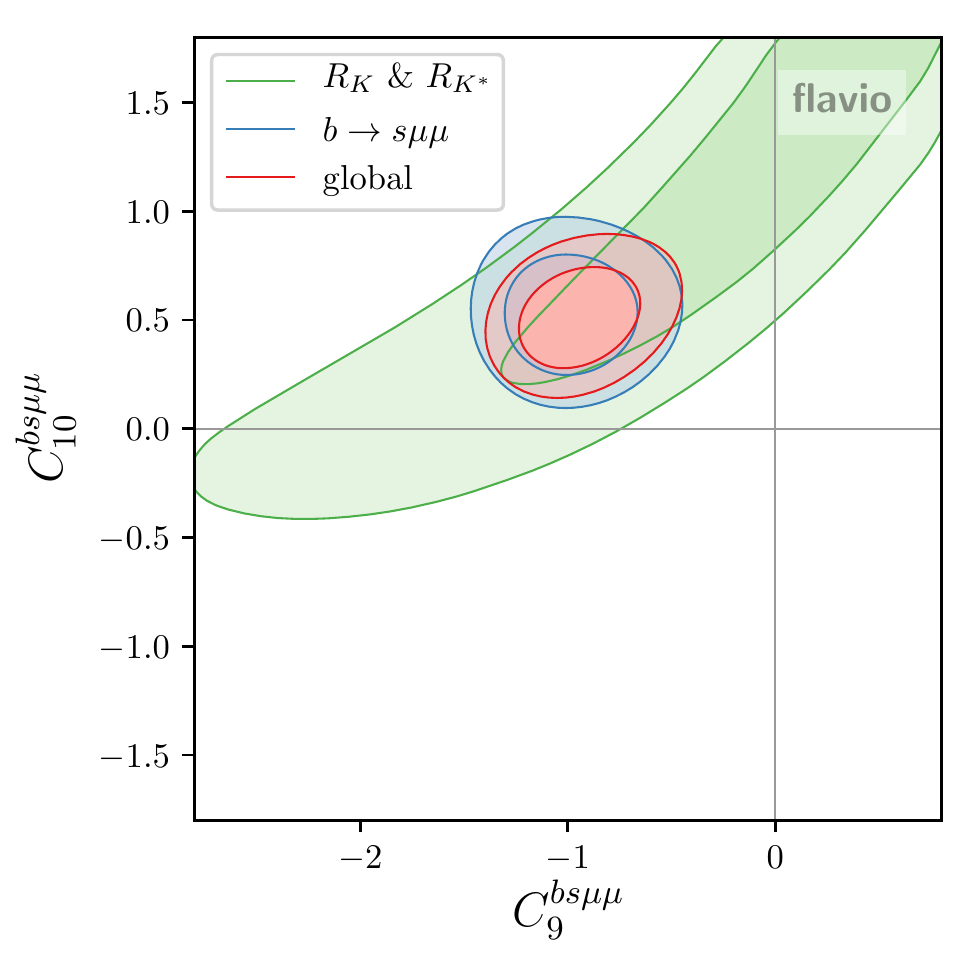}%
\includegraphics[width=0.5\textwidth]{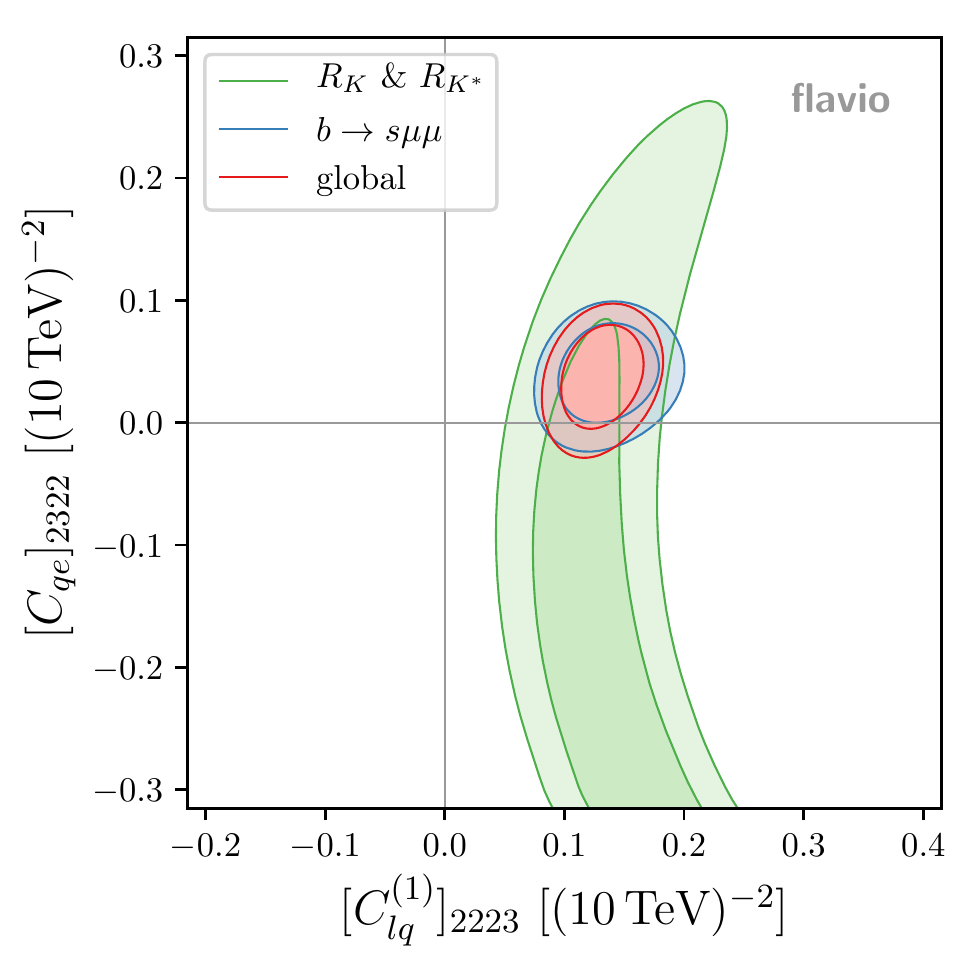}
\caption{Likelihood contours from $b\to s\mu\mu$ transitions and from $R_K$ and
$R_{K^*}$ in the space of the two WET Wilson coefficients $C_9^{bs\mu\mu}$ and $C_{10}^{bs\mu\mu}$
at the $b$ quark scale (left)
and the two SMEFT Wilson coefficients $[C_{lq}^{(1)}]_{2223}$ and $[C_{qe}]_{2322}$
at the scale 10~TeV.
All other Wilson coefficients are assumed to vanish.}
\label{fig:bsmumu_C9-C10}
\end{figure}

Model-independent fits of the WET Wilson coefficients $C_9^{bs\mu\mu}$ and $C_{10}^{bs\mu\mu}$ of the operators\footnote{Throughout, we use the WCxf convention \cite{Aebischer:2017ugx} of writing the effective
Lagrangian as $\mathcal L_\text{eff} = -\mathcal H_\text{eff} =\sum_{O_i= O_i^\dagger} C_i \, O_i + \sum_{O_i\neq O_i^\dagger} \left( C_i \, O_i + C^*_i \, O^\dagger_i\right)$ and include normalization factors directly in the definition of the operators.}
\begin{align}
	O_{9}^{bs\mu\mu}&=
  \frac{4G_F}{\sqrt{2}}V_{tb}V_{ts}^*\frac{e^2}{16\pi^2}
  (\bar s_L \gamma^\rho b_L) (\bar \mu \gamma_\rho \mu)
  \,,&
  O_{10}^{bs\mu\mu}&=
  \frac{4G_F}{\sqrt{2}}V_{tb}V_{ts}^*\frac{e^2}{16\pi^2}
  (\bar s_L \gamma^\rho b_L) (\bar \mu \gamma_\rho \gamma_5 \mu)\,,
\end{align}
play an important role in the NP interpretation of the $B\to K^*\mu^+\mu^-$,
$R_K$, and $R_{K^*}$ anomalies and have been performed by several groups
(for recent examples see
\cite{Altmannshofer:2017fio, Altmannshofer:2017yso, Ciuchini:2017mik,
Hurth:2017hxg, Capdevila:2017bsm}).
Since all relevant $b\to s\ell\ell$ observables are part of our global likelihood,
we can plot the well-known likelihood contour plots in the space of two WET
Wilson coefficients as a two-dimensional slice of the global likelihood.
In fig.~\ref{fig:bsmumu_C9-C10} left we plot contours in the $C_9^{bs\mu\mu}$-$C_{10}^{bs\mu\mu}$ plane,
assuming them to be real and
setting all other Wilson coefficients to zero.
The result is equivalent to \cite{Altmannshofer:2017fio, Altmannshofer:2017yso}
apart from the addition of the $\Lambda_b\to\Lambda\mu^+\mu^-$ decay.
In fig.~\ref{fig:bsmumu_C9-C10} right, we show the analogous plot for the SMEFT
Wilson coefficients $[C_{lq}^{(1)}]_{2223}$ and $[C_{qe}]_{2322}$ of the operators
\begin{equation}
	[O_{lq}^{(1)}]_{2223}= (\bar \ell_2 \gamma^\mu \ell_2) (\bar q_2 \gamma_\mu q_3)\,,\quad [O_{qe}]_{2322}= (\bar q_2 \gamma^\mu q_3) (\bar e_2 \gamma_\mu e_2)\,,
\end{equation}
that match at tree level onto $C_9^{bs\mu\mu}$ and $C_{10}^{bs\mu\mu}$ (cf.~\cite{Celis:2017doq}).

While the plot of the real parts of $C_9^{bs\mu\mu}$ and $C_{10}^{bs\mu\mu}$ is well known,
the global likelihood allows to explore arbitrary scenarios with real or complex
contributions to several Wilson coefficients.

\subsection{Model-independent analysis of $b\to c\tau\nu$ transitions}\label{sec:bctaunu}

Model-independent EFT analyses of $b\to c\tau\nu$ transitions relevant for
the $R_D$ and $R_{D^*}$ anomalies have been performed within the WET
\cite{Freytsis:2015qca,Celis:2016azn,Bhattacharya:2018kig,Celis:2012dk}
and SMEFT \cite{Feruglio:2018fxo,Hu:2018veh}.

\begin{figure}
\centering
\includegraphics[width=0.5\textwidth]{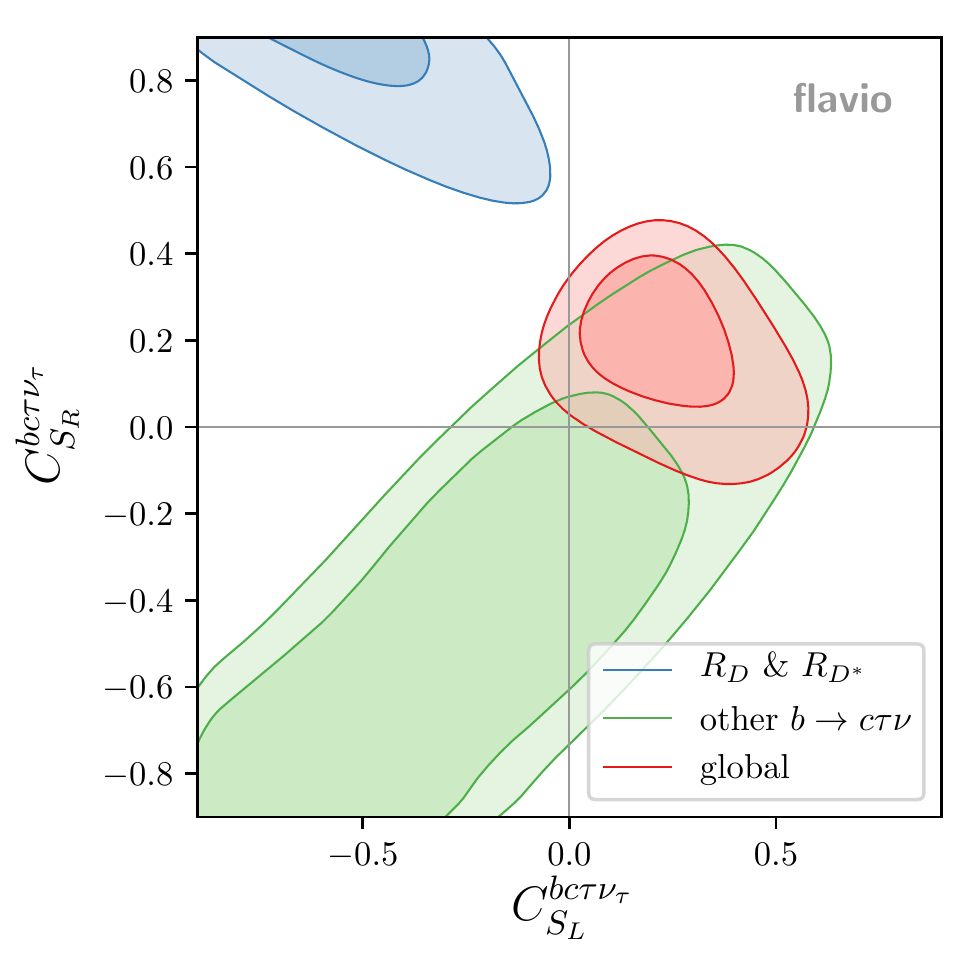}%
\includegraphics[width=0.5\textwidth]{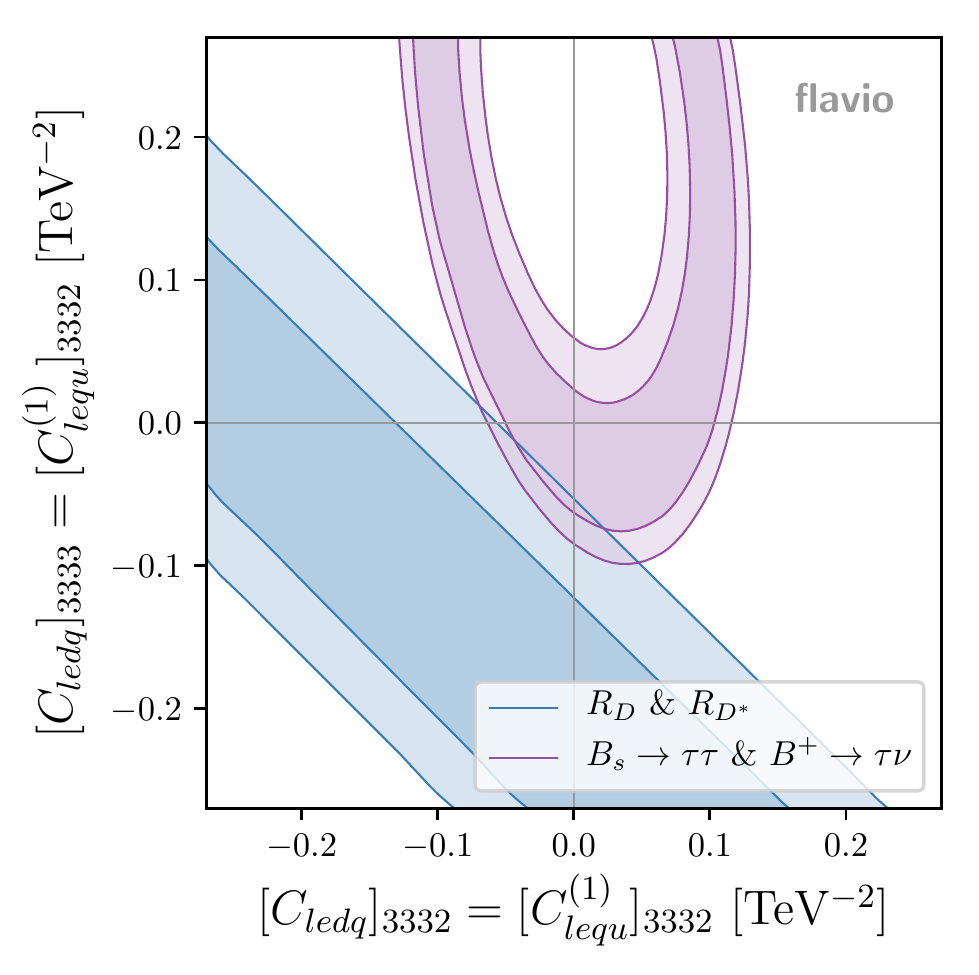}%
\caption{Left: Likelihood contours in the space of the $b\to c\tau\nu_\tau$ WET
scalar operators from $R_D$ and $R_{D^*}$ (blue), the combination of
$B_c\to\tau\nu$, $B\to D^{(*)}\tau\nu$ differential rates and $F_L(B\to D^*\tau\nu)$
(green) and the global likelihood (red).
Right: Likelihood contours for the SMEFT Wilson coefficients matching onto the WET scalar
operators for two choices of flavour indices, imposing the relation between
coefficients \eqref{eq:evade} that evades the $B_c\to\tau\nu$ constraint.
The purple region is allowed by $B_s\to\tau^+\tau^-$ and $B^+\to\tau\nu$.}
\label{fig:bctaunu_CSL-CSR}
\end{figure}

Within simple two-coefficient scenarios, an interesting case is the one with
new physics in the two WET Wilson coefficients $C_{S_L}^{bc\tau\nu_\tau}$ and $C_{S_R}^{bc\tau\nu_\tau}$.
The corresponding operators are defined by
\begin{align}
O_{S_L}^{bc\tau\nu_\tau}&=
-\frac{4G_F}{\sqrt{2}}V_{cb}
(\bar c_R b_L) (\bar \tau_R \nu_{\tau L})\,,
&
O_{S_R}^{bc\tau\nu_\tau} &=
-\frac{4G_F}{\sqrt{2}}V_{cb}
(\bar c_L b_R) (\bar \tau_R \nu_{\tau L})\,.
\end{align}
The constraint from $B_c\to\tau\nu$ \cite{Li:2016vvp, Akeroyd:2017mhr}
allows a solution to the $R_D$ anomaly only for $C_{S_L}^{bc\tau\nu_\tau}\approx C_{S_R}^{bc\tau\nu_\tau}$ and precludes a solution of the $R_{D^*}$ anomaly
\cite{Alonso:2016oyd}. Additional disjoint solutions in the 2D Wilson coefficient
space are excluded by the $B\to D\tau\nu$ differential distributions \cite{Celis:2016azn}.
Both effects are visible in figure~\ref{fig:bctaunu_CSL-CSR} left.
The preferred region is only improved slightly more than $2\sigma$ compared to
the SM, signaling that the $R_D$ and $R_{D^*}$ anomalies, that have a combined
significance of around $4\sigma$, cannot be solved simultaneously.

Even this less-than-perfect solution turns out to be very difficult to realize
in SMEFT. In fact, the immediate choice for SMEFT Wilson coefficients matching onto
$C_{S_L}^{bc\tau\nu_\tau}$ and $C_{S_R}^{bc\tau\nu_\tau}$
would be $[C_{ledq}]_{3332}$ and $[C_{lequ}^{(1)}]_{3332}$, respectively,
defined by the operators
\begin{equation}
	[O_{ledq}]_{3332}= (\bar \ell_3 e_3) (\bar d_3 q_2)\,,\quad [O_{lequ}^{(1)}]_{3332}= (\bar \ell_3^j e_3)\epsilon_{jk} (\bar q_3^k u_2)\,.
\end{equation}
However, $[C_{ledq}]_{3332}$ also generates the FCNC decay $B_s\to\tau^+\tau^-$,
and even though this has not been observed yet, the existing bound puts strong
constraints. Choosing instead $[C_{ledq}]_{3333}$, the Wilson coefficient has
to be larger by a factor $1/V_{cb}$ and leads to a sizable NP effect in the
decay $B^+\to\tau\nu_\tau$ based on the $b\to u\tau\nu$ transition.
These effects are demonstrated in fig.~\ref{fig:bctaunu_CSL-CSR} right,
where the relation between the left- and right-handed coefficients
that evades the $B_c\to\tau\nu$ constraint,
\begin{equation}
[C_{lequ}^{(1)}]_{3332}=[C_{ledq}]_{3332} + V_{cb} \, [C_{ledq}]_{3332} \,,
\label{eq:evade}
\end{equation}
has been imposed.

\begin{figure}
\centering
\includegraphics[width=0.5\textwidth]{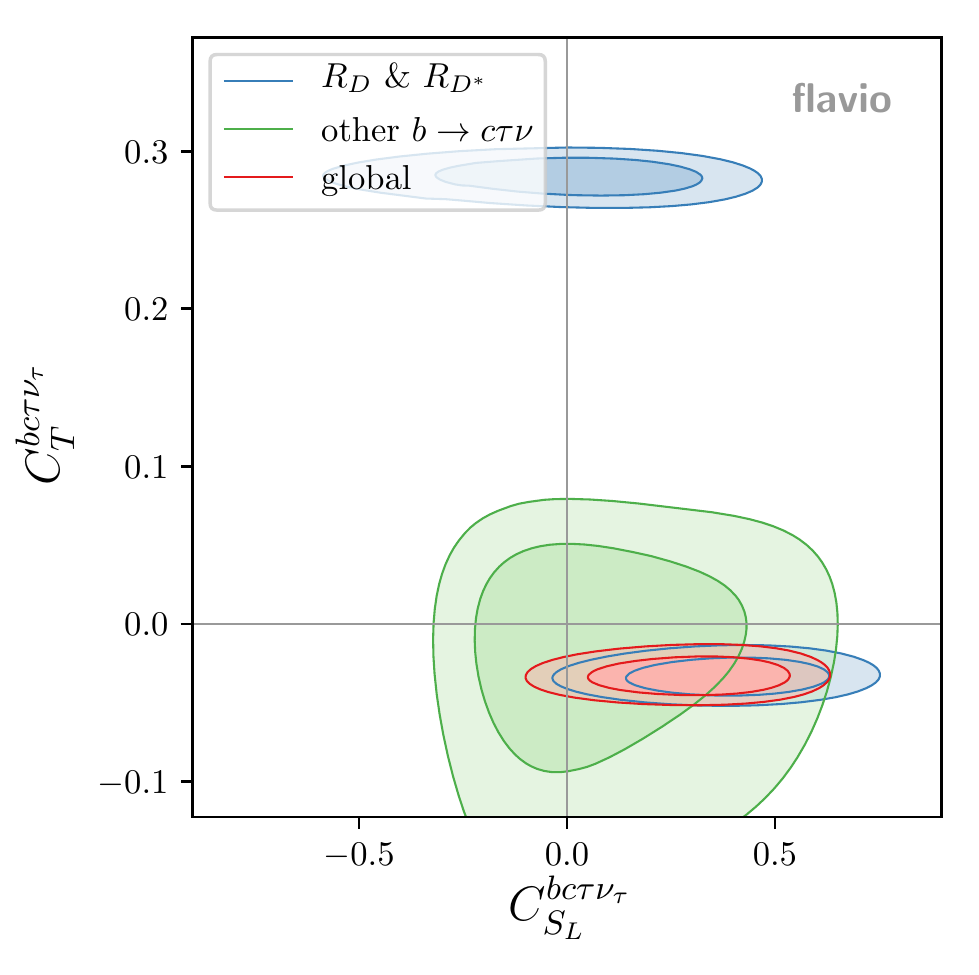}%
\includegraphics[width=0.5\textwidth]{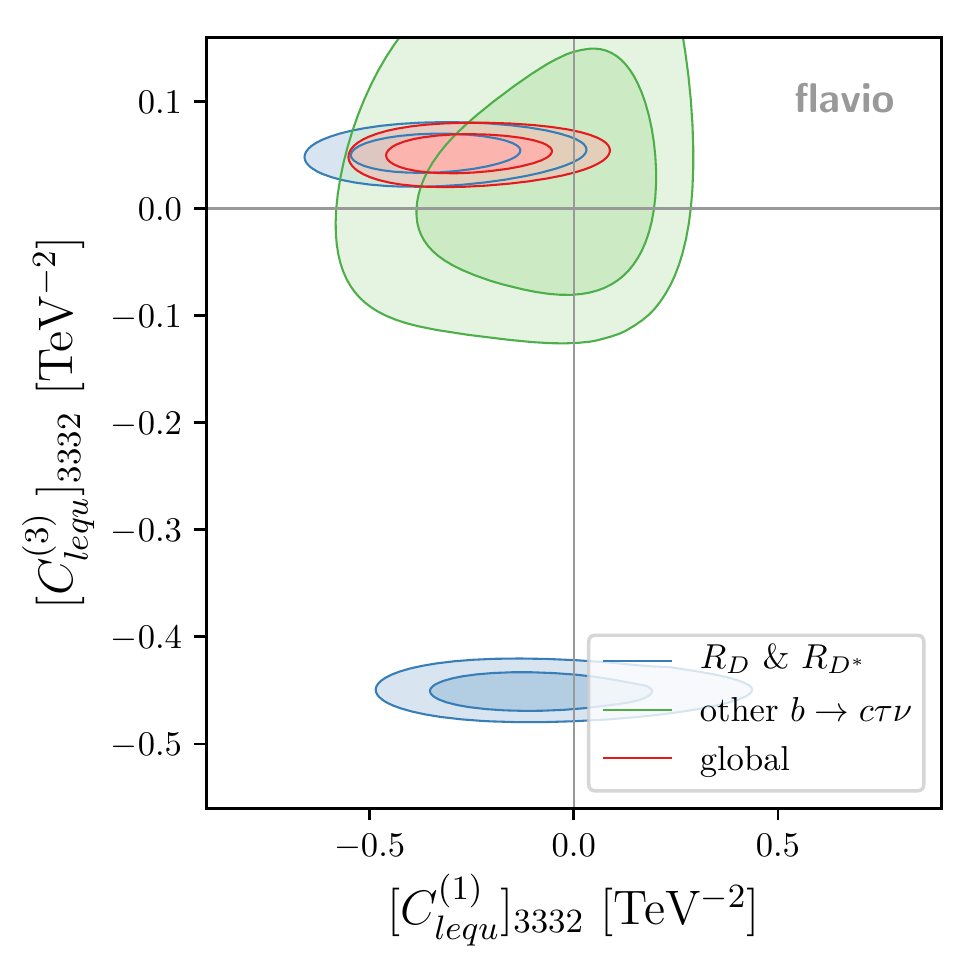}%
\caption{Left: Likelihood contours in the space of  $b\to c\tau\nu_\tau$ WET
scalar and tensor operator from $R_D$ and $R_{D^*}$ (blue), the combination of
$B_c\to\tau\nu$, $B\to D^{(*)}\tau\nu$ differential rates and $F_L(B\to D^*\tau\nu)$
(green) and the global likelihood (red).
Right: Likelihood contours for the SMEFT Wilson coefficients matching onto the WET scalar and tensor
operators.}
\label{fig:bctaunu_CSL-T}
\end{figure}

Another interesting two-coefficient scenario is the one with new physics in  $C_{S_L}^{bc\tau\nu_\tau}$ and the tensor Wilson coefficient $C_{T}^{bc\tau\nu_\tau}$, that are generated with the relation $C_{S_L}^{bc\tau\nu_\tau}=-4 C_T^{bc\tau\nu_\tau}$ at the matching scale in the scalar singlet leptoquark $S_1$
scenario\footnote{See also \cite{Becirevic:2018afm,Dekens:2018bci} for the $R_2$ leptoquark scenario with complex couplings, which generates the Wilson coefficients with the relation $C_{S_L}^{bc\tau\nu_\tau}= 4 C_T^{bc\tau\nu_\tau}$.} \cite{Freytsis:2015qca}.
In fig.~\ref{fig:bctaunu_CSL-T} left, we show the constraints on this scenario. A new finding, that to our knowledge has not been discussed in the literature before, is that a second, disjoint
solution with large tensor Wilson coefficient is excluded by
the new, preliminary Belle measurement of the longitudinal polarization fraction $F_L$ in $B\to D^*\tau\nu$ \cite{Adamczyk:2019wyt}, which is included in our likelihood and enters the green contour in the plot.

The analogous scenario in SMEFT with the Wilson coefficients
$[C_{lequ}^{(1)}]_{3332}$ and $[C_{lequ}^{(3)}]_{3332}$
 does not suffer from the constraints of the scenario with $C_{S_R}$ as the operator involves a right-handed up-type quark, so is not related by $SU(2)_L$ rotations to any FCNC operator in the down-type sector. Here the Wilson coefficient $[C_{lequ}^{(3)}]_{3332}$ is defined by the operator
 \begin{equation}
	 [O_{lequ}^{(3)}]_{3332}= (\bar \ell_3^j \sigma_{\mu\nu}e_3)\epsilon_{jk} (\bar q_3^k \sigma^{\mu\nu} u_2)\,.
\end{equation}
Consequently, the constraints are qualitatively similar as for WET, as shown in fig.~\ref{fig:bctaunu_CSL-T} right.
Note that we have included the anomalous magnetic muon and tau in our likelihood, but do not find a relevant constraint for this simple scenario (cf.~\cite{Feruglio:2018fxo}).

\subsection{$B$ anomalies from new physics in top}

\begin{figure}
\centering
\includegraphics[width=0.5\textwidth]{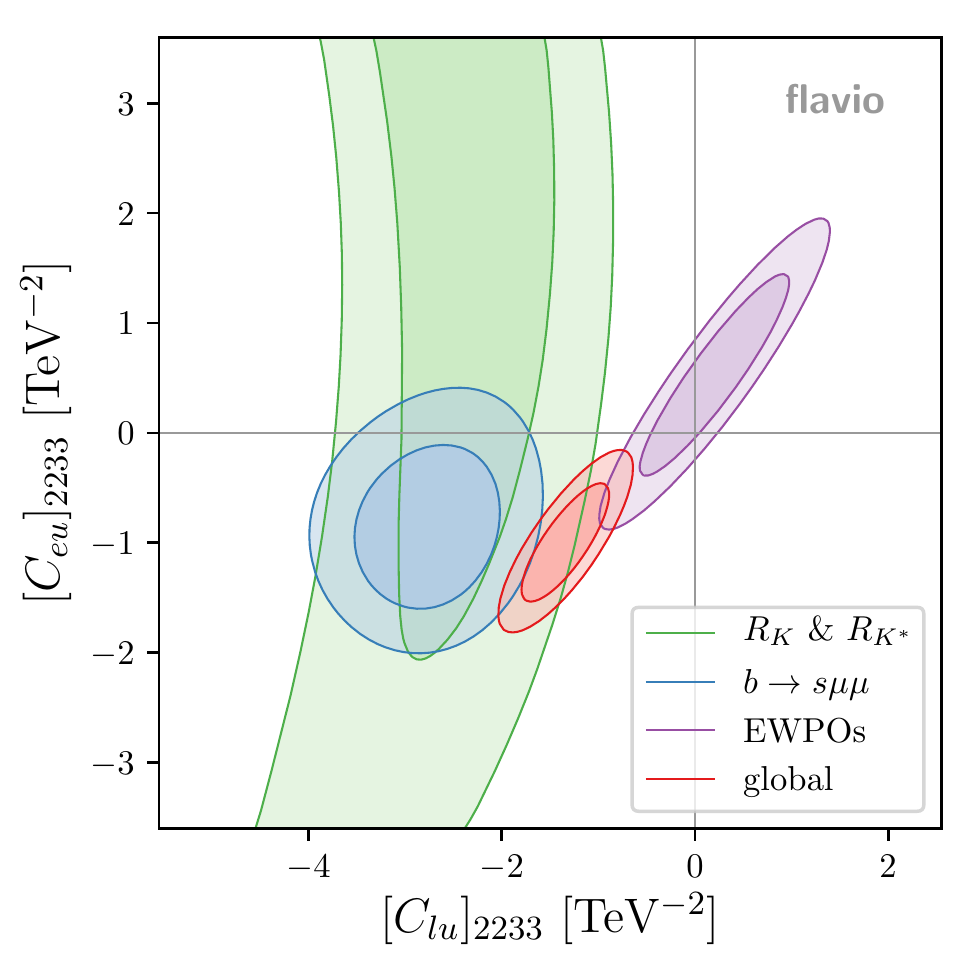}%
\includegraphics[width=0.5\textwidth]{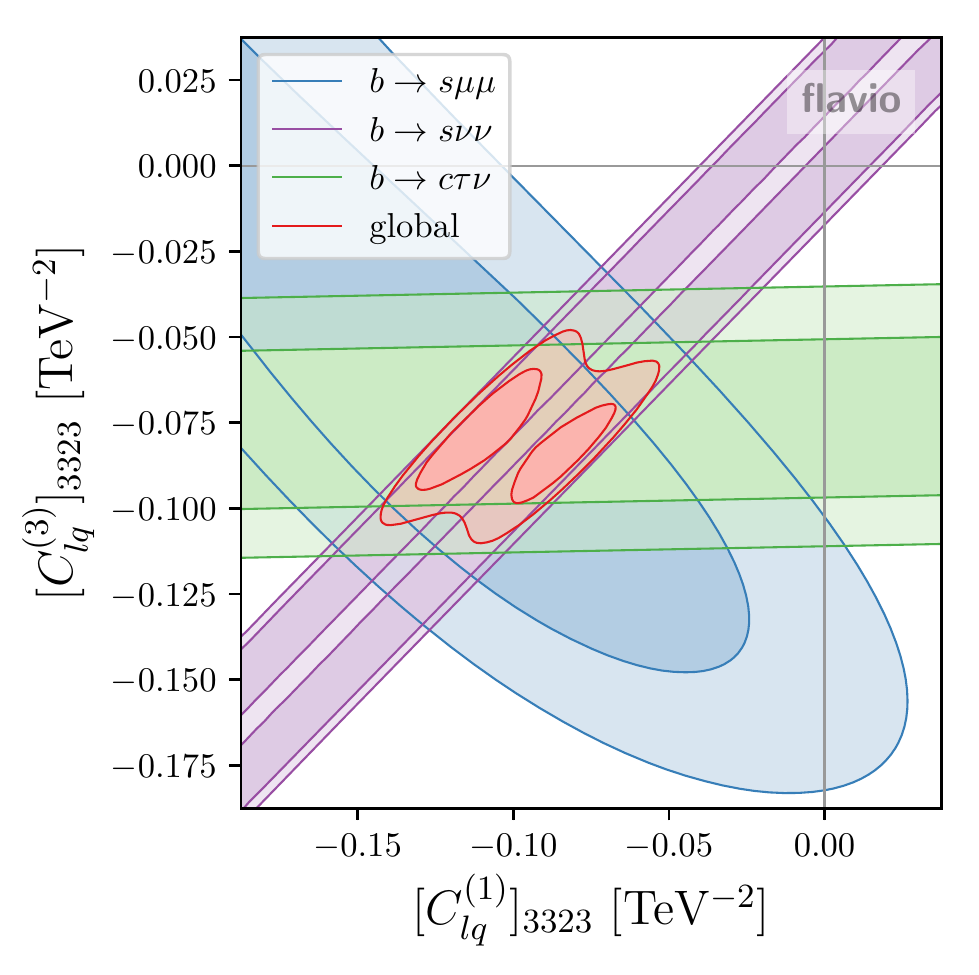}
\caption{Left:
Likelihood contours in the plane of the SMEFT Wilson coefficients $[C_{lu}]_{2233}$ and $[C_{eu}]_{2233}$ at 1~TeV.
Right:
Likelihood contours in the plane of the SMEFT Wilson coefficients $[C_{lq}^{(1)}]_{3323}$ and $[C_{lq}^{(3)}]_{3323}$ at 1~TeV.}
\label{fig:lq1_lq3}
\label{fig:Ceu-Clu}
\end{figure}

A new physics effect in the semi-leptonic SMEFT operator $[C_{lu}]_{2233}$
involving two left-handed muons and two right-handed top quarks
was suggested in \cite{Celis:2017doq} as a solution to the neutral-current
$B$ anomalies, as it induces a $b\to s\mu\mu$ transition at low-energies
via electroweak renormalization effects. This effect can be realized in
$Z'$ models \cite{Kamenik:2017tnu}.
It was subsequently shown however that the effect is strongly constrained
by the effects it induces in $Z\to\mu^+\mu^-$
\cite{Camargo-Molina:2018cwu},
which can be cancelled by a simultaneous contribution to $[C_{eu}]_{2233}$.
The result obtained there can be reproduced with our likelihood by plotting
likelihood contours in the plane of these two Wilson coefficients at
1~TeV, see fig.~\ref{fig:Ceu-Clu} left. Here the operators for the Wilson coefficients
$[C_{eu}]_{2233}$ and $[C_{lu}]_{2233}$ are given by
\begin{equation}
        [O_{eu}]_{2233}= (\bar e_2 \gamma_\mu e_2) (\bar u_3 \gamma^\mu u_3)\,,\quad [O_{lu}]_{2233}
	= (\bar \ell_2 \gamma_\mu \ell_2) (\bar u_3 \gamma^\mu u_3)\,.
\end{equation}

At $2\sigma$, the two constraints cannot be brought into agreement and the global likelihood is optimized at an intermediate point.

\subsection{Tauonic vector operators for charged-current anomalies}

The SMEFT operator $[C_{lq}^{(3)}]_{3323}$ can
interfere coherently with the SM contribution to the $b\to c\tau\nu_\tau$
process, does not suffer from any CKM suppression
and is thus a good candidate to explain the $R_D$ and $R_{D^*}$
anomalies.
However, a strong constraint is given by the limits on the
$B\to K^{(*)}\nu\bar\nu$ decays, which can
receive contributions from tau neutrinos
\cite{Buras:2014fpa}.
At tree level and in the absence of RG effects, this constraint can be
avoided in models that predict $[C_{lq}^{(3)}]_{3323} = [C_{lq}^{(1)}]_{3323}$.
The modification of this constrain in the presence of SMEFT RG effects above
the EW scale can be seen in fig.~\ref{fig:lq1_lq3} right. The Wilson coefficients
$[C_{lq}^{(1)}]_{3323}$ and $[C_{lq}^{(3)}]_{3323}$ are defined by the operators
\begin{equation}
	[O_{lq}^{(1)}]_{3323}= (\bar \ell_3 \gamma_\mu \ell_3) (\bar q_2 \gamma^\mu q_3)\,,\quad [O_{lq}^{(3)}]_{3323}
        = (\bar \ell_3 \gamma_\mu \tau^I \ell_3) (\bar q_2 \gamma^\mu \tau^I q_3)\,.
\end{equation}
Recently, it has been pointed out that the large value of the tauonic Wilson
coefficient required to accommodate $R_D$ and $R_{D^*}$ induces a
LFU contribution to the $b\to s\ell\ell$ Wilson coefficient $C_9$ at the
one loop level \cite{Crivellin:2018yvo},
an effect discussed for the first time in \cite{Bobeth:2011st}.
This effect can be reproduced by taking into account the
SMEFT and QED running. In agreement with \cite{Crivellin:2018yvo}, fig.~\ref{fig:lq1_lq3} right
shows that the $b\to s\mu\mu$ anomalies as well as $R_D$ and $R_{D^*}$
can be explained simultaneously
without violating the $B\to K^{(*)}\nu\bar\nu$ constraint.
Note that $R_K$ and $R_{K^*}$ are SM-like in this simple scenario.

\subsection{Flavour vs. electroweak constraints on modified top couplings}

\begin{figure}
\centering
\includegraphics[width=0.5\textwidth]{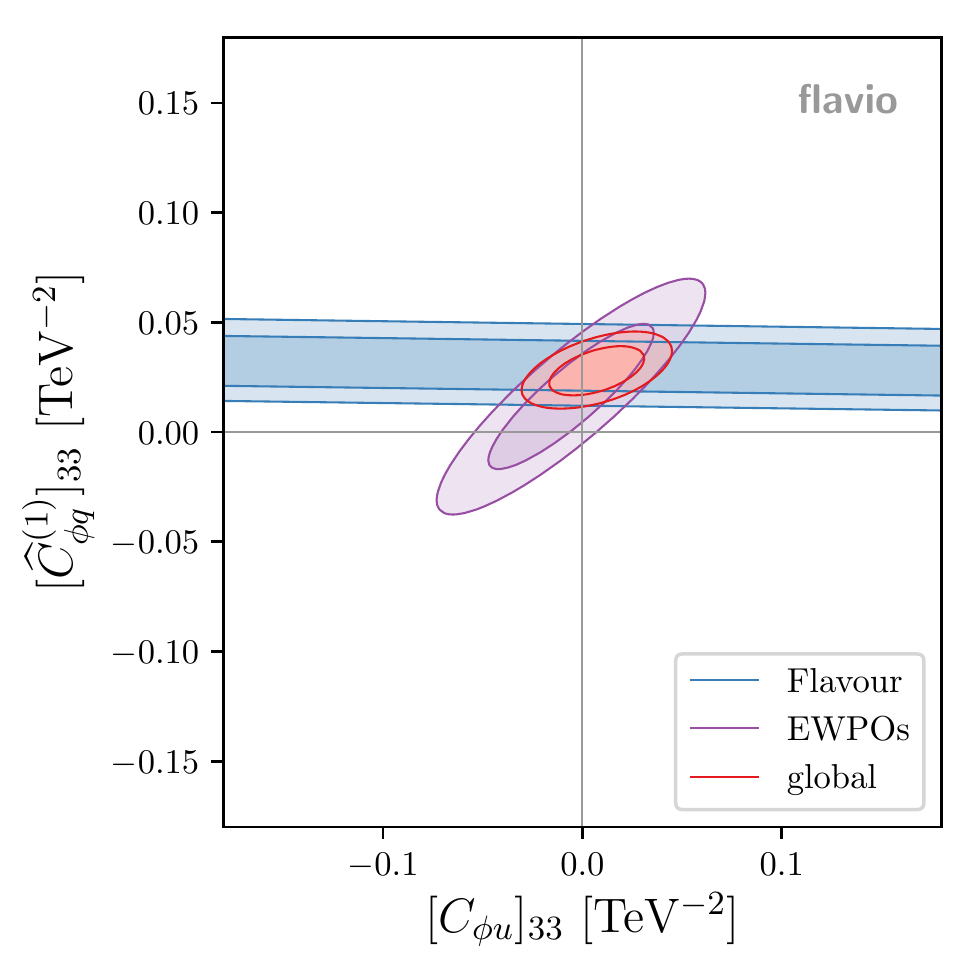}%
\includegraphics[width=0.5\textwidth]{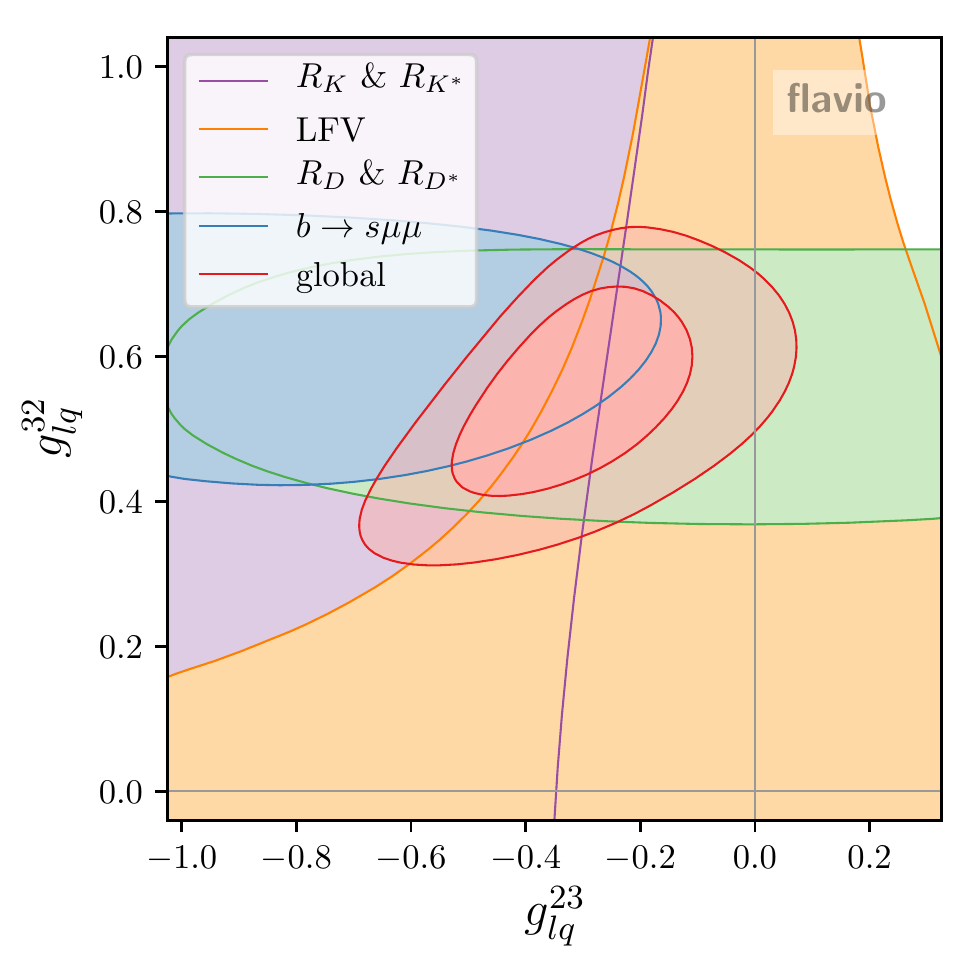}
\caption{Left:
Likelihood contours in the space of the two SMEFT Wilson coefficients that correspond to modified $Z$ couplings to left- or right-handed top quarks. The constraints from flavour physics (dominated by $B_s\to\mu^+\mu^-$) and EWPOs are complementary.
Right:
Likelihood contours in the plane of the couplings
$g_{lq}^{23}$ and $g_{lq}^{32}$ of the $U_1$ vector leptoquark model at $1\sigma$ level. }
\label{fig:U1}
\label{fig:Ztt}
\end{figure}

Another nice example of the interplay between flavour and EW precision constraints was presented in \cite{Brod:2014hsa}.
The Wilson coefficients corresponding to modified couplings of the $Z$ boson to left- and right-handed top quarks,
$[\widehat{C}_{\phi q}^{(1)}]_{33}$ (in the Warsaw-up basis where the up-type quark mass matrix is diagonal, see appendix~\ref{app:conv})
and $[C_{\phi u}]_{33}$, defined by
\begin{equation}
	[O_{\phi q}^{(1)}]_{33}= (\phi^\dagger i \overset{\text{$\leftrightarrow$}}{D_\mu} \phi) (\bar q_3 \gamma^\mu q_3)\,,\quad [O_{\phi u}]_{33}
        = (\phi^\dagger i \overset{\text{$\leftrightarrow$}}{D_\mu} \phi) (\bar u_3 \gamma^\mu u_3)\,,
\end{equation}
induce on the one hand effects in flavour-changing neutral currents in $K$ and $B$ physics such as $B_s\to\mu^+\mu^-$ and $K^+\to\pi^+\nu\bar\nu$, on the other hand radiatively induce a correction to the Wilson coefficient of the bosonic operator $O_{\phi D}$ that corresponds to the oblique $T$ parameter. This interplay is reproduced in fig.~\ref{fig:Ztt} left.

\subsection{Vector leptoquark solution to the $B$ anomalies}

The $U_1$ vector leptoquark transforming as $(3, 1)_{\frac{2}{3}}$
under the SM gauge group is the phenomenologically most
successful single-multiplet scenario that
simultaneously solves the charged- and neutral-current
$B$ anomalies \cite{Barbieri:2015yvd} as it does not give rise to $b\to s\nu\bar\nu$ at
tree level \cite{Buras:2014fpa} and is still allowed by direct searches \cite{Buttazzo:2017ixm}.

Writing the leptoquark's couplings to left-handed fermions as
\begin{equation}
\mathcal L_{U_1} \supset g_{lq}^{ji} \,\left(\bar q_L^i \gamma^\mu l_L^j\right) \, U_{\mu}
+\text{h.c.}\,,
\end{equation}
the solution of the neutral-current $B$ anomalies depends on
the coupling combination $g_{lq}^{22}g_{lq}^{23*}$,
while the charged-current anomalies require a sizable
$g_{lq}^{32}g_{lq}^{33*}$.\footnote{While the coupling
$g_{lq}^{33}$ would be sufficient to enhance $R_D$ and $R_{D^*}$, this
solution is disfavoured by direct searches \cite{Faroughy:2016osc}.}

Fig.~\ref{fig:U1} right shows the likelihood contours for the $U_1$ scenario in the
plane $g_{lq}^{32}$ vs. $g_{lq}^{23}$ where we have fixed
\begin{align}
m_{U_1} &= 2\,\text{TeV}
\,,&
g_{lq}^{33} &= 1
\,,&
g_{lq}^{22} &= 0.04^2 \approx V_{cb}^2
\,.
\end{align}

The LFV decays are important constraints to determine the allowed
pattern of the couplings $g_{lq}^{ij}$ \cite{Kumar:2018kmr}.
This can be seen from the orange contour in Fig.~\ref{fig:U1} right,
which shows constraints from BR$(B \to K \tau^+ \mu^-)$,
BR$(B\to K\mu^+ \tau^-)$, and BR$(\tau \to \phi \mu)$.
The former two depend on the coupling combinations
$g_{lq}^{33} g_{lq}^{22}$ and $g_{lq}^{23} g_{lq}^{32}$ respectively,
whereas the latter is controlled by $g_{lq}^{32} g_{lq}^{22}$.

\subsection{$B$ anomalies from third generation couplings}

An interesting EFT scenario for the combined explanation of the $B$ anomalies in the
neutral and charged currents is to assume TeV-scale NP in the purely third generation
operators $[O_{lq}^{(1)}]_{3333}$ and $[O_{lq}^{(3)}]_{3333}$ in the interaction
basis \cite{Bhattacharya:2014wla}.
The effective Lagrangian in the Warsaw basis (as defined in  WCxf \cite{Aebischer:2017ugx}) can be written as
\begin{equation}
	\mathcal{L}_\text{eff} \supset \frac{\lambda^{ij}_\ell \lambda^{kl}_q }{\Lambda^2}\left (C_1 	\bar \ell_{iL} \gamma_\mu \ell_{jL} \bar q_{kL} \gamma^\mu q_{lL}  +
	C_3 \bar \ell_{iL} \gamma_\mu \tau^I \ell_{jL} \bar q_{kL} \gamma^\mu\tau^I q_{lL}  \right ),
\label{eq:3333}
\end{equation}
where $\lambda_\ell$ and $\lambda_q$ parameterize the mismatch between the interaction
basis and the basis where the down-type quark mass matrix is diagonal.

As required by the data,
purely third generation operators induce a large NP contribution in $b\to c\tau \bar\nu$,
whereas in $b\to s\mu^+\mu^-$ comparatively smaller effects arise
due to mixing on rotating to the mass basis.

In this context, ref.~\cite{Feruglio:2017rjo} found that electroweak corrections can lead to important
effects in $Z$ pole observables and $\tau$ decays challenging this simultaneous
solution for the $B$ anomalies.
Since all the relevant observables as well as the SMEFT RG evolution are included in
our global likelihood, we can reproduce these conclusions.

\begin{figure}
\centering
\includegraphics[width=0.5\textwidth]{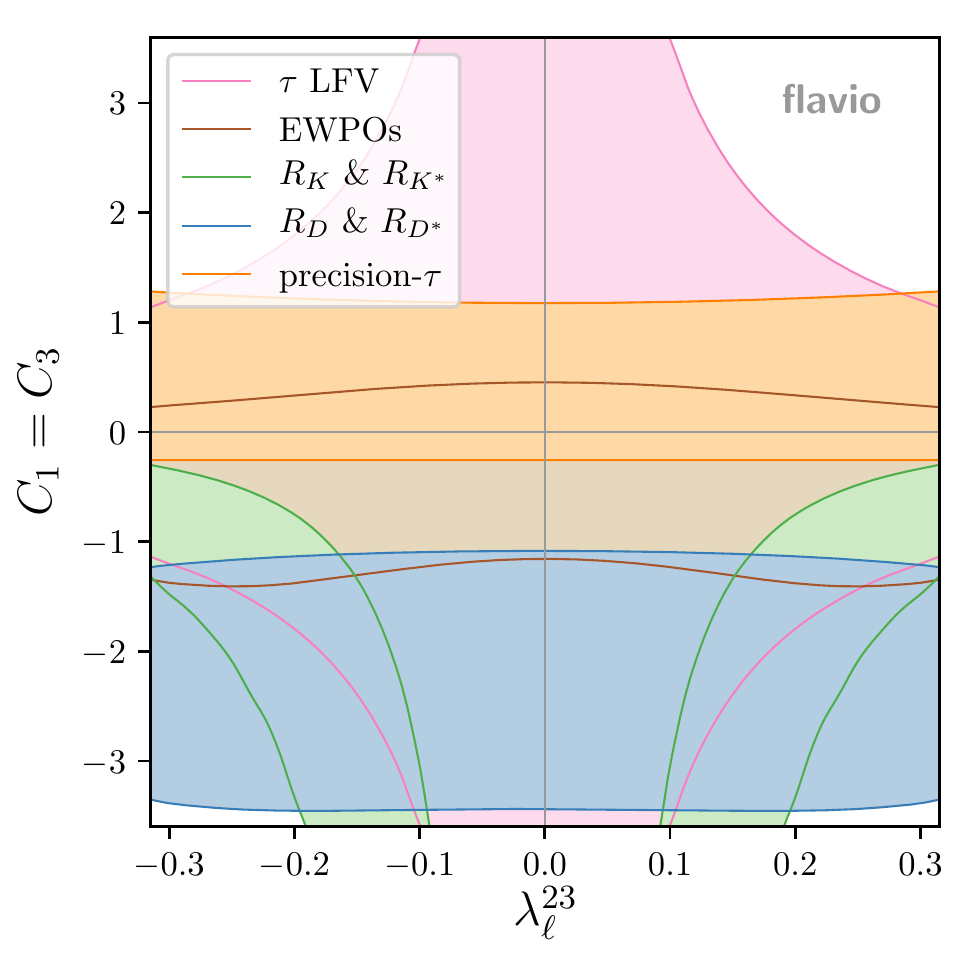}
	\caption{Likelihood contours at $2\sigma$ for various sets of observables for the scenario with mostly third generation couplings defined in eq.~\eqref{eq:3333}.}
\label{fig:3333}
\end{figure}

\begin{figure}
\centering
\includegraphics[width=0.6\textwidth]{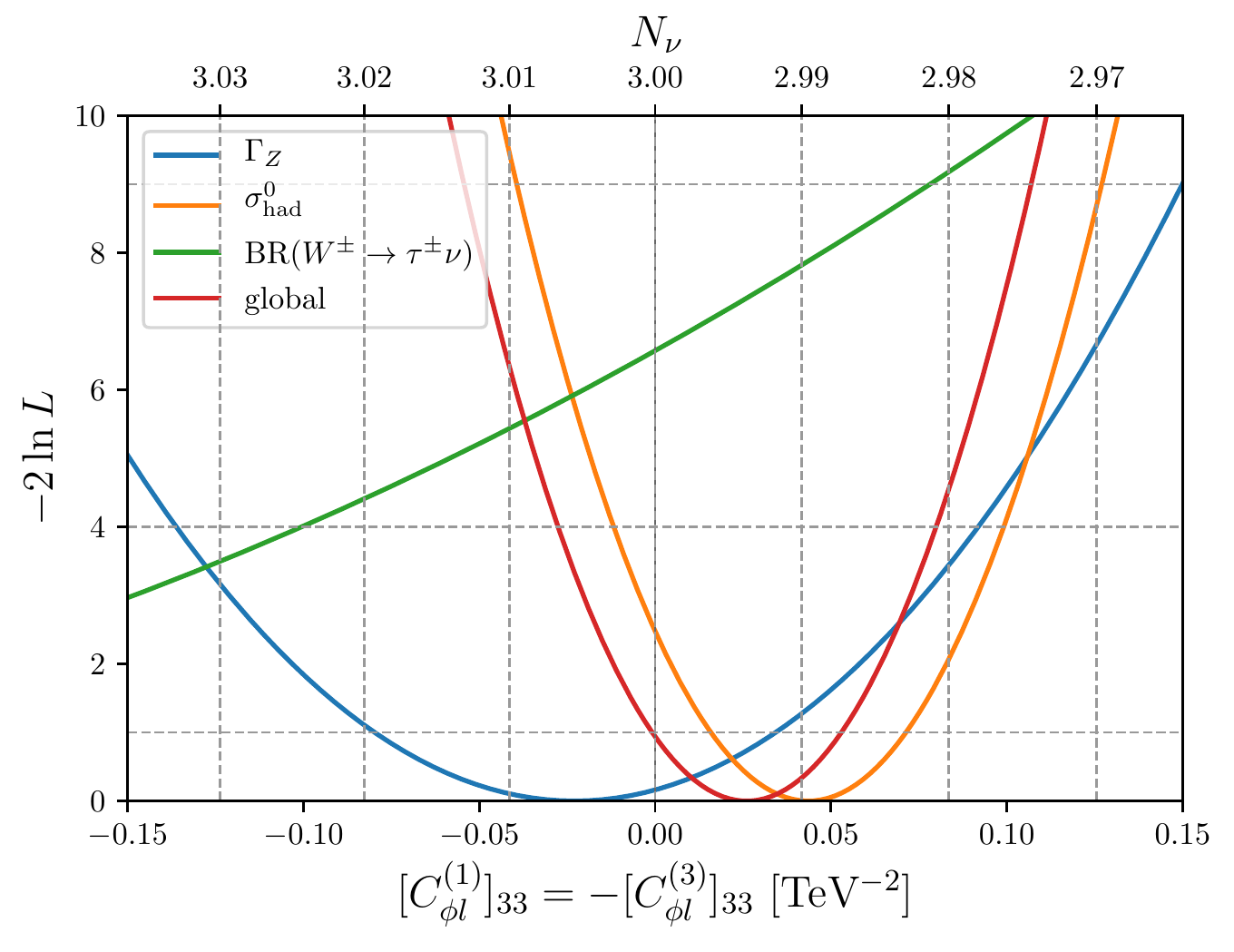}
\caption{Contributions to the log-likelihood $\ln L$ from the observables sensitive to a shift in the tau neutrino's electroweak coupling and their combination, relative to their respective extrema. The axis on top shows the effective number of neutrino species that would correspond to the relative modification of the $Z$ boson's invisible width.}
\label{fig:gLnu33}
\end{figure}

In figure~\ref{fig:3333} we show likelihood contours of the
various observables in the plane of $C_1=C_3$ and $\lambda_\ell^{23}$. We have set
$\Lambda=1$ TeV, $\lambda_q^{23}=-0.008$ and the relations
$\lambda_{\ell,q}^{22}=(\lambda_{\ell, q}^{23})^2$,
$\lambda_{\ell}^{33} = \lambda_{q}^{33} =1$ are imposed\footnote{The overall conclusion are unchanged even if we vary the parameter $\lambda_q^{23}$.}. Like \cite{Feruglio:2017rjo},
we find  that the $2\sigma$ region for the precision $\tau$ decays
does not overlap with
the $2\sigma$ regions preferred by $R_{D^{(*)}}$ and $R_{K^{(*)}}$.
Furthermore, the $2\sigma$ region from EWPOs has only a very small
overlap with the $2\sigma$ region preferred by $R_{D^{(*)}}$.
Compared to \cite{Feruglio:2017rjo}, we find a stronger constraint on the
shift in the tau neutrino's electroweak coupling. We have traced this difference
back to the treatment of the LEP constraint in the invisible $Z$ width.
\cite{Feruglio:2017rjo} uses the invisible $Z$ width extracted by LEP \cite{ALEPH:2005ab},
corresponding to the effective number of neutrino species $N_\nu=2.984\pm0.008$, which favours a destructive interference with the SM at $2\sigma$. This number is obtained exclusively from $\sigma_\text{had}$, using the measured value of $R_l$ (assuming lepton flavour universality). Our treatment differs in two respects. First, since both $\sigma_\text{had}$ and $R_{e,\mu,\tau}$ are among the observables in the likelihood, we
effectively use the SM values of $R_{e,\mu,\tau}$ rather than the measured ones when shifting only the neutrino coupling.
This leads to a value $N_\nu=2.990\pm0.007$, in better agreement with the SM value.
Second, we include additional observables sensitive to the electroweak coupling of the
tau neutrino, notably the total $Z$ width $\Gamma_Z$ and the $W\to\tau\nu$ branching ratio\footnote{We find the total $W$ width to not give a relevant constraint.}. Figure~\ref{fig:gLnu33} shows the contributions of these three observables to the likelihood as well as their combination.
While $\sigma_\text{had}$ alone favours a slightly shifted coupling (less significant than $2\sigma$ due to the different treatment of $R_l$), the combined constraints are in agreement with the SM at $1\sigma$ and more strongly disfavour a positive shift in $[C_{\phi l}^{(1)}]_{33}=-[C_{\phi l}^{(3)}]_{33}$.

\section{Usage}\label{sec:py}

The global likelihood is accessed via the Python package \smelli{}
(\uline{SME}FT \uline{l}ike\uline{li}hood).
Given a working installation of Python version 3.5 or above, the package
can be installed with the simple command
\\\noindent
\begin{minipage}{\linewidth}
\begin{lstlisting}[language=iPython, basicstyle=\normalsize\ttfamily]
python3 -m pip install smelli --user
\end{lstlisting}
\end{minipage}
that downloads it from the Python package archive (PyPI) along with all required
dependencies and installs it in the user's home directory (no administrator
privileges required).
The source code of the package can be browsed via a public Github repository\footnote{\url{https://github.com/smelli/smelli}}.

As with any Python package, \smelli{} can be used as library imported from
other scripts, directly in the command line interpreter,
 or in an interactive session. For interactive use, we recommend
the Jupyter notebook\footnote{See \url{https://jupyter.org}.} that runs in a web browser. In all cases,
the first step is to import the package and to initialize the class
\texttt{GlobalLikelihood},
\\\noindent
\begin{minipage}{\linewidth}
\begin{lstlisting}[language=iPython, basicstyle=\normalsize\ttfamily]
import smelli
gl = smelli.GlobalLikelihood()
\end{lstlisting}
\end{minipage}
The initialization function takes two optional arguments:
\begin{itemize}
\item The argument \texttt{eft} (default value: \texttt{\textquotesingle SMEFT\textquotesingle})
can be set to \texttt{\textquotesingle WET\textquotesingle} to obtain a likelihood in the parameter
space of WET rather than SMEFT Wilson coefficients. In this case
EWPOs are ignored.
\item The argument \texttt{basis} allows to select a different WCxf basis
(default: \texttt{\textquotesingle Warsaw\textquotesingle} in the case of SMEFT,
\texttt{\textquotesingle flavio\textquotesingle} in the case of WET).
\end{itemize}

By default, \smelli{} uses the leading logarithmic approximation
for the SMEFT RG evolution, since it is faster than the full
numerical solution of the coupled RGEs. This behaviour can be
changed by setting the corresponding option of the \wilson{}
package \textit{after} importing \smelli{}, e.g.
\begin{lstlisting}[language=iPython, basicstyle=\normalsize\ttfamily]
import smelli, wilson
wilson.Wilson.set_default_option('smeft_accuracy', 'integrate')
\end{lstlisting}

The next step is to select a point in Wilson coefficient space
by using the \texttt{parameter\char`_point} method.
The Wilson coefficients must be provided in the EFT and basis fixed in the first step.
There are three possible input formats:
\begin{itemize}
\item a Python dictionary (containing Wilson coefficient name/value pairs) and an input scale,
\item as a \wcxf{} data file in YAML or JSON format (specified by its file path as a string),
\item as an instance of \texttt{wilson.Wilson} defined by the \texttt{wilson} package.
\end{itemize}
Using the first option,
fixing the Wilson coefficient $[C_{lq}^{(1)}]_{2223}$ to $10^{-8}\,\text{GeV}^{-2}$
at the scale 1\,TeV is achieved with
\begin{lstlisting}[language=iPython, basicstyle=\normalsize\ttfamily]
glp = gl.parameter_point({'lq1_2223': 1e-8}, scale=1000)
\end{lstlisting}
Note that, consistently with the WCxf format, all dimensionful values are  expected to be in appropriate powers of GeV.
The same result could be achieved with a \wcxf{} file in YAML format,
\begin{lstlisting}[language=yaml, basicstyle=\normalsize\ttfamily\color{red!70!black}]
eft: SMEFT
basis: Warsaw
scale: 1000
values:
  lq1_2223: 1e-8
\end{lstlisting}
that is imported as
\begin{lstlisting}[language=iPython, basicstyle=\normalsize\ttfamily]
glp = gl.parameter_point('my_wc.yaml')
\end{lstlisting}

The variable \texttt{glp} defined above holds an instance of the
\texttt{GlobalLikelihoodPoint} class that gives access to the results
for the chosen parameter point. Its most important methods are
\begin{itemize}
\item \texttt{glp.log\char`_likelihood\char`_global()} returns the numerical value of the
logarithm of the likelihood minus its SM value $\ln \Delta L$, i.e. the logarithm of the
likelihood ratio or $-\Delta\chi^2/2$ when writing the likelihood as
$L=e^{-\chi^2/2}$.
\item \texttt{glp.log\char`_likelihood\char`_dict()} returns
a dictionary with the contributions to  $\ln\Delta L$ from the individual
products in \eqref{eq:nfSL}.
\item \texttt{glp.obstable()} returns a \texttt{pandas.DataFrame} table-like
object that lists all the individual observables with their experimental
and theoretical central values and uncertainties ordered by their ``pull''
that is defined by $\sqrt{|\Delta \chi^2_i|}$ where $-\chi^2_i/2$ is their
individual contribution to the log-likelihood neglecting all correlations.
This table can be useful to get a better understanding of the likelihood value
at a given point. However it should be used with caution. In particular,
the log-likelihood is {\em not} equal to the sum of the individual contributions
obtained from the pulls, as there can be significant correlations between them.
Also, the uncertainties listed in this table can be inaccurate in the case of
strongly non-Gaussian probability distributions.
\end{itemize}

The observables with the highest pulls in the SM as obtained by this method are shown for illustration in table~\ref{tab:pulls_SM}.
A few comments are in order.
\begin{itemize}
\item The largest deviation is in the branching ratio of $B_s\to\phi\mu^+\mu^-$ at low $q^2$, where the prediction relies strongly on the form factors from \cite{Straub:2015ica}.
\item The observable $R_{\tau \ell}(B\to D^{\ast}\ell^+\nu)$
is nothing but $R_{D^\ast}$\footnote{The observable $R_D$ is found to have a pull of $2.1\sigma$ and thus does not appear in table~\ref{tab:pulls_SM}.}, while $\langle R_{\mu e} \rangle(B^\pm\to K^\pm \ell^+\ell^-)^{[1.0,6.0]}$ and $\langle R_{\mu e} \rangle(B^0\to K^{\ast 0}\ell^+\ell^-)^{[a, b]}$
are $R_K$ and $R_{K^\ast}$, respectively.
$\langle A_\text{FB}^{\ell h}\rangle(\Lambda_b\to\Lambda \mu^+\mu^-)$ is denoted $K_6$ in \cite{Aaij:2018gwm}.
We use the full observable names as defined in \flavio{} here.
\item The SM uncertainties in $\epsilon'/\epsilon$ are entirely due to matrix elements from lattice QCD \cite{Blum:2015ywa,Bai:2015nea}.
\end{itemize}

\begin{table}[t]
\resizebox{\textwidth}{!}{
\renewcommand{\arraystretch}{1.3}
\begin{tabularx}{1.36\textwidth}{lllll}
\hline
Observable ~~~~~~~~~~~~~~~~~~~~~~~~~~~~ & Prediction ~~~~~~~~~~~~~~~~~~~~~~~~~~~ & Measurement ~~~~~~~~~~~~~~~~~~~~~~~ & ~~~~~~~~~~~~~~~~~~ & Pull\\
\hline
$\langle \frac{d\overline{\text{BR}}}{dq^2} \rangle(B_s\to \phi \mu^+\mu^-)^{[1.0,6.0]}$ & $\left(5.39 \pm 0.66\right) \times 10^{-8}\ \frac{1}{\text{GeV}^2}$ & $\left(2.57 \pm 0.37\right) \times 10^{-8}\ \frac{1}{\text{GeV}^2}$ & \cite{Aaij:2015esa,CDF:2012qwd} & $3.7\sigma$\\
$a_\mu$ & $\left(1.1659182 \pm 0.0000004\right) \times 10^{-3}$ & $\left(1.1659209 \pm 0.0000006\right) \times 10^{-3}$ & \cite{Tanabashi:2018oca} & $3.4\sigma$\\
$\langle P_5^\prime\rangle(B^0\to K^{\ast 0}\mu^+\mu^-)^{[4,6]}$ & $-0.757 \pm 0.074$ & $-0.21 \pm 0.15$ & \cite{Aaboud:2018krd,Aaij:2015oid} & $3.3\sigma$\\
$R_{\tau \ell}(B\to D^{\ast}\ell^+\nu)$ & $0.255$ & $0.306 \pm 0.018$ & \cite{Hirose:2016wfn,Huschle:2015rga,Lees:2013uzd,Sato:2016svk} & $2.9\sigma$\\
$\langle A_\text{FB}^{\ell h}\rangle(\Lambda_b\to\Lambda \mu^+\mu^-)^{[15,20]}$ & $0.1400 \pm 0.0074$ & $0.249 \pm 0.040$ & \cite{Aaij:2018gwm} & $2.7\sigma$\\
$\epsilon^\prime/\epsilon$ & $\left(-0.3 \pm 5.9\right) \times 10^{-4}$ & $\left(1.66 \pm 0.23\right) \times 10^{-3}$ & \cite{Patrignani:2016xqp} & $2.7\sigma$\\
$\langle R_{\mu e} \rangle(B^\pm\to K^\pm \ell^+\ell^-)^{[1.0,6.0]}$ & $1.001$ & $0.745 \pm 0.097$ & \cite{Aaij:2014ora} & $2.6\sigma$\\
$\text{BR}(W^\pm\to \tau^\pm\nu)$ & $0.1084$ & $0.1138 \pm 0.0021$ & \cite{Schael:2013ita} & $2.6\sigma$\\
$\langle R_{\mu e} \rangle(B^0\to K^{\ast 0}\ell^+\ell^-)^{[1.1,6.0]}$ & $1.00$ & $0.68 \pm 0.12$ & \cite{Aaij:2017vbb} & $2.5\sigma$\\
$A_\text{FB}^{0, b}$ & $10.31 \times 10^{-2}$ & $\left(9.92 \pm 0.16\right) \times 10^{-2}$ & \cite{ALEPH:2005ab} & $2.4\sigma$\\
$\langle R_{\mu e} \rangle(B^0\to K^{\ast 0}\ell^+\ell^-)^{[0.045,1.1]}$ & $0.93$ & $0.65 \pm 0.12$ & \cite{Aaij:2017vbb} & $2.4\sigma$\\
\hline
\end{tabularx}

}
\caption{Observables with highest pulls in the SM.}
\label{tab:pulls_SM}
\end{table}

\section{Conclusions}\label{sec:concl}

In this paper we have presented a likelihood function in the space of dimension-6
Wilson coefficients of the SMEFT. This function is made publicly available in the
form of the Python package \smelli{}, building on the existing public codes
\flavio{} and \wilson{}.
At present, the likelihood includes numerous observables from $B$ and $K$ decays, EWPOs, neutral meson mixing, LFV and CP violating processes and many more, counting a total of 265 observables.
We have demonstrated its validity and usefulness by reproducing various results given in the literature. In passing, we have also pointed out new results, in particular the
fact that one of the two possible solutions to the $R_D$ and $R_{D^*}$
anomalies involving the tensor operator is excluded by the
recent Belle measurement of the longitudinal polarization fraction in $B\to D^*\tau\nu$, which is included in our likelihood (see section~\ref{sec:bctaunu}).

Clearly, the 265 observables do not constrain the entire 2499-dimensional
parameter space of SMEFT Wilson coefficients yet.
Observables that are still missing include
\begin{itemize}
\item Higgs production and decay \cite{Khachatryan:2016vau,Butter:2016cvz,Ellis:2018gqa} including $h\to\gamma\gamma$ \cite{Hartmann:2015aia,Hartmann:2015oia,Dedes:2018seb},
\item top physics \cite{AguilarSaavedra:2010zi,Buckley:2015nca,deBlas:2015aea,AguilarSaavedra:2018nen},
\item further low-energy observables \cite{Falkowski:2017pss}, such as neutrino scattering, parity violation in atoms, and quark pair production in $e^+e^-$ collisions,
\item non-leptonic $B$ decays \cite{Bobeth:2014rra},
\item rare $D$ decays \cite{Fajfer:2015mia,deBoer:2015boa,Petrov:2016kmb,deBoer:2017que},
\item further hadronic tau decays \cite{Celis:2013xja,Cirigliano:2018dyk},
\item beta decay
\cite{Cirigliano:2012ab,Alioli:2017ces,Gonzalez-Alonso:2018omy},
\item paramagnetic EDMs
\cite{Cirigliano:2016nyn,Dekens:2018bci},
\end{itemize}
among others. Furthermore, as discussed at the end of section~\ref{sec:setup},
a major limitation of the nuisance-free likelihood we have constructed is that
several classes of observables cannot be incorporated consistently without
scanning over nuisance parameters. The next step in generalizing our results would
be to allow the 4 parameters of the CKM matrix to vary in addition to the Wilson
coefficients. This would make it possible to consistently include
semi-leptonic charged-current $B$ and $K$ decays with general NP effects.

We hope that the groundwork laid by us will allow the community to build a
more and more global likelihood as a powerful tool to constrain UV models from
precision measurements.

\section*{Note added}

After our preprint was published, ref.~\cite{Descotes-Genon:2018foz} appeared that proposes a procedure for a consistent treatment of the CKM matrix in the presence of dimension-6 contributions. Implemented in our framework, this would allow to include semi-leptonic charged-current decays without the need to scan over nuisance parameters.

\section*{Acknowledgements}

\noindent
We thank Wolfgang Altmannshofer, Christoph Bobeth, Ilaria Brivio, Andreas Crivellin, Martin Jung, Aneesh Manohar,
and Jordy de Vries for discussions. We thank Alejandro Celis, M{\'e}ril Reboud, and Olcyr Sumensari for pointing out typos.
We thank Martín González-Alonso, Admir Greljo, and Marco Nardecchia for useful comments.
The work of D.\,S. and J.\,A. is supported by the DFG cluster of excellence
``Origin and Structure of the Universe''.  The work of J.\,K. is
financially supported by NSERC of Canada.

\appendix

\section{Conventions and caveats}\label{app:conv}

In this appendix, we fix some of our conventions necessary for a consistent usage of the likelihood function and recall a few caveats when dealing with different bases of Wilson coefficients.

\subsection{SMEFT flavour basis}

Within SMEFT, a complete basis of gauge-invariant operators has to be chosen. Here we adopt the ``Warsaw basis'', as defined in \cite{Grzadkowski:2010es}. This basis is defined in the interaction basis above the electroweak scale.
Having fixed this basis, there remains a continuous choice for the basis in flavour space, parameterized by the $U(3)^5$ flavour symmetry of unitary fermion field rotations.
 Anticipating spontaneous symmetry breaking at the EW scale motivates the choice of basis closely related to the mass eigenbasis. Due to the misalignment of the up- and down sector, a choice has to be made concerning the diagonality of the mass matrices. Above the electroweak scale, only five instead of the usual six fermion-field rotation matrices can be used to diagonalize the three mass matrices of the SM. This is because left-handed up- and down-type quarks form doublets of the unbroken $SU(2)_L$ symmetry and therefore have to be rotated by the same matrix. Denoting the quark rotations by
\begin{equation}
  \psi \to U_{\psi} \psi\,, \hspace{1cm} \psi=q,u,d,
\end{equation}
leads to the following quark masses including dimension-6 corrections \cite{Dedes:2017zog}:

\begin{align}
  M_u&=\frac{v}{\sqrt{2}}U_q^{\dag}\left(Y_u-\frac{v^2}{2}C_{u\phi}\right)U_u\,, \\
  M_d&=\frac{v}{\sqrt{2}}U_q^{\dag}\left(Y_d-\frac{v^2}{2}C_{d\phi}\right)U_d\,.
\end{align}
  Choosing the up-type mass matrix to be diagonal results in the ``Warsaw-up'' basis, such defined in the Wilson coefficient exchange format (WCxf) \cite{Aebischer:2017ugx}. This is equivalent of choosing $U_q=U_{u_L}=U_{d_L}V^{\dag}$, where $U_{u_L},U_{d_L}$ are the rotation matrices of the left-handed up- and down-quarks, which diagonalize the corresponding mass matrices, and $V$ is the CKM matrix. Therefore, in the Warsaw-up basis, the mass matrices read:
  \begin{align}
    M_u&=\hat M_u\,, \\
    M_d&=V\hat M_d\,,
  \end{align}
with the diagonal matrices $\hat M_u,\hat M_d$.

Furthermore, all operators containing left-handed down-type quarks are rotated by $V$ compared to the usual Warsaw basis, after having absorbed factors of $U_{u_L}$ in the Wilson coefficients. For example the operator
$O_{qe}^{ijkl} = (\bar q_i \gamma_\mu q_j)(\bar e_k \gamma^\mu e_l)$ in the Warsaw basis
\begin{equation}
  C_{qe}^{ijkl}O_{qe}^{ijkl}=C_{qe}^{ijkl}(\bar u_L^i\gamma_\mu u_L^j +\bar d_L^i\gamma_\mu d_L^j)(\bar e_R^k\gamma^\mu e_R^l)\,,
\end{equation}
will read after performing quark rotations and choosing the Warsaw-up basis (denoted by a hat):
\begin{align}
  C_{qe}^{ijkl}O_{qe}^{ijkl}\to &\,C_{qe}^{ijkl}\left((U^{\dag}_{u_L})_{ia}(U_{u_L})_{jb}\bar u_L^a\gamma_\mu u_L^b\, +(U^{\dag}_{d_L})_{ia}(U_{d_L})_{jb}(\bar d_L^a\gamma_\mu d_L^b)\right)(\bar e_R^k\gamma^\mu e_R^l) \\\notag
  = &\underbrace{C_{qe}^{ijkl}(U^{\dag}_{d_L})_{ia}(U_{d_L})_{jb}}_{=:\,\widehat C_{qe}^{ijkl}}\underbrace{(\bar u_L^a\gamma_\mu u_L^b +(V^{\dag})_{af}(V)_{bg}\bar d_L^f\gamma_\mu d_L^g)(\bar e_R^k\gamma^\mu e_R^l)}_{=:\,\widehat O_{qe}^{ijkl}} \\\notag
  =& \,\widehat C_{qe}^{ijkl}\widehat O_{qe}^{ijkl}\,.
\end{align}

\subsection{Non-redundant SMEFT basis}

To derive the complete anomalous dimension matrix \cite{Alonso:2013hga,Jenkins:2013zja,Jenkins:2013wua}  as well as the complete tree-level matching \cite{Jenkins:2017dyc} of the SMEFT onto WET it is convenient to allow for all possible flavour combinations in the SMEFT operators. Nevertheless, many operators are symmetric under the exchange of flavour indices. This is for example the case for four-fermi operators consisting of two identical fermion currents, like the operator $O_{dd}^{ijkl}$:
\begin{equation}
O_{dd}^{ijkl}=(\bar d^i_R\gamma_{\mu}d^j_R)(\bar d^k_R\gamma^{\mu}d^l_R)\,,
\end{equation}
for which clearly
\begin{equation}
O_{dd}^{abcd}=O_{dd}^{cdab}\,.
\end{equation}
For the computation of physical processes it can however be more convenient to choose a minimal basis, in which all operators are independent of each other. Such a choice avoids unwanted symmetry factors in the Lagrangian. For example the Lagrangian written in a redundant basis featuring the operator $O_{dd}$ would contain terms of the form
\begin{align}
\mathcal{L}_\text{red}\supset& \,C^{1122}_{dd} O^{1122}_{dd}+C^{2211}_{dd} O^{2211}_{dd} \\\notag
=& \,C^{1122}_{dd} O^{1122}_{dd}+C^{2211}_{dd} O^{1122}_{dd} \\\notag
=& \,\left(C^{1122}_{dd}+C^{2211}_{dd}\right)O^{1122}_{dd} \\\notag
=& \,2C^{1122}_{dd} O^{1122}_{dd}\\\notag
=& \,2C^{2211}_{dd} O^{2211}_{dd}\,,
\end{align}
whereas in a non-redundant basis only one flavour combination is taken into account:
\begin{equation}
  \mathcal{L}\supset C^{1122}_{dd} O^{1122}_{dd}\,,
\end{equation}
and the redundant contribution is not part of the Lagrangian.

 Furthermore, such symmetry factors can also enter the beta functions of the Wilson coefficients, since contributions from operators that are not linearly independent are counted individually. For example the beta function of the Wilson coefficient $C_{dd}$ in a redundant SMEFT basis contains terms of the form \cite{Alonso:2013hga}:

 \begin{equation}
   \dot C_{dd}^{prst} =\frac{2}{3}g_1^2N_cy_d^2(C_{dd}^{prww}+C_{dd}^{wwpr})\delta_{st}+\ldots\,\,\,.
 \end{equation}
Therefore, operators with symmetric index combinations, like f.e. $prst=aabb,\, a\neq b$, get the same contribution from $C_{dd}^{aaww}$ and $C_{dd}^{wwaa}$, whereas in a non-redundant basis, only one of such contributions is present. The operator corresponding to the second contribution is not included in the Lagrangian.

This issue has to be taken into account when using the results of \cite{Alonso:2013hga,Jenkins:2013zja,Jenkins:2013wua,Jenkins:2017jig,Jenkins:2017dyc} together with a non-redundant basis, like the one defined in \cite{Celis:2017hod}. All operators of the non-redundant basis exhibiting such symmetries have to be divided by their corresponding symmetry factor $S$ before the running and multiplied by $S$ after the running to cancel the effect of the redundant operators in the RGEs. Similar comments apply to the matching at the EW scale and the running below the EW scale.

Moreover, the choice of basis has to be made before making it minimal by discarding redundant operators, since a basis change can reintroduce redundant operators. Looking at the example of $O^{(1),prst}_{qq}$ in the Warsaw basis with diagonal up quark mass matrix (denoted with a hat) and diagonal down quark mass matrix (no hat), respectively, one finds for the index combination $prst=1122$ \cite{Aebischer:2015fzz}:
\begin{equation}
  \widehat O_{qq}^{(1),1122}=V_{ui}\,V^*_{uj}\,V_{ck}\,V^*_{cl}\, O_{qq}^{(1),ijkl}\,.
\end{equation}
The operator $\widehat O_{qq}^{(1),1122}$ in the Warsaw-up basis therefore depends in particular on the operator $O_{qq}^{(1),1122}$ and its redundant counterpart $O_{qq}^{(1),2211}$.

We stress that, being based on WCxf, the input to our likelihood function always refers to the basis {\em without} any redundant operators.

\subsection{Definitions}

A frequently overlooked ambiguity is the sign convention for the covariant derivative, that affects the overall sign of all dipole and triple gauge boson operators in both SMEFT and WET (see e.g.~\cite{Grzadkowski:2010es}). For definiteness, we specify our conventions here:
\begin{align}
  D_\mu\psi&=\partial_\mu+ieQ_\psi A_\mu+ig_sT^AG^A_\mu\,,\label{eq:cov} \\
  F_{\mu\nu}&=\partial_\mu A_\nu - \partial_\nu A_\mu\,, \\
  G_{\mu\nu}^A &= \partial_\mu G_\nu^A - \partial_\nu G_\mu^A-g_s f^{ABC}G_\mu^B G_\nu^C\,,\\
  \sigma_{\mu\nu}&= \frac{i}{2}\left[\gamma_\mu,\gamma_\nu\right]\,.
\end{align}
This sign convention for the covariant derivative is prevalent in the flavour physics literature and corresponds to the ``usual'' sign of the $b\to s\gamma$ dipole Wilson coefficient in the SM, but differs from several textbooks, see \cite{Romao:2012pq} for an overview. The convention for $\sigma_{\mu\nu}$ is also the most common one, but there are notable exceptions, e.g.~\cite{Buras:2001ra}.

With these conventions, one obtains the following relation between the  effective Lagrangian in the WCxf flavio basis
 \begin{equation}
 \mathcal{L}\supset i \bar \psi \slashed{D} \psi
 + \left[
 C_\gamma^\psi ~(\bar \psi \sigma_{\mu\nu}P_R\psi)~ F^{\mu\nu}
 + C_g^\psi ~(\bar \psi \sigma_{\mu\nu}T^AP_R \psi)~ G^{A\,\mu\nu}
 +\text{h.c.}\right]
 \end{equation}
 and the the anomalous magnetic moment of a fermion $\psi$ with electric charge $Q_\psi$,
\begin{equation}
  a_\psi = - \frac{4m}{eQ_\psi} \, \text{Re}(C_\gamma^\psi)\,.
\end{equation}

\section{$\tau\to \ell V$ decays}\label{app:tau_lV}

In the following, we summarize the full tree-level results of the $\tau\to \ell V$ decay width $\Gamma_{\tau\to \ell V}$ in the WET, where $V\in\{\rho,\phi\}$ is a vector meson and $\ell\in\{e,\mu\}$ is a lepton.
The decay width can be expressed in terms of the squared amplitude $\overline{|\mathcal{M}_{\tau\to \ell V}|}^2$, which has been averaged over initial spins and summed over final spins and polarizations.
One finds (cf.~\cite{Agashe:2014kda})
\begin{equation}
 \Gamma_{\tau\to \ell V}=\frac{\sqrt{\lambda(m_\tau^2,m_\ell^2,m_V^2)}}{16\,\pi\,m_\tau^3}\,\overline{|\mathcal{M}_{\tau\to \ell V}|}^2,
\end{equation}
where
\begin{equation}\label{eq:triangle_lambda}
\lambda(a,b,c)=a^2+b^2+c^2-2(ab+ac+bc)
\end{equation}
is the Källén function\cite{Kallen1964ElementaryParticlePhysics}.

\subsection{Squared amplitudes}

The $\tau\to\ell V$ matrix element due to generic couplings of the vector meson to the leptonic vector current can be written as
\begin{equation}\label{eq:matrix_element_vector}
\mathcal{M}^\text{VC}_{\tau\to\ell V} = \epsilon_{\mu}^*(p_V)\, \bar{\ell}(p_\ell)\, \gamma^{\mu} \left(g_{L}^{\tau \ell V}\, P_L + g_{R}^{\tau \ell V}\, P_R\right) \tau(p_\tau),
\end{equation}
where $p_\tau$, $p_\ell$, and $p_V$ are the momenta of $\tau$, $\ell$, and~$V$, respectively, and $g_{L}^{\tau \ell V}$ and $g_{R}^{\tau \ell V}$ are effective coupling constants.
Squaring this matrix element, averaging over initial spins, and summing over final spins and polarizations yields
\begin{equation}
 \begin{aligned}
\overline{|\mathcal{M}^\text{VC}_{\tau\to \ell V}|}^2=
 \frac{1}{2} \Bigg\{
&
\left(\left| g_{L}^{\tau \ell V}\right| ^2+\left| g_{R}^{\tau \ell V}\right| ^2\right) \left(\frac{\left(m_\tau^2-m_\ell^2\right)^2}{m_V^2}+m_\tau^2+m_\ell^2-2 \,m_V^2\right)
\\&
-12\,m_\tau\,m_\ell \,
\Re\Bigg(
g_{R}^{\tau \ell V}\,\left(g_{L}^{\tau \ell V}\right)^{\!*}
\Bigg)
\Bigg\}.
 \end{aligned}
\end{equation}
The $\tau\to\ell V$ matrix element due to generic couplings of the vector meson to the leptonic tensor current can be written as\footnote{Our convention for the epsilon tensor is $\epsilon^{0123}=-\epsilon_{0123}=1$.}
\begin{equation}\label{eq:matrix_element_tensor}
\begin{aligned}
\mathcal{M}^\text{TC}_{\tau\to\ell V}
&=
i\,p_V^\alpha\, \epsilon^{*\beta}(p_V)\,
\\
&\times
\bar{\ell}(p_\ell)\, \sigma^{\mu\nu} \left(
g_{\alpha\mu}\,g_{\beta\nu}
\left(g_{TL}^{\tau \ell V}\, P_L + g_{TR}^{\tau \ell V}\, P_R\right)
+
\frac{i}{2}\,\epsilon_{\alpha\beta\mu\nu}
\left(\tilde{g}_{TL}^{\tau \ell V}\, P_L + \tilde{g}_{TR}^{\tau \ell V}\, P_R\right)
\right) \tau(p_\tau),
\end{aligned}
\end{equation}
where $g_{TL}^{\tau \ell V}$, $g_{TR}^{\tau \ell V}$, $\tilde{g}_{TL}^{\tau \ell V}$, and $\tilde{g}_{TR}^{\tau \ell V}$ are effective coupling constants.
Squaring this matrix element, averaging over initial spins, and summing over final spins and polarizations yields
\begin{equation}
 \begin{aligned}
\overline{|\mathcal{M}^\text{TC}_{\tau\to \ell V}|}^2
=
 \frac{m_V^2}{2} \Bigg\{
 &
\left(\frac{2\left(m_\tau^2-m_\ell^2\right)^2}{m_V^2}-m_\tau^2-m_\ell^2-m_V^2\right)
\left(
 \left| g_{TL}^{\tau \ell V}-\tilde{g}_{TL}^{\tau \ell V}\right| ^2
+\left| g_{TR}^{\tau \ell V}+\tilde{g}_{TR}^{\tau \ell V}\right| ^2
\right)
\\&
-12\,m_\tau\,m_\ell \,
\Re\Bigg(
    \left( g_{TR}^{\tau \ell V}+\tilde{g}_{TR}^{\tau \ell V}\right)
    \left( g_{TL}^{\tau \ell V}-\tilde{g}_{TL}^{\tau \ell V}\right)^{\!\!*}
\Bigg)
\Bigg\}.
 \end{aligned}
\end{equation}
The full $\tau\to\ell V$ amplitude $\mathcal{M}^\text{full}_{\tau\to \ell V}$ is given by the sum of the vector current amplitude $\mathcal{M}^\text{VC}_{\tau\to\ell V}$ and the tensor current amplitude $\mathcal{M}^\text{TC}_{\tau\to\ell V}$,
\begin{equation}
 \mathcal{M}^\text{full}_{\tau\to \ell V}
 =
 \mathcal{M}^\text{VC}_{\tau\to \ell V}
 +
 \mathcal{M}^\text{TC}_{\tau\to \ell V}.
\end{equation}
Squaring the full amplitude, averaging over initial spins, and summing over final spins and polarizations yields
\begin{equation}
  \overline{|\mathcal{M}^\text{full}_{\tau\to \ell V}|}^2
  =
  \overline{|\mathcal{M}^\text{VC}_{\tau\to \ell V}|}^2
  +
  \overline{|\mathcal{M}^\text{TC}_{\tau\to \ell V}|}^2
  +
  \mathcal{I},
\end{equation}
where the interference term $\mathcal{I}$ is given by
\begin{equation}
 \begin{aligned}
  \mathcal{I}
  =\,\,&
  3\,m_\tau\left(m_\tau^2-m_\ell^2-m_V^2\right)
  \!\!\!\!\!\!
  &&\Re\Bigg(
    g_{L}^{\tau \ell V}
    \left( g_{TR}^{\tau \ell V}+\tilde{g}_{TR}^{\tau \ell V}\right)^*
    +
    g_{R}^{\tau \ell V}
    \left( g_{TL}^{\tau \ell V}-\tilde{g}_{TL}^{\tau \ell V}\right)^*
  \Bigg)
  \\
  +\,&
  3\,m_\ell\left(m_\ell^2-m_\tau^2-m_V^2\right)
  \!\!\!\!\!\!
  &&\Re\Bigg(
    g_{R}^{\tau \ell V}
    \left( g_{TR}^{\tau \ell V}+\tilde{g}_{TR}^{\tau \ell V}\right)^*
    +
    g_{L}^{\tau \ell V}
    \left( g_{TL}^{\tau \ell V}-\tilde{g}_{TL}^{\tau \ell V}\right)^*
  \Bigg).
 \end{aligned}
\end{equation}

\subsection{Effective coupling constants in the WET}
\subsubsection{Vector operators}
The semi-leptonic vector operators
\begin{equation}\label{eq:operators_vector}
\begin{aligned}
 \mathcal{L}_\text{eff}\subset\
 \sum_{\ell\in\{e,\mu\}, q\in\{u,d,s\}}
 \Big\{\,
 &
 C_{VLL}^{\tau\ell qq}
 &&\!\!
 (\bar{\ell}_L\gamma^\mu \tau_L)(\bar{q}_L\gamma_\mu q_L)
 \\
 +\,&
 C_{VLR}^{\tau\ell qq}
 &&\!\!
(\bar{\ell}_L\gamma^\mu \tau_L)(\bar{q}_R\gamma_\mu q_R)
 \\
 +\,&
 C_{VLR}^{qq\tau\ell}
 &&\!\!
\vphantom{\Bigg\{}
 (\bar{\ell}_R\gamma^\mu \tau_R)(\bar{q}_L\gamma_\mu q_L)
 \\
 +\,&
 C_{VRR}^{\tau\ell qq}
 &&\!\!
(\bar{\ell}_R\gamma^\mu \tau_R)(\bar{q}_R\gamma_\mu q_R)
 \Big\}+\text{h.c.}
\end{aligned}
\end{equation}
contribute to the vector current amplitude $\mathcal{M}^\text{VC}_{\tau\to \ell V}$.
Using the vacuum to vector meson matrix element of the quark vector current for the case $V=\phi$ (cf. e.g.~\cite{Becirevic:2003pn}),
\begin{equation}\label{eq:matrix_elem_meson_prod}
 \left<\phi|\bar{s}\gamma_\mu s|0\right> = m_\phi\,f_\phi\,\epsilon_\mu^*,
\end{equation}
where $f_\phi$ is the $\phi$ decay constant and $m_\phi$ the $\phi$ mass, the effective couplings $g_{L}^{\tau \ell \phi}$ and $g_{R}^{\tau \ell \phi}$ are given by
\begin{equation}\label{eq:gLgR}
 \begin{aligned}
  g_{L}^{\tau \ell \phi} &=
  \frac{1}{2}\, m_\phi\,f_\phi\,\left(C_{VLL}^{\tau\ell ss} + C_{VLR}^{\tau\ell ss}\right),
  \\
  g_{R}^{\tau \ell \phi} &=
  \frac{1}{2}\, m_\phi\,f_\phi\,\left(C_{VRR}^{\tau\ell ss} + C_{VLR}^{ss\tau\ell}\right).
 \end{aligned}
\end{equation}
In the case $V=\rho$, the vacuum to vector meson matrix element is
\begin{equation}\label{eq:matrix_elem_meson_prod_rho}
 \left<\rho\left|\frac{\bar{u}\gamma_\mu u-\bar{d}\gamma_\mu d}{\sqrt{2}}\right|0\right> = m_\rho\,f_\rho\,\epsilon_\mu^*,
\end{equation}
and the effective couplings $g_{L}^{\tau \ell \rho}$ and $g_{R}^{\tau \ell \rho}$ are given by
\begin{equation}\label{eq:gLgR_rho}
 \begin{aligned}
  g_{L}^{\tau \ell \rho} &=
  \frac{1}{2}\, m_\rho\,f_\rho\,\left(
  \frac{C_{VLL}^{\tau\ell uu}-C_{VLL}^{\tau\ell dd}}{\sqrt{2}}
  +
  \frac{C_{VLR}^{\tau\ell uu}-C_{VLR}^{\tau\ell dd}}{\sqrt{2}}
  \right),
  \\
  g_{R}^{\tau \ell \rho} &=
  \frac{1}{2}\, m_\rho\,f_\rho\,\left(
  \frac{C_{VRR}^{\tau\ell u} - C_{VRR}^{\tau\ell dd}}{\sqrt{2}}
  +
  \frac{C_{VLR}^{uu\tau\ell} - C_{VLR}^{dd\tau\ell}}{\sqrt{2}}
  \right),
 \end{aligned}
\end{equation}
where $f_\rho$ and $m_\rho$ are the $\rho$'s decay constant and mass.

\subsubsection{Dipole and tensor operators}
The leptonic dipole operators
\begin{equation}\label{eq:operators_dipole}
\begin{aligned}
 \mathcal{L}_\text{eff}\subset\
 \sum_{\ell\in\{e,\mu\}}
 \Big\{\,
 &
 C_{\gamma}^{\tau\ell}
 &&\!\!\!\!
 (\bar{\ell}_L\sigma^{\mu\nu} \tau_R)\,F_{\mu\nu}
 \\
 +\,&
 C_{\gamma}^{\ell\tau}
 &&\!\!\!\!
 (\bar{\tau}_L\sigma^{\mu\nu} \ell_R)\,F_{\mu\nu}
  \Big\}+\text{h.c.}\,,
\end{aligned}
\end{equation}
as well as the semi-leptonic tensor operators
\begin{equation}\label{eq:operators_tensor}
\begin{aligned}
 \mathcal{L}_\text{eff}\subset\
 \sum_{\ell\in\{e,\mu\}, q\in\{u,d,s\}}
 \Big\{\,
 &
 C_{TRR}^{\tau\ell qq}
 &&\!\!\!\!
 (\bar{\ell}_L\sigma^{\mu\nu} \tau_R)\,(\bar{q}_L\sigma_{\mu\nu} q_R)
 \\
 +\,&
 C_{TRR}^{\ell\tau qq}
 &&\!\!\!\!
 (\bar{\tau}_L\sigma^{\mu\nu} \ell_R)\,(\bar{q}_L\sigma_{\mu\nu} q_R)
  \Big\}+\text{h.c.}\,,
\end{aligned}
\end{equation}
contribute to the tensor current amplitude $\mathcal{M}^\text{TC}_{\tau\to \ell V}$.
Following~\cite{Brignole:2004ah}, the vacuum to vector meson matrix element of the electromagnetic field strength tensor $F_{\mu\nu}$ can be written as
\begin{equation}
\begin{aligned}
 \left<V\left|F_{\mu\nu}\right|0\right>
 &=
 \frac{i\,f_V\,K_V}{m_V}
 \left(
 {p_V}_\mu\,\epsilon_\nu^*
 -
 {p_V}_\nu\,\epsilon_\mu^*
 \right),
\end{aligned}
\end{equation}
where ${p_V}_\mu$ is the outgoing momentum of the vector meson and the constant $K_V$ depends on the fermion content of the meson $V$ and the electric charges $Q_f$ of its constituent fermions.
For $V=\phi$ and $V=\rho$, one finds\footnote{%
The overall sign of $K_V$ depends on the convention used for the covariant derivative.
Our choice in eq.~(\ref{eq:cov}) yields the result in eq.~(\ref{eq:K_V}).
The sign of $K_V$ is flipped if the sign of the second term in eq.~(\ref{eq:cov}) is chosen to be negative.
}
\begin{equation}\label{eq:K_V}
 K_\phi=-e\,Q_s=\frac{1}{3}\,e,
 \quad\quad
 K_\rho=-e\frac{Q_u-Q_d}{\sqrt{2}}=-\frac{1}{\sqrt{2}}\,e.
\end{equation}
The vacuum to vector meson matrix element of the quark tensor current  for the case $V=\phi$ is given by (cf. e.g.~\cite{Becirevic:2003pn})
\begin{equation}
 \left<\phi|\bar{s}\sigma_{\mu\nu} s|0\right> = i\,{f_T}_\phi(\mu)
 \left(
 {p_\phi}_\mu\,\epsilon_\nu^*
 -
 {p_\phi}_\nu\,\epsilon_\mu^*
 \right),
\end{equation}
where ${p_\phi}_\mu$ is the outgoing momentum of the $\phi$ and ${f_T}_\phi(\mu)$ is its transverse decay constant, which depends on the scale $\mu$ at which the corresponding operator is renormalized.
For $\tau$ decays, we set $\mu=1.8\,\text{GeV}$ and define
\begin{equation}
 {f_T}_V^\tau={f_T}_V(1.8\,\text{GeV}),\quad V\in\{\phi,\rho\}.
\end{equation}
The contributions from dipole and tensor Wilson coefficients to the coupling constants $g_{TL}^{\tau \ell \phi}$, $g_{TR}^{\tau \ell \phi}$, $\tilde{g}_{TL}^{\tau \ell \phi}$, and $\tilde{g}_{TR}^{\tau \ell \phi}$ are thus given by
\begin{equation}\label{eq:gT}
 \begin{aligned}
  g_{TL}^{\tau \ell \phi} &=
  {f_T}_\phi^\tau\,{C_{TRR}^{\ell\tau ss}}^*
  +\frac{2\,f_\phi\,K_\phi}{m_\phi}
  {C_{\gamma}^{\ell\tau}}^*,
  &\quad\quad
  \tilde{g}_{TL}^{\tau \ell \phi} &= -{f_T}_\phi^\tau\,{C_{TRR}^{\ell\tau ss}}^*,
  \\
  g_{TR}^{\tau \ell \phi} &=
  {f_T}_\phi^\tau\,C_{TRR}^{\tau\ell ss}
  +\frac{2\,f_\phi\,K_\phi}{m_\phi}
  C_{\gamma}^{\tau\ell},
  &\quad\quad
  \tilde{g}_{TR}^{\tau \ell \phi} &= {f_T}_\phi^\tau\,C_{TRR}^{\tau\ell ss}.
 \end{aligned}
\end{equation}
In the case $V=\rho$, the vacuum to vector meson matrix element of the quark tensor current is
\begin{equation}
 \left<\rho\left|\frac{\bar{u}\sigma_{\mu\nu} u-\bar{d}\sigma_{\mu\nu} d}{\sqrt{2}}\right|0\right> = i\,{f_T}_\rho(\mu)
 \left(
 {p_\rho}_\mu\,\epsilon_\nu^*
 -
 {p_\rho}_\nu\,\epsilon_\mu^*
 \right),
\end{equation}
where ${f_T}_\rho(\mu)$ is the $\rho$ transverse decay constant and ${p_\rho}_\mu$ is its outgoing momentum.
The effective couplings $g_{TL}^{\tau \ell \rho}$, $g_{TR}^{\tau \ell \rho}$, $\tilde{g}_{TL}^{\tau \ell \rho}$, and $\tilde{g}_{TR}^{\tau \ell \rho}$ are thus given by
\begin{equation}
 \begin{aligned}
  g_{TL}^{\tau \ell \rho} &=
  {f_T}_\rho^\tau\,\frac{{C_{TRR}^{\ell\tau uu}}^*-{C_{TRR}^{\ell\tau dd}}^*}{\sqrt{2}}
  +\frac{2\,f_\rho\,K_\rho}{m_\rho}
  {C_{\gamma}^{\ell\tau}}^*,
  &\quad
  \tilde{g}_{TL}^{\tau \ell \rho} &= -{f_T}_\rho^\tau\,\frac{{C_{TRR}^{\ell\tau uu}}^*-{C_{TRR}^{\ell\tau dd}}^*}{\sqrt{2}},
  \\
  g_{TR}^{\tau \ell \rho} &=
  {f_T}_\rho^\tau\,\frac{C_{TRR}^{\tau\ell uu}-C_{TRR}^{\tau\ell dd}}{\sqrt{2}}
  +\frac{2\,f_\rho\,K_\rho}{m_\rho}
  C_{\gamma}^{\tau\ell},
  &\quad
  \tilde{g}_{TR}^{\tau \ell \rho} &= {f_T}_\rho^\tau\,\frac{C_{TRR}^{\tau\ell uu}-C_{TRR}^{\tau\ell dd}}{\sqrt{2}}.
 \end{aligned}
\end{equation}

\section{$\tau\to P \ell $ decays}

\subsection{$\tau \to \ell \pi^0$: Effective coupling constants in the WET}
The matrix elements in this case can be defined as\cite{Celis:2014asa}
\bea \label{eq:vector mat element}
\left < \pi^0 |  \bar u \gamma_\mu \gamma_5 u |0  \right >  =  \frac{i f_\pi p_{\pi\mu}}{\sqrt 2} ,\ \ \ \ \
\left < \pi^0 |  \bar d \gamma_\mu \gamma_5 d |0  \right >  = - \frac{i f_\pi p_{\pi\mu}}{\sqrt 2},
\eea
\bea \label{eq:scalar mat element}
\left < \pi^0 |  \bar u  \gamma_5 u |0  \right >  =    \frac{i f_\pi m_\pi^2}{ \sqrt  2 (m_u+m_d)}, \ \ \ \ \
\left < \pi^0 |  \bar d  \gamma_5 d |0  \right >  =    -\frac{i f_\pi m_\pi^2}{ \sqrt 2 (m_u+m_d)}.
\eea
here $f_\pi=130.2$MeV.
For the process $\tau\to \pi^0 \ell$, the relevant part of the WET Lagrangian reads
\bea
\mathcal{L}_{eff} \label{eq: leff}
& \supset & C_{VLL}^{\tau \ell qq}  (\bar \ell_L \gamma^\mu \tau_L)
(\bar q_L \gamma_\mu q_L)
 +  C_{VLR}^{\tau \ell qq}  (\bar \ell_L \gamma^\mu \tau_L)
(\bar q_R \gamma_\mu q_R)   \\
& + & C_{VLR}^{qq \tau \ell}  (\bar \ell_R \gamma^\mu \tau_R)
(\bar q_L \gamma_\mu q_L)   \nonumber
 +  C_{VRR}^{\tau \ell qq}  (\bar \ell_R \gamma^\mu \tau_R)
(\bar q_R \gamma_\mu q_R) \nonumber \\
& + & C_{SRL}^{\tau \ell qq}  (\bar \ell_L \tau_R)
(\bar q_R  q_L)
 +  C_{SRR}^{\tau \ell qq}  (\bar \ell_L \tau_R)
(\bar q_L q_R) \nonumber \\
& + & C_{SRR}^{*\ell \tau qq}  (\bar \ell_R \tau_L)
(\bar q_R  q_L)
 +  C_{SRL}^{* \ell\tau qq}  (\bar \ell_R \tau_L)
(\bar q_L  q_R) + \text{h.c.} \ . \nonumber
\eea

From $\mathcal{L}_{eff}$ using eqs. \eqref{eq:vector mat element} and \eqref{eq:scalar mat element} and the momentum conservation,
$p_\pi^\mu=p_\tau^\mu - p_\ell^\mu$, we can define the matrix element as
\bea
{\mathcal{M}}_{\tau\ell\pi^0} &=& i (g_L^{\tau\ell \pi^0} \bar \ell  P_L \tau + g_R^{\tau\ell \pi^0} \bar \ell  P_R \tau )\,.
\eea
Here the couplings $g_L^{\tau\ell\pi^0}$ and $g_R^{\tau\ell\pi^0}$ are given by
\bea
g_L^{\tau\ell\pi^0} = s_L^{\tau\ell\pi^0} + (-v_L^{\tau\ell\pi^0} m_l +v_R^{\tau\ell\pi^0} m_\tau), \ \ g_R^{\tau\ell\pi^0} = s_R^{\tau\ell\pi^0} + (-v_R^{\tau\ell\pi^0} m_l +v_L^{\tau\ell\pi^0} m_\tau).
\eea
with the vector $v_L^{\tau\ell\pi^0}$, $v_R^{\tau\ell \pi^0}$ and scalar $s_L^{\tau\ell\pi^0}$, $s_R^{\tau\ell \pi^0}$ couplings
\bea
v_L^{\tau \ell \pi^0} &=& \frac{f_\pi}{\sqrt 2} \left (\frac{C_{VLR}^{\tau\ell uu} - C_{VLL}^{\tau \ell uu}}{2} - \frac{ C_{VLR}^{\tau \ell dd} -C_{VLL}^{\tau \ell dd}}{2}  \right ),  \\
v_R^{\tau\ell \pi^0} &=& \frac{f_\pi}{\sqrt 2} \left (\frac{C_{VRR}^{\tau\ell uu} - C_{VLR}^{uu\tau \ell}}{2} - \frac{ C_{VRR}^{\tau \ell dd } -C_{VLR}^{dd\tau \ell }}{2 }  \right ),   \\
s_R^{\tau\ell \pi^0} &=& \frac{f_\pi m_\pi^2}{\sqrt 2(m_u+m_d)} \left (\frac{C_{SRR}^{ \tau \ell uu} - C_{SRL}^{ \tau \ell uu}}{2} - \frac{ C_{SRR}^{\tau \ell dd} -C_{SRL}^{\tau \ell dd}}{2}  \right ),  \\
s_L^{\tau \ell \pi^0} &=& \frac{f_\pi m_\pi^2}{\sqrt 2(m_u+m_d)} \left (\frac{C_{SRL}^{*\ell \tau uu} - C_{SRR}^{* \ell \tau uu}}{2} - \frac{ C_{SRL}^{*  \ell \tau dd} - C_{SRR}^{* \ell \tau dd}}{2 }  \right ).
\eea

\subsection{$\tau \to \ell K^0$: Effective coupling constants in the WET}
For $K^0$ the pseudo vector matrix element is defined as\cite{Black:2002wh}\footnote{Note: For the scalar matrix element we have got a different sign from \cite{Black:2002wh}.}
\bea \label{eq:vecmeK}
\left <K^0(p)|\bar d \gamma_\mu \gamma_5 s|0\right >= -i f_K p_{K\mu},
\eea
and for the scalar current
\bea \label{eq:sclrmeK}
\left <K^0(p)|\bar d  \gamma_5 s|0\right >=  -\frac{if_K m_K^2 }{m_d+m_s}.
\eea
The relevant part of the WET Lagrangian reads
\bea
\mathcal{L}_{eff} \label{eq: leffK}
& \supset &
K_v C_{9}^{sd\tau \ell} (\bar d_L \gamma_\mu s_L)  (\bar \ell \gamma^\mu \tau)  +K_v C_{10}^{sd\tau \ell} (\bar d_L \gamma_\mu s_L)  (\bar \ell \gamma^\mu\gamma_5 \tau) \\
&+& K_v C_{9}^{'sd\tau \ell} (\bar d_R \gamma_\mu s_R)  (\bar \ell \gamma^\mu \tau)  +K_vC_{10}^{'sd\tau \ell} (\bar d_R \gamma_\mu s_R)  (\bar \ell \gamma^\mu\gamma_5 \tau)  \nonumber \\
&+&K_s C_{S}^{sd\tau \ell} (\bar d_L s_R)  (\bar \ell \tau)  +K_sC_{S}^{'sd\tau \ell} (\bar d_R  s_L)  (\bar \ell  \tau)  \nonumber \\
&+&K_s C_{P}^{sd\tau \ell} (\bar d_L s_R)  (\bar \ell\gamma_5 \tau)  +K_s C_{P}^{'sd\tau \ell} (\bar d_R  s_L)  (\bar \ell \gamma_5 \tau),  \nonumber
\eea
here $K_v = \frac{4 G_F}{\sqrt 2} V_{ts} V_{td}^* \frac{e^2}{16\pi^2 }$ and
$K_s=m_s K_v$.
The matrix element is given by
\bea
\mathcal{M}_{\tau \ell K^0} \label{eq:meK}
& = &   i p_{K\mu} \left ( g_V^{\tau \ell K^0} \bar \ell \gamma^\mu \tau  +  g_A^{\tau \ell K^0} \bar \ell \gamma^\mu\gamma_5 \tau  \right)
+ i \left( g_S^{\tau \ell K^0} \bar \ell \tau   + g_P^{\tau \ell K^0} \bar \ell\gamma_5 \tau \right),
\eea
with
\vskip-17pt
\bea
g_V^{\tau\ell K^0} &=& \frac{-f_K K_v ( -C_{9}^{sd\tau \ell} + C_{9}^{'sd\tau \ell})}{2}, \ \ \ \ g_A^{\tau \ell K^0}=\frac{-f_K K_v (-C_{10}^{sd\tau \ell} +C_{10}^{'sd\tau \ell} )}{2},  \\
g_S^{\tau \ell K^0}&=& \frac{-K_s f_K m_K^2 (C_{S}^{sd\tau \ell}-C_{S}^{'sd\tau \ell} )} {2(m_s+m_d)},  \ \ \ \ g_P^{\tau \ell K^0}= \frac{-K_s f_K m_K^2(C_{P}^{sd\tau \ell}-C_{P}^{'sd\tau \ell})}{2(m_s+m_d)}.
\eea

Using the momentum conservation, $p_K^\mu=p_\tau^\mu - p_\ell^\mu$, in Eq. \ref{eq:meK}, we can redefine the matrix element as
\begin{equation}
{\mathcal{M}}_{\tau\ell K^0} = i \left (g_{L}^{\tau\ell K^0} \bar \ell P_L  \tau + g_{R}^{\tau\ell K^0} \bar \ell  P_R \tau \right)
\end{equation}
here
\begin{equation}
g_L^{\tau\ell K^0}=\left (g_{VS}^{\tau\ell K^0}-g_{AP}^{\tau\ell K^0} \right ) \ \ \ \ g_R^{\tau\ell K^0}=\left (g_{VS}^{\tau\ell K^0}+ g_{AP}^{\tau\ell K^0} \right )
\end{equation}
and the couplings $g_{VS}^{\tau\ell K^0}$ and $g_{AP}^{\tau\ell K^0}$ are given by
\bea
g_{VS}^{\tau\ell K^0} = g_S^{\tau\ell K^0} + g_V^{\tau\ell K^0} \left ( m_\tau -  m_l \right), \ \ g_{AP}^{\tau\ell K^0} = g_P^{\tau\ell K^0} -  g_A^{\tau \ell K^0} \left (m_l + m_\tau \right ).
\eea

\subsection{Squared amplitude}
The squared matrix element, summed over the final states and averaged over the initial states, is given by
\begin{align} \label{eq:mmod}
|{\mathcal {M}}_{\tau\ell P}|^2   =& \frac{1}{2} \left (|g_L^{\tau\ell P}|^2+ |g_R^{\tau\ell P}|^2 \right) \left (m_\tau^2+m_l^2-m_P^2 \right) \\ \nonumber
&+2 m_l m_ \tau \left( \Im(g_L^{\tau\ell P}) \Im(g_R^{\tau\ell P}) + \Re(g_L^{\tau\ell P}) \Re(g_R^{\tau\ell P}) \right ) .
\end{align}

\section{List of observables}\label{app:obstable}

In this appendix we collect the SM predictions, uncertainties,
experimental measurements (combinations in case of multiple measurements) and uncertainties for each individual observable.  This table roughly corresponds to the output of the \texttt{GlobalLikelihoodPoint.obstable} method. It is only approximate in several cases, e.g.\ in case of non-Gaussian uncertainties present in the code. The ``pull'' ignores any correlations with other observables and is just meant as an indication of the agreement of an individual measurement with the SM.
For observables where we neglect theory uncertainties, the predictions are just given as numbers.
In the case of upper limits, we give the 95\% confidence level limits in all cases.

\begin{table}[ht]
\resizebox{\textwidth}{!}{
\renewcommand{\arraystretch}{1.3}
\begin{tabularx}{1.36\textwidth}{lllll}
\hline
Observable ~~~~~~~~~~~~~~~~~~~~~~~~~~~~ & Prediction ~~~~~~~~~~~~~~~~~~~~~~~~~~~ & Measurement ~~~~~~~~~~~~~~~~~~~~~~~ & ~~~~~~~~~~~~~~~~~~ & Pull\\
\hline
$\langle A_\text{FB}\rangle(B^0\to K^{\ast 0}\mu^+\mu^-)^{[0,2]}$ & $-0.104 \pm 0.011$ & $0.07 \pm 0.31$ & \cite{CDF:2012qwd} & $0.6\sigma$\\
$\langle A_\text{FB}\rangle(B^0\to K^{\ast 0}\mu^+\mu^-)^{[2,4.3]}$ & $\left(-2.6 \pm 3.1\right) \times 10^{-2}$ & $-0.11 \pm 0.12$ & \cite{CDF:2012qwd,Khachatryan:2015isa} & $0.7\sigma$\\
$\langle A_\text{FB}\rangle(B^0\to K^{\ast 0}\mu^+\mu^-)^{[1,2]}$ & $-0.156 \pm 0.031$ & $-0.28 \pm 0.17$ & \cite{Khachatryan:2015isa} & $0.7\sigma$\\
$\langle A_\text{FB}\rangle(B^0\to K^{\ast 0}\mu^+\mu^-)^{[4.3,6]}$ & $0.133 \pm 0.042$ & $0.01 \pm 0.15$ & \cite{Khachatryan:2015isa} & $0.8\sigma$\\
$\langle A_\text{FB}^h\rangle(\Lambda_b\to\Lambda \mu^+\mu^-)^{[15,20]}$ & $-0.272 \pm 0.011$ & $-0.299 \pm 0.053$ & \cite{Aaij:2018gwm} & $0.5\sigma$\\
$\langle A_\text{FB}^\ell\rangle(\Lambda_b\to\Lambda \mu^+\mu^-)^{[15,20]}$ & $-0.353 \pm 0.021$ & $-0.390 \pm 0.041$ & \cite{Aaij:2018gwm} & $0.8\sigma$\\
$\langle A_\text{FB}^{\ell h}\rangle(\Lambda_b\to\Lambda \mu^+\mu^-)^{[15,20]}$ & $0.1400 \pm 0.0074$ & $0.249 \pm 0.040$ & \cite{Aaij:2018gwm} & $2.7\sigma$\\
$\langle A_T^\text{Im}\rangle(B^0\to K^{\ast 0}e^+e^-)^{[0.002,1.12]}$ & $\left(3.2 \pm 6.4\right) \times 10^{-4}$ & $0.14 \pm 0.23$ & \cite{Aaij:2015dea} & $0.6\sigma$\\
$\langle \text{BR} \rangle(B\to X_se^+e^-)^{[1.0,6.0]}$ & $\left(1.74 \pm 0.18\right) \times 10^{-6}$ & $\left(2.01 \pm 0.53\right) \times 10^{-6}$ & \cite{Lees:2013nxa} & $0.5\sigma$\\
$\langle \text{BR} \rangle(B\to X_se^+e^-)^{[14.2,25.0]}$ & $\left(3.01 \pm 0.34\right) \times 10^{-7}$ & $\left(5.7 \pm 1.9\right) \times 10^{-7}$ & \cite{Lees:2013nxa} & $1.4\sigma$\\
$\langle \text{BR} \rangle(B\to X_s\mu^+\mu^-)^{[1.0,6.0]}$ & $\left(1.67 \pm 0.17\right) \times 10^{-6}$ & $\left(7.5 \pm 8.2\right) \times 10^{-7}$ & \cite{Lees:2013nxa} & $1.1\sigma$\\
$\langle \text{BR} \rangle(B\to X_s\mu^+\mu^-)^{[14.2,25.0]}$ & $\left(3.46 \pm 0.39\right) \times 10^{-7}$ & $\left(6.3 \pm 3.0\right) \times 10^{-7}$ & \cite{Lees:2013nxa} & $0.9\sigma$\\
$\langle F_L\rangle(B^0\to K^{\ast 0}\mu^+\mu^-)^{[0.04,2]}$ & $0.388 \pm 0.059$ & $0.44 \pm 0.11$ & \cite{Aaboud:2018krd} & $0.4\sigma$\\
$\langle F_L\rangle(B^0\to K^{\ast 0}\mu^+\mu^-)^{[2,4]}$ & $0.799 \pm 0.036$ & $0.64 \pm 0.12$ & \cite{Aaboud:2018krd} & $1.3\sigma$\\
$\langle F_L\rangle(B^0\to K^{\ast 0}\mu^+\mu^-)^{[4,6]}$ & $0.711 \pm 0.049$ & $0.596 \pm 0.053$ & \cite{Aaboud:2018krd,Aaij:2015oid} & $1.6\sigma$\\
$\langle F_L\rangle(B^0\to K^{\ast 0}\mu^+\mu^-)^{[0,2]}$ & $0.388 \pm 0.059$ & $0.27 \pm 0.14$ & \cite{CDF:2012qwd} & $0.8\sigma$\\
$\langle F_L\rangle(B^0\to K^{\ast 0}\mu^+\mu^-)^{[2,4.3]}$ & $0.793 \pm 0.036$ & $0.758 \pm 0.080$ & \cite{CDF:2012qwd,Khachatryan:2015isa} & $0.4\sigma$\\
$\langle F_L\rangle(B^0\to K^{\ast 0}\mu^+\mu^-)^{[1,2]}$ & $0.725 \pm 0.050$ & $0.63 \pm 0.10$ & \cite{Khachatryan:2015isa} & $0.8\sigma$\\
$\langle F_L\rangle(B^0\to K^{\ast 0}\mu^+\mu^-)^{[4.3,6]}$ & $0.703 \pm 0.050$ & $0.63 \pm 0.12$ & \cite{Khachatryan:2015isa} & $0.6\sigma$\\
$\langle F_L\rangle(B^0\to K^{\ast 0}\mu^+\mu^-)^{[1.1,2.5]}$ & $0.760 \pm 0.045$ & $0.660 \pm 0.083$ & \cite{Aaij:2015oid} & $1.1\sigma$\\
$\langle F_L\rangle(B^0\to K^{\ast 0}\mu^+\mu^-)^{[2.5,4]}$ & $0.796 \pm 0.036$ & $0.87 \pm 0.10$ & \cite{Aaij:2015oid} & $0.7\sigma$\\
$\langle F_L\rangle(B^0\to K^{\ast 0}\mu^+\mu^-)^{[15,19]}$ & $0.340 \pm 0.022$ & $0.345 \pm 0.030$ & \cite{Aaij:2015oid} & $0.1\sigma$\\
$\langle \overline{F_L}\rangle(B_s\to \phi \mu^+\mu^-)^{[2.0,5.0]}$ & $0.811 \pm 0.019$ & $0.68 \pm 0.15$ & \cite{Aaij:2015esa} & $0.9\sigma$\\
$\langle \overline{F_L}\rangle(B_s\to \phi \mu^+\mu^-)^{[15.0,19.0]}$ & $0.341 \pm 0.015$ & $0.288 \pm 0.068$ & \cite{Aaij:2015esa} & $0.8\sigma$\\
$\langle P_1\rangle(B^0\to K^{\ast 0}e^+e^-)^{[0.002,1.12]}$ & $\left(3.6 \pm 2.3\right) \times 10^{-2}$ & $-0.24 \pm 0.24$ & \cite{Aaij:2015dea} & $1.2\sigma$\\
$\langle P_1\rangle(B^0\to K^{\ast 0}\mu^+\mu^-)^{[0.04,2]}$ & $\left(4.3 \pm 2.9\right) \times 10^{-2}$ & $-0.06 \pm 0.31$ & \cite{Aaboud:2018krd} & $0.3\sigma$\\
$\langle P_1\rangle(B^0\to K^{\ast 0}\mu^+\mu^-)^{[2,4]}$ & $\left(-9.5 \pm 3.8\right) \times 10^{-2}$ & $-0.78 \pm 0.65$ & \cite{Aaboud:2018krd} & $1.1\sigma$\\
$\langle P_1\rangle(B^0\to K^{\ast 0}\mu^+\mu^-)^{[4,6]}$ & $-0.178 \pm 0.048$ & $0.12 \pm 0.30$ & \cite{Aaboud:2018krd,Aaij:2015oid} & $1.0\sigma$\\
$\langle P_1\rangle(B^0\to K^{\ast 0}\mu^+\mu^-)^{[1,2]}$ & $\left(4.4 \pm 4.8\right) \times 10^{-2}$ & $0.12 \pm 0.47$ & \cite{CMS:2017ivg} & $0.2\sigma$\\
$\langle P_1\rangle(B^0\to K^{\ast 0}\mu^+\mu^-)^{[2,4.3]}$ & $-0.106 \pm 0.038$ & $-0.44 \pm 0.45$ & \cite{CMS:2017ivg} & $0.7\sigma$\\
$\langle P_1\rangle(B^0\to K^{\ast 0}\mu^+\mu^-)^{[4.3,6]}$ & $-0.180 \pm 0.049$ & $0.45 \pm 0.34$ & \cite{CMS:2017ivg} & $1.8\sigma$\\
$\langle P_1\rangle(B^0\to K^{\ast 0}\mu^+\mu^-)^{[1.1,2.5]}$ & $\left(2.3 \pm 4.7\right) \times 10^{-2}$ & $-0.45 \pm 0.58$ & \cite{Aaij:2015oid} & $0.8\sigma$\\
$\langle P_1\rangle(B^0\to K^{\ast 0}\mu^+\mu^-)^{[2.5,4]}$ & $-0.116 \pm 0.039$ & $0.6 \pm 2.0$ & \cite{Aaij:2015oid} & $0.4\sigma$\\
$\langle P_1\rangle(B^0\to K^{\ast 0}\mu^+\mu^-)^{[15,19]}$ & $-0.623 \pm 0.043$ & $-0.50 \pm 0.10$ & \cite{Aaij:2015oid} & $1.1\sigma$\\
$\langle P_2\rangle(B^0\to K^{\ast 0}\mu^+\mu^-)^{[1.1,2.5]}$ & $-0.451 \pm 0.014$ & $-0.38 \pm 0.17$ & \cite{Aaij:2015oid} & $0.4\sigma$\\
$\langle P_2\rangle(B^0\to K^{\ast 0}\mu^+\mu^-)^{[2.5,4]}$ & $-0.06 \pm 0.10$ & $-0.63 \pm 0.87$ & \cite{Aaij:2015oid} & $0.6\sigma$\\
$\langle P_2\rangle(B^0\to K^{\ast 0}\mu^+\mu^-)^{[4,6]}$ & $0.292 \pm 0.074$ & $\left(3.2 \pm 8.8\right) \times 10^{-2}$ & \cite{Aaij:2015oid} & $2.3\sigma$\\
$\langle P_2\rangle(B^0\to K^{\ast 0}\mu^+\mu^-)^{[15,19]}$ & $0.373 \pm 0.017$ & $0.361 \pm 0.027$ & \cite{Aaij:2015oid} & $0.4\sigma$\\
$\langle P_4^\prime\rangle(B^0\to K^{\ast 0}\mu^+\mu^-)^{[0.04,2]}$ & $0.150 \pm 0.018$ & $0.36 \pm 0.57$ & \cite{Aaboud:2018krd} & $0.4\sigma$\\
$\langle P_4^\prime\rangle(B^0\to K^{\ast 0}\mu^+\mu^-)^{[2,4]}$ & $-0.349 \pm 0.049$ & $-0.98 \pm 0.46$ & \cite{Aaboud:2018krd} & $1.4\sigma$\\
$\langle P_4^\prime\rangle(B^0\to K^{\ast 0}\mu^+\mu^-)^{[4,6]}$ & $-0.504 \pm 0.027$ & $-0.30 \pm 0.16$ & \cite{Aaboud:2018krd,Aaij:2015oid} & $1.2\sigma$\\
$\langle P_4^\prime\rangle(B^0\to K^{\ast 0}\mu^+\mu^-)^{[1.1,2.5]}$ & $\left(-6.4 \pm 4.4\right) \times 10^{-2}$ & $-0.17 \pm 0.24$ & \cite{Aaij:2015oid} & $0.4\sigma$\\
\hline
\end{tabularx}

}
\caption{Quark flavour observables where theory uncertainties are taken into account (part~1/3).}
\end{table}

\begin{table}[ht]
\resizebox{\textwidth}{!}{
\renewcommand{\arraystretch}{1.3}
\begin{tabularx}{1.36\textwidth}{lllll}
\hline
Observable ~~~~~~~~~~~~~~~~~~~~~~~~~~~~ & Prediction ~~~~~~~~~~~~~~~~~~~~~~~~~~~ & Measurement ~~~~~~~~~~~~~~~~~~~~~~~ & ~~~~~~~~~~~~~~~~~~ & Pull\\
\hline
$\langle P_4^\prime\rangle(B^0\to K^{\ast 0}\mu^+\mu^-)^{[2.5,4]}$ & $-0.393 \pm 0.046$ & $-0.72 \pm 0.74$ & \cite{Aaij:2015oid} & $0.4\sigma$\\
$\langle P_4^\prime\rangle(B^0\to K^{\ast 0}\mu^+\mu^-)^{[15,19]}$ & $-0.6351 \pm 0.0088$ & $-0.598 \pm 0.084$ & \cite{Aaij:2015oid} & $0.4\sigma$\\
$\langle P_5^\prime\rangle(B^0\to K^{\ast 0}\mu^+\mu^-)^{[0.04,2]}$ & $0.513 \pm 0.036$ & $0.67 \pm 0.30$ & \cite{Aaboud:2018krd} & $0.5\sigma$\\
$\langle P_5^\prime\rangle(B^0\to K^{\ast 0}\mu^+\mu^-)^{[2,4]}$ & $-0.41 \pm 0.11$ & $-0.33 \pm 0.33$ & \cite{Aaboud:2018krd} & $0.2\sigma$\\
$\langle P_5^\prime\rangle(B^0\to K^{\ast 0}\mu^+\mu^-)^{[4,6]}$ & $-0.757 \pm 0.074$ & $-0.21 \pm 0.15$ & \cite{Aaboud:2018krd,Aaij:2015oid} & $3.3\sigma$\\
$\langle P_5^\prime\rangle(B^0\to K^{\ast 0}\mu^+\mu^-)^{[1,2]}$ & $0.288 \pm 0.068$ & $0.11 \pm 0.34$ & \cite{CMS:2017ivg} & $0.5\sigma$\\
$\langle P_5^\prime\rangle(B^0\to K^{\ast 0}\mu^+\mu^-)^{[2,4.3]}$ & $-0.45 \pm 0.10$ & $-0.54 \pm 0.36$ & \cite{CMS:2017ivg} & $0.2\sigma$\\
$\langle P_5^\prime\rangle(B^0\to K^{\ast 0}\mu^+\mu^-)^{[4.3,6]}$ & $-0.769 \pm 0.072$ & $-0.96 \pm 0.26$ & \cite{CMS:2017ivg} & $0.7\sigma$\\
$\langle P_5^\prime\rangle(B^0\to K^{\ast 0}\mu^+\mu^-)^{[1.1,2.5]}$ & $0.139 \pm 0.084$ & $0.29 \pm 0.21$ & \cite{Aaij:2015oid} & $0.7\sigma$\\
$\langle P_5^\prime\rangle(B^0\to K^{\ast 0}\mu^+\mu^-)^{[2.5,4]}$ & $-0.50 \pm 0.10$ & $-0.07 \pm 0.35$ & \cite{Aaij:2015oid} & $1.2\sigma$\\
$\langle P_5^\prime\rangle(B^0\to K^{\ast 0}\mu^+\mu^-)^{[15,19]}$ & $-0.594 \pm 0.035$ & $-0.684 \pm 0.082$ & \cite{Aaij:2015oid} & $1.0\sigma$\\
$\langle \overline{S_3}\rangle(B_s\to \phi \mu^+\mu^-)^{[2.0,5.0]}$ & $\left(-8.7 \pm 3.9\right) \times 10^{-3}$ & $-0.06 \pm 0.21$ & \cite{Aaij:2015esa} & $0.3\sigma$\\
$\langle \overline{S_3}\rangle(B_s\to \phi \mu^+\mu^-)^{[15.0,19.0]}$ & $-0.2098 \pm 0.0067$ & $-0.09 \pm 0.12$ & \cite{Aaij:2015esa} & $1.0\sigma$\\
$\langle \overline{S_4}\rangle(B_s\to \phi \mu^+\mu^-)^{[2.0,5.0]}$ & $-0.148 \pm 0.018$ & $-0.46 \pm 0.36$ & \cite{Aaij:2015esa} & $0.9\sigma$\\
$\langle \overline{S_4}\rangle(B_s\to \phi \mu^+\mu^-)^{[15.0,19.0]}$ & $-0.3017 \pm 0.0044$ & $-0.14 \pm 0.11$ & \cite{Aaij:2015esa} & $1.4\sigma$\\
$\langle \frac{d\text{BR}}{dq^2} \rangle(B^+\to K^{\ast +}\mu^+\mu^-)^{[2.0,4.0]}$ & $\left(4.87 \pm 0.74\right) \times 10^{-8}\ \frac{1}{\text{GeV}^2}$ & $\left(5.6 \pm 1.6\right) \times 10^{-8}\ \frac{1}{\text{GeV}^2}$ & \cite{Aaij:2014pli} & $0.5\sigma$\\
$\langle \frac{d\text{BR}}{dq^2} \rangle(B^+\to K^{\ast +}\mu^+\mu^-)^{[4.0,6.0]}$ & $\left(5.43 \pm 0.82\right) \times 10^{-8}\ \frac{1}{\text{GeV}^2}$ & $\left(2.6 \pm 1.0\right) \times 10^{-8}\ \frac{1}{\text{GeV}^2}$ & \cite{Aaij:2014pli} & $2.1\sigma$\\
$\langle \frac{d\text{BR}}{dq^2} \rangle(B^+\to K^{\ast +}\mu^+\mu^-)^{[15.0,19.0]}$ & $\left(6.44 \pm 0.68\right) \times 10^{-8}\ \frac{1}{\text{GeV}^2}$ & $\left(4.01 \pm 0.83\right) \times 10^{-8}\ \frac{1}{\text{GeV}^2}$ & \cite{Aaij:2014pli} & $2.3\sigma$\\
$\langle \frac{d\text{BR}}{dq^2} \rangle(B^+\to K^{\ast +}\mu^+\mu^-)^{[0,2]}$ & $\left(8.7 \pm 1.2\right) \times 10^{-8}\ \frac{1}{\text{GeV}^2}$ & $\left(7.3 \pm 4.8\right) \times 10^{-8}\ \frac{1}{\text{GeV}^2}$ & \cite{CDF:2012qwd} & $0.3\sigma$\\
$\langle \frac{d\text{BR}}{dq^2} \rangle(B^+\to K^{\ast +}\mu^+\mu^-)^{[2,4.3]}$ & $\left(4.90 \pm 0.74\right) \times 10^{-8}\ \frac{1}{\text{GeV}^2}$ & $\left(5.0 \pm 3.7\right) \times 10^{-8}\ \frac{1}{\text{GeV}^2}$ & \cite{CDF:2012qwd} & $0.0\sigma$\\
$\langle \frac{d\text{BR}}{dq^2} \rangle(B^\pm\to K^\pm \mu^+\mu^-)^{[1.1,2.0]}$ & $\left(3.53 \pm 0.62\right) \times 10^{-8}\ \frac{1}{\text{GeV}^2}$ & $\left(2.33 \pm 0.19\right) \times 10^{-8}\ \frac{1}{\text{GeV}^2}$ & \cite{Aaij:2014pli} & $1.8\sigma$\\
$\langle \frac{d\text{BR}}{dq^2} \rangle(B^\pm\to K^\pm \mu^+\mu^-)^{[2.0,3.0]}$ & $\left(3.51 \pm 0.61\right) \times 10^{-8}\ \frac{1}{\text{GeV}^2}$ & $\left(2.82 \pm 0.21\right) \times 10^{-8}\ \frac{1}{\text{GeV}^2}$ & \cite{Aaij:2014pli} & $1.1\sigma$\\
$\langle \frac{d\text{BR}}{dq^2} \rangle(B^\pm\to K^\pm \mu^+\mu^-)^{[3.0,4.0]}$ & $\left(3.50 \pm 0.60\right) \times 10^{-8}\ \frac{1}{\text{GeV}^2}$ & $\left(2.54 \pm 0.20\right) \times 10^{-8}\ \frac{1}{\text{GeV}^2}$ & \cite{Aaij:2014pli} & $1.5\sigma$\\
$\langle \frac{d\text{BR}}{dq^2} \rangle(B^\pm\to K^\pm \mu^+\mu^-)^{[4.0,5.0]}$ & $\left(3.47 \pm 0.59\right) \times 10^{-8}\ \frac{1}{\text{GeV}^2}$ & $\left(2.21 \pm 0.18\right) \times 10^{-8}\ \frac{1}{\text{GeV}^2}$ & \cite{Aaij:2014pli} & $2.0\sigma$\\
$\langle \frac{d\text{BR}}{dq^2} \rangle(B^\pm\to K^\pm \mu^+\mu^-)^{[5.0,6.0]}$ & $\left(3.45 \pm 0.59\right) \times 10^{-8}\ \frac{1}{\text{GeV}^2}$ & $\left(2.31 \pm 0.18\right) \times 10^{-8}\ \frac{1}{\text{GeV}^2}$ & \cite{Aaij:2014pli} & $1.9\sigma$\\
$\langle \frac{d\text{BR}}{dq^2} \rangle(B^\pm\to K^\pm \mu^+\mu^-)^{[15.0,22.0]}$ & $\left(1.51 \pm 0.17\right) \times 10^{-8}\ \frac{1}{\text{GeV}^2}$ & $\left(1.210 \pm 0.072\right) \times 10^{-8}\ \frac{1}{\text{GeV}^2}$ & \cite{Aaij:2014pli} & $1.6\sigma$\\
$\langle \frac{d\text{BR}}{dq^2} \rangle(B^\pm\to K^\pm \mu^+\mu^-)^{[0,2]}$ & $\left(3.54 \pm 0.63\right) \times 10^{-8}\ \frac{1}{\text{GeV}^2}$ & $\left(2.56 \pm 0.41\right) \times 10^{-8}\ \frac{1}{\text{GeV}^2}$ & \cite{CDF:2012qwd} & $1.3\sigma$\\
$\langle \frac{d\text{BR}}{dq^2} \rangle(B^\pm\to K^\pm \mu^+\mu^-)^{[2,4.3]}$ & $\left(3.50 \pm 0.61\right) \times 10^{-8}\ \frac{1}{\text{GeV}^2}$ & $\left(3.14 \pm 0.56\right) \times 10^{-8}\ \frac{1}{\text{GeV}^2}$ & \cite{CDF:2012qwd} & $0.4\sigma$\\
$\langle \frac{d\text{BR}}{dq^2} \rangle(B^0\to K^{\ast 0}\mu^+\mu^-)^{[1.1,2.5]}$ & $\left(4.66 \pm 0.68\right) \times 10^{-8}\ \frac{1}{\text{GeV}^2}$ & $\left(3.27 \pm 0.40\right) \times 10^{-8}\ \frac{1}{\text{GeV}^2}$ & \cite{Aaij:2016flj} & $1.8\sigma$\\
$\langle \frac{d\text{BR}}{dq^2} \rangle(B^0\to K^{\ast 0}\mu^+\mu^-)^{[2.5,4.0]}$ & $\left(4.49 \pm 0.69\right) \times 10^{-8}\ \frac{1}{\text{GeV}^2}$ & $\left(3.33 \pm 0.41\right) \times 10^{-8}\ \frac{1}{\text{GeV}^2}$ & \cite{Aaij:2016flj} & $1.4\sigma$\\
$\langle \frac{d\text{BR}}{dq^2} \rangle(B^0\to K^{\ast 0}\mu^+\mu^-)^{[4.0,6.0]}$ & $\left(5.02 \pm 0.76\right) \times 10^{-8}\ \frac{1}{\text{GeV}^2}$ & $\left(3.54 \pm 0.37\right) \times 10^{-8}\ \frac{1}{\text{GeV}^2}$ & \cite{Aaij:2016flj} & $1.7\sigma$\\
$\langle \frac{d\text{BR}}{dq^2} \rangle(B^0\to K^{\ast 0}\mu^+\mu^-)^{[15.0,19.0]}$ & $\left(5.94 \pm 0.63\right) \times 10^{-8}\ \frac{1}{\text{GeV}^2}$ & $\left(4.36 \pm 0.36\right) \times 10^{-8}\ \frac{1}{\text{GeV}^2}$ & \cite{Aaij:2016flj} & $2.2\sigma$\\
$\langle \frac{d\text{BR}}{dq^2} \rangle(B^0\to K^{\ast 0}\mu^+\mu^-)^{[0,2]}$ & $\left(8.3 \pm 1.2\right) \times 10^{-8}\ \frac{1}{\text{GeV}^2}$ & $\left(9.4 \pm 2.0\right) \times 10^{-8}\ \frac{1}{\text{GeV}^2}$ & \cite{CDF:2012qwd} & $0.5\sigma$\\
$\langle \frac{d\text{BR}}{dq^2} \rangle(B^0\to K^{\ast 0}\mu^+\mu^-)^{[1,2]}$ & $\left(4.86 \pm 0.70\right) \times 10^{-8}\ \frac{1}{\text{GeV}^2}$ & $\left(4.71 \pm 0.70\right) \times 10^{-8}\ \frac{1}{\text{GeV}^2}$ & \cite{Khachatryan:2015isa} & $0.2\sigma$\\
$\langle \frac{d\text{BR}}{dq^2} \rangle(B^0\to K^{\ast 0}\mu^+\mu^-)^{[2,4.3]}$ & $\left(4.50 \pm 0.69\right) \times 10^{-8}\ \frac{1}{\text{GeV}^2}$ & $\left(3.54 \pm 0.46\right) \times 10^{-8}\ \frac{1}{\text{GeV}^2}$ & \cite{CDF:2012qwd,Khachatryan:2015isa} & $1.2\sigma$\\
$\langle \frac{d\text{BR}}{dq^2} \rangle(B^0\to K^{\ast 0}\mu^+\mu^-)^{[4.3,6]}$ & $\left(5.07 \pm 0.77\right) \times 10^{-8}\ \frac{1}{\text{GeV}^2}$ & $\left(3.40 \pm 0.58\right) \times 10^{-8}\ \frac{1}{\text{GeV}^2}$ & \cite{Khachatryan:2015isa} & $1.7\sigma$\\
$\langle \frac{d\text{BR}}{dq^2} \rangle(B^0\to K^0\mu^+\mu^-)^{[2.0,4.0]}$ & $\left(3.25 \pm 0.56\right) \times 10^{-8}\ \frac{1}{\text{GeV}^2}$ & $\left(1.93 \pm 0.53\right) \times 10^{-8}\ \frac{1}{\text{GeV}^2}$ & \cite{Aaij:2014pli} & $1.7\sigma$\\
$\langle \frac{d\text{BR}}{dq^2} \rangle(B^0\to K^0\mu^+\mu^-)^{[4.0,6.0]}$ & $\left(3.21 \pm 0.55\right) \times 10^{-8}\ \frac{1}{\text{GeV}^2}$ & $\left(1.77 \pm 0.51\right) \times 10^{-8}\ \frac{1}{\text{GeV}^2}$ & \cite{Aaij:2014pli} & $1.9\sigma$\\
$\langle \frac{d\text{BR}}{dq^2} \rangle(B^0\to K^0\mu^+\mu^-)^{[15.0,22.0]}$ & $\left(1.39 \pm 0.15\right) \times 10^{-8}\ \frac{1}{\text{GeV}^2}$ & $\left(9.6 \pm 1.6\right) \times 10^{-9}\ \frac{1}{\text{GeV}^2}$ & \cite{Aaij:2014pli} & $1.9\sigma$\\
$\langle \frac{d\text{BR}}{dq^2} \rangle(B^0\to K^0\mu^+\mu^-)^{[0,2]}$ & $\left(3.28 \pm 0.58\right) \times 10^{-8}\ \frac{1}{\text{GeV}^2}$ & $\left(2.4 \pm 1.6\right) \times 10^{-8}\ \frac{1}{\text{GeV}^2}$ & \cite{CDF:2012qwd} & $0.5\sigma$\\
$\langle \frac{d\text{BR}}{dq^2} \rangle(B^0\to K^0\mu^+\mu^-)^{[2,4.3]}$ & $\left(3.25 \pm 0.56\right) \times 10^{-8}\ \frac{1}{\text{GeV}^2}$ & $\left(2.6 \pm 1.8\right) \times 10^{-8}\ \frac{1}{\text{GeV}^2}$ & \cite{CDF:2012qwd} & $0.4\sigma$\\
$\langle \frac{d\overline{\text{BR}}}{dq^2} \rangle(B_s\to \phi \mu^+\mu^-)^{[1.0,6.0]}$ & $\left(5.39 \pm 0.66\right) \times 10^{-8}\ \frac{1}{\text{GeV}^2}$ & $\left(2.57 \pm 0.37\right) \times 10^{-8}\ \frac{1}{\text{GeV}^2}$ & \cite{Aaij:2015esa,CDF:2012qwd} & $3.7\sigma$\\
\hline
\end{tabularx}

}
\caption{Quark flavour observables where theory uncertainties are taken into account (part~2/3).}
\end{table}

\begin{table}[ht]
\resizebox{\textwidth}{!}{
\renewcommand{\arraystretch}{1.3}
\begin{tabularx}{1.36\textwidth}{lllll}
\hline
Observable ~~~~~~~~~~~~~~~~~~~~~~~~~~~~ & Prediction ~~~~~~~~~~~~~~~~~~~~~~~~~~~ & Measurement ~~~~~~~~~~~~~~~~~~~~~~~ & ~~~~~~~~~~~~~~~~~~ & Pull\\
\hline
$\langle \frac{d\overline{\text{BR}}}{dq^2} \rangle(B_s\to \phi \mu^+\mu^-)^{[15.0,19.0]}$ & $\left(5.57 \pm 0.46\right) \times 10^{-8}\ \frac{1}{\text{GeV}^2}$ & $\left(4.05 \pm 0.50\right) \times 10^{-8}\ \frac{1}{\text{GeV}^2}$ & \cite{Aaij:2015esa} & $2.2\sigma$\\
$\langle \frac{d\text{BR}}{dq^2} \rangle(\Lambda_b\to\Lambda \mu^+\mu^-)^{[1.1,6]}$ & $\left(1.04 \pm 0.56\right) \times 10^{-8}\ \frac{1}{\text{GeV}^2}$ & $\left(9.7 \pm 6.0\right) \times 10^{-9}\ \frac{1}{\text{GeV}^2}$ & \cite{Aaij:2015xza} & $0.1\sigma$\\
$\langle \frac{d\text{BR}}{dq^2} \rangle(\Lambda_b\to\Lambda \mu^+\mu^-)^{[15,20]}$ & $\left(7.11 \pm 0.77\right) \times 10^{-8}\ \frac{1}{\text{GeV}^2}$ & $\left(1.19 \pm 0.27\right) \times 10^{-7}\ \frac{1}{\text{GeV}^2}$ & \cite{Aaij:2015xza} & $1.7\sigma$\\
$A_\text{CP}(B\to X_{s+d}\gamma)$ & $\left(-3.7 \pm 2.5\right) \times 10^{-18}$ & $\left(3.2 \pm 3.4\right) \times 10^{-2}$ & \cite{Amhis:2014hma} & $0.9\sigma$\\
$\text{BR}(B^+\to K^{*+}\gamma)$ & $\left(4.25 \pm 0.89\right) \times 10^{-5}$ & $\left(4.21 \pm 0.18\right) \times 10^{-5}$ & \cite{Amhis:2014hma} & $0.0\sigma$\\
$\text{BR}(B^+\to e^+\nu_e)$ & $\left(9.46 \pm 0.83\right) \times 10^{-12}$ & $\left(4.7 \pm 3.6\right) \times 10^{-7}$ & \cite{Amhis:2016xyh} & $1.3\sigma$\\
$\text{BR}(B^+\to \mu^+\nu_\mu)$ & $\left(4.04 \pm 0.36\right) \times 10^{-7}$ & $\left(4.9 \pm 3.7\right) \times 10^{-7}$ & \cite{Amhis:2016xyh} & $0.2\sigma$\\
$\text{BR}(B^+\to \tau^+\nu_\tau)$ & $\left(8.99 \pm 0.79\right) \times 10^{-5}$ & $\left(1.09 \pm 0.24\right) \times 10^{-4}$ & \cite{Tanabashi:2018oca} & $0.7\sigma$\\
$\text{BR}(B\to X_s\gamma)$ & $\left(3.29 \pm 0.22\right) \times 10^{-4}$ & $\left(3.27 \pm 0.14\right) \times 10^{-4}$ & \cite{Misiak:2017bgg} & $0.1\sigma$\\
$\text{BR}(B^0\to K^{*0}\gamma)$ & $\left(4.18 \pm 0.85\right) \times 10^{-5}$ & $\left(4.34 \pm 0.15\right) \times 10^{-5}$ & \cite{Amhis:2014hma} & $0.2\sigma$\\
$\text{BR}(B^0\to \mu^+\mu^-)$ & $\left(1.17 \pm 0.13\right) \times 10^{-10}$ & $\left(1.5 \pm 1.1\right) \times 10^{-10}$ & \cite{Aaboud:2018mst,Aaij:2017vad,Chatrchyan:2013bka} & $0.3\sigma$\\
$\text{BR}(B^0\to \pi^- \tau^+\nu_\tau)$ & $\left(8.42 \pm 0.92\right) \times 10^{-5}$ & $\left(1.51 \pm 0.73\right) \times 10^{-4}$ & \cite{Hamer:2015jsa} & $0.9\sigma$\\
$\overline{\text{BR}}(B_s\to \mu^+\mu^-)$ & $\left(3.61 \pm 0.19\right) \times 10^{-9}$ & $\left(2.88 \pm 0.42\right) \times 10^{-9}$ & \cite{Aaboud:2018mst,Aaij:2017vad,Chatrchyan:2013bka} & $1.6\sigma$\\
$\overline{\text{BR}}(B_s\to \phi\gamma)$ & $\left(4.01 \pm 0.52\right) \times 10^{-5}$ & $\left(3.52 \pm 0.36\right) \times 10^{-5}$ & \cite{Aaij:2012ita,Dutta:2014sxo} & $0.8\sigma$\\
$\text{BR}(K^+\to\pi^+\nu\bar\nu)$ & $\left(9.24 \pm 0.83\right) \times 10^{-11}$ & $\left(1.8 \pm 1.1\right) \times 10^{-10}$ & \cite{Artamonov:2009sz} & $0.8\sigma$\\
$\text{BR}(K_L\to e^+e^-)$ & $\left(1.93 \pm 0.34\right) \times 10^{-13}$ & $\left(1.06 \pm 0.51\right) \times 10^{-11}$ & \cite{Tanabashi:2018oca} & $2.1\sigma$\\
$\text{BR}(K_L\to \mu^+\mu^-)$ & $\left(7.5 \pm 1.3\right) \times 10^{-9}$ & $\left(6.84 \pm 0.11\right) \times 10^{-9}$ & \cite{Tanabashi:2018oca} & $0.5\sigma$\\
$\text{BR}(K_L\to\pi^0\nu\bar\nu)$ & $\left(3.32 \pm 0.37\right) \times 10^{-11}$ & $\left(1.4 \pm 1.1\right) \times 10^{-9}$ & \cite{Ahn:2018mvc} & $1.3\sigma$\\
$\text{BR}(K_S\to e^+e^-)$ & $\left(1.625 \pm 0.016\right) \times 10^{-16}$ & $\left(4.4 \pm 3.3\right) \times 10^{-9}$ & \cite{Tanabashi:2018oca} & $1.3\sigma$\\
$\text{BR}(K_S\to \mu^+\mu^-)$ & $\left(5.193 \pm 0.053\right) \times 10^{-12}$ & $\left(3.9 \pm 2.9\right) \times 10^{-10}$ & \cite{Tanabashi:2018oca} & $1.3\sigma$\\
$\Delta M_d$ & $\left(0.617 \pm 0.083\right)\, \frac{1}{\text{ps}}$ & $\left(0.5054 \pm 0.0020\right)\, \frac{1}{\text{ps}}$ & \cite{Amhis:2014hma} & $1.3\sigma$\\
$\Delta M_s$ & $\left(18.7 \pm 1.3\right)\, \frac{1}{\text{ps}}$ & $\left(17.76 \pm 0.02\right)\, \frac{1}{\text{ps}}$ & \cite{Amhis:2014hma} & $0.7\sigma$\\
$S_{K^{*}\gamma}$ & $\left(-2.3 \pm 1.5\right) \times 10^{-2}$ & $-0.16 \pm 0.22$ & \cite{Amhis:2014hma} & $0.6\sigma$\\
$S_{\psi K_S}$ & $0.706 \pm 0.025$ & $0.679 \pm 0.020$ & \cite{Amhis:2014hma} & $0.8\sigma$\\
$S_{\psi\phi}$ & $\left(3.87 \pm 0.23\right) \times 10^{-2}$ & $\left(3.3 \pm 3.3\right) \times 10^{-2}$ & \cite{Amhis:2014hma} & $0.2\sigma$\\
$\vert\epsilon_K\vert$ & $\left(1.81 \pm 0.20\right) \times 10^{-3}$ & $\left(2.228 \pm 0.011\right) \times 10^{-3}$ & \cite{Tanabashi:2018oca} & $2.1\sigma$\\
$\epsilon^\prime/\epsilon$ & $\left(-0.3 \pm 5.9\right) \times 10^{-4}$ & $\left(1.66 \pm 0.23\right) \times 10^{-3}$ & \cite{Patrignani:2016xqp} & $2.7\sigma$\\
$x_{12}^{\text{Im},D}$ & $\left(0.0 \pm 5.9\right) \times 10^{-6}$ & $\left(0.0 \pm 2.4\right) \times 10^{-4}$ & \cite{Amhis:2016xyh} & $0.0\sigma$\\
\hline
\end{tabularx}

}
\caption{Quark flavour observables where theory uncertainties are taken into account (part~3/3).}
\end{table}

\begin{table}[ht]
\resizebox{\textwidth}{!}{
\renewcommand{\arraystretch}{1.3}
\begin{tabularx}{1.36\textwidth}{lllll}
\hline
Observable ~~~~~~~~~~ & Prediction ~~~~~~~~~~~~~~~~~~~~~~~~~~~~~~~~~~~~ & Measurement ~~~~~~~~~~~~~~~~~~~~~~~~~~~~~~~~ & ~~~~~~~~~~~~~~~~~~ & Pull\\
\hline
$\text{BR}(\tau^+\to K^+\bar\nu)$ & $\left(7.119 \pm 0.083\right) \times 10^{-3}$ & $\left(6.961 \pm 0.099\right) \times 10^{-3}$ & \cite{Patrignani:2016xqp} & $1.2\sigma$\\
$\text{BR}(\tau^-\to e^- \nu\bar\nu)$ & $0.1778 \pm 0.0003$ & $0.1782 \pm 0.0004$ & \cite{Patrignani:2016xqp} & $0.8\sigma$\\
$\text{BR}(\tau^-\to \mu^- \nu\bar\nu)$ & $0.1729 \pm 0.0003$ & $0.1739 \pm 0.0004$ & \cite{Patrignani:2016xqp} & $2.0\sigma$\\
$\text{BR}(\tau^+\to \pi^+\bar\nu)$ & $0.1090 \pm 0.0013$ & $0.1082 \pm 0.0005$ & \cite{Patrignani:2016xqp} & $0.5\sigma$\\
$a_e$ & $\left(1.1596521816 \pm 0.0000000002\right) \times 10^{-3}$ & $\left(1.1596521809 \pm 0.0000000003\right) \times 10^{-3}$ & \cite{Tanabashi:2018oca} & $1.9\sigma$\\
$a_\mu$ & $\left(1.1659182 \pm 0.0000004\right) \times 10^{-3}$ & $\left(1.1659209 \pm 0.0000006\right) \times 10^{-3}$ & \cite{Tanabashi:2018oca} & $3.4\sigma$\\
$a_\tau$ & $\left(1.17721 \pm 0.00005\right) \times 10^{-3}$ & $\left(-1.8 \pm 1.7\right) \times 10^{-2}$ & \cite{Tanabashi:2018oca} & $1.1\sigma$\\
\hline
\end{tabularx}

}
\caption{Leptonic observables where theory uncertainties are taken into account.}
\end{table}

\begin{table}[ht]
\resizebox{\textwidth}{!}{
\renewcommand{\arraystretch}{1.3}
\begin{tabularx}{1.36\textwidth}{lllll}
\hline
Observable ~~~~~~~~~~~~~~~~~~~~~~~~~~~~ & Prediction ~~~~~~~~~~~~~~~~~~~~~~~~~~~ & Measurement ~~~~~~~~~~~~~~~~~~~~~~~ & ~~~~~~~~~~~~~~~~~~ & Pull\\
\hline
$\sigma_\text{trident}/\sigma_\text{trident}^\text{SM}$ & $1.00$ & $0.97 \pm 0.25$ & \cite{Geiregat:1990gz,Mishra:1991bv} & $0.1\sigma$\\
\hline
\end{tabularx}

}
\caption{Observable: neutrino trident production.}
\end{table}

\begin{table}[ht]
\resizebox{\textwidth}{!}{
\renewcommand{\arraystretch}{1.3}
\begin{tabularx}{1.36\textwidth}{lllll}
\hline
Observable ~~~~~~~~~~~~~~~~~~~~~~~~~~~~ & Prediction ~~~~~~~~~~~~~~~~~~~~~~~~~~~ & Measurement ~~~~~~~~~~~~~~~~~~~~~~~ & ~~~~~~~~~~~~~~~~~~ & Pull\\
\hline
$\frac{\langle \text{BR} \rangle}{\text{BR}}(B\to D^\ast\tau^+\nu)^{[4.0,4.53]}$ & $2.6 \times 10^{-2}$ & $\left(3.0 \pm 5.5\right) \times 10^{-2}$ & \cite{Huschle:2015rga} & $0.1\sigma$\\
$\frac{\langle \text{BR} \rangle}{\text{BR}}(B\to D^\ast\tau^+\nu)^{[4.53,5.07]}$ & $4.4 \times 10^{-2}$ & $\left(1.9 \pm 4.7\right) \times 10^{-2}$ & \cite{Huschle:2015rga} & $0.5\sigma$\\
$\frac{\langle \text{BR} \rangle}{\text{BR}}(B\to D^\ast\tau^+\nu)^{[5.07,5.6]}$ & $6.0 \times 10^{-2}$ & $\left(-2.1 \pm 4.0\right) \times 10^{-2}$ & \cite{Huschle:2015rga} & $2.0\sigma$\\
$\frac{\langle \text{BR} \rangle}{\text{BR}}(B\to D^\ast\tau^+\nu)^{[5.6,6.13]}$ & $7.4 \times 10^{-2}$ & $\left(6.6 \pm 4.9\right) \times 10^{-2}$ & \cite{Huschle:2015rga} & $0.2\sigma$\\
$\frac{\langle \text{BR} \rangle}{\text{BR}}(B\to D^\ast\tau^+\nu)^{[6.13,6.67]}$ & $8.8 \times 10^{-2}$ & $\left(4.4 \pm 5.5\right) \times 10^{-2}$ & \cite{Huschle:2015rga} & $0.8\sigma$\\
$\frac{\langle \text{BR} \rangle}{\text{BR}}(B\to D^\ast\tau^+\nu)^{[6.67,7.2]}$ & $0.095$ & $0.135 \pm 0.048$ & \cite{Huschle:2015rga} & $0.8\sigma$\\
$\frac{\langle \text{BR} \rangle}{\text{BR}}(B\to D^\ast\tau^+\nu)^{[7.2,7.73]}$ & $10.2 \times 10^{-2}$ & $\left(1.5 \pm 4.6\right) \times 10^{-2}$ & \cite{Huschle:2015rga} & $1.9\sigma$\\
$\frac{\langle \text{BR} \rangle}{\text{BR}}(B\to D^\ast\tau^+\nu)^{[7.73,8.27]}$ & $0.108$ & $0.157 \pm 0.054$ & \cite{Huschle:2015rga} & $0.9\sigma$\\
$\frac{\langle \text{BR} \rangle}{\text{BR}}(B\to D^\ast\tau^+\nu)^{[8.27,8.8]}$ & $0.106$ & $0.168 \pm 0.049$ & \cite{Huschle:2015rga} & $1.3\sigma$\\
$\frac{\langle \text{BR} \rangle}{\text{BR}}(B\to D^\ast\tau^+\nu)^{[8.8,9.33]}$ & $0.101$ & $0.136 \pm 0.045$ & \cite{Huschle:2015rga} & $0.8\sigma$\\
$\frac{\langle \text{BR} \rangle}{\text{BR}}(B\to D^\ast\tau^+\nu)^{[9.33,9.86]}$ & $9.1 \times 10^{-2}$ & $\left(8.2 \pm 4.0\right) \times 10^{-2}$ & \cite{Huschle:2015rga} & $0.2\sigma$\\
$\frac{\langle \text{BR} \rangle}{\text{BR}}(B\to D^\ast\tau^+\nu)^{[9.86,10.4]}$ & $7.1 \times 10^{-2}$ & $\left(8.1 \pm 3.0\right) \times 10^{-2}$ & \cite{Huschle:2015rga} & $0.3\sigma$\\
$\frac{\langle \text{BR} \rangle}{\text{BR}}(B\to D^\ast\tau^+\nu)^{[10.4,10.93]}$ & $2.0 \times 10^{-2}$ & $\left(8.9 \pm 3.0\right) \times 10^{-2}$ & \cite{Huschle:2015rga} & $2.3\sigma$\\
$\frac{\langle \text{BR} \rangle}{\text{BR}}(B\to D^\ast\tau^+\nu)^{[4.0,4.5]}$ & $2.4 \times 10^{-2}$ & $\left(0.1 \pm 2.1\right) \times 10^{-2}$ & \cite{Lees:2013uzd} & $1.1\sigma$\\
$\frac{\langle \text{BR} \rangle}{\text{BR}}(B\to D^\ast\tau^+\nu)^{[4.5,5.0]}$ & $4.0 \times 10^{-2}$ & $\left(5.7 \pm 2.9\right) \times 10^{-2}$ & \cite{Lees:2013uzd} & $0.6\sigma$\\
$\frac{\langle \text{BR} \rangle}{\text{BR}}(B\to D^\ast\tau^+\nu)^{[5.0,5.5]}$ & $5.4 \times 10^{-2}$ & $\left(5.4 \pm 2.4\right) \times 10^{-2}$ & \cite{Lees:2013uzd} & $0.0\sigma$\\
$\frac{\langle \text{BR} \rangle}{\text{BR}}(B\to D^\ast\tau^+\nu)^{[5.5,6.0]}$ & $6.7 \times 10^{-2}$ & $\left(5.0 \pm 2.4\right) \times 10^{-2}$ & \cite{Lees:2013uzd} & $0.7\sigma$\\
$\frac{\langle \text{BR} \rangle}{\text{BR}}(B\to D^\ast\tau^+\nu)^{[6.0,6.5]}$ & $7.8 \times 10^{-2}$ & $\left(4.8 \pm 2.3\right) \times 10^{-2}$ & \cite{Lees:2013uzd} & $1.3\sigma$\\
$\frac{\langle \text{BR} \rangle}{\text{BR}}(B\to D^\ast\tau^+\nu)^{[6.5,7.0]}$ & $8.7 \times 10^{-2}$ & $\left(9.3 \pm 2.6\right) \times 10^{-2}$ & \cite{Lees:2013uzd} & $0.2\sigma$\\
$\frac{\langle \text{BR} \rangle}{\text{BR}}(B\to D^\ast\tau^+\nu)^{[7.0,7.5]}$ & $0.094$ & $0.107 \pm 0.028$ & \cite{Lees:2013uzd} & $0.5\sigma$\\
$\frac{\langle \text{BR} \rangle}{\text{BR}}(B\to D^\ast\tau^+\nu)^{[7.5,8.0]}$ & $0.099$ & $0.119 \pm 0.031$ & \cite{Lees:2013uzd} & $0.7\sigma$\\
$\frac{\langle \text{BR} \rangle}{\text{BR}}(B\to D^\ast\tau^+\nu)^{[8.0,8.5]}$ & $10.0 \times 10^{-2}$ & $\left(9.6 \pm 2.9\right) \times 10^{-2}$ & \cite{Lees:2013uzd} & $0.1\sigma$\\
$\frac{\langle \text{BR} \rangle}{\text{BR}}(B\to D^\ast\tau^+\nu)^{[8.5,9.0]}$ & $9.9 \times 10^{-2}$ & $\left(9.0 \pm 2.9\right) \times 10^{-2}$ & \cite{Lees:2013uzd} & $0.3\sigma$\\
$\frac{\langle \text{BR} \rangle}{\text{BR}}(B\to D^\ast\tau^+\nu)^{[9.0,9.5]}$ & $9.3 \times 10^{-2}$ & $\left(9.2 \pm 3.0\right) \times 10^{-2}$ & \cite{Lees:2013uzd} & $0.0\sigma$\\
$\frac{\langle \text{BR} \rangle}{\text{BR}}(B\to D^\ast\tau^+\nu)^{[9.5,10.0]}$ & $8.1 \times 10^{-2}$ & $\left(7.6 \pm 3.3\right) \times 10^{-2}$ & \cite{Lees:2013uzd} & $0.2\sigma$\\
$\frac{\langle \text{BR} \rangle}{\text{BR}}(B\to D^\ast\tau^+\nu)^{[10.0,10.5]}$ & $5.9 \times 10^{-2}$ & $\left(7.7 \pm 3.1\right) \times 10^{-2}$ & \cite{Lees:2013uzd} & $0.6\sigma$\\
$\frac{\langle \text{BR} \rangle}{\text{BR}}(B\to D^\ast\tau^+\nu)^{[10.5,11.0]}$ & $1.0 \times 10^{-2}$ & $\left(4.0 \pm 3.1\right) \times 10^{-2}$ & \cite{Lees:2013uzd} & $0.9\sigma$\\
$\frac{\langle \text{BR} \rangle}{\text{BR}}(B\to D\tau^+\nu)^{[4.0,4.53]}$ & $3.9 \times 10^{-2}$ & $\left(7.8 \pm 5.7\right) \times 10^{-2}$ & \cite{Huschle:2015rga} & $0.7\sigma$\\
$\frac{\langle \text{BR} \rangle}{\text{BR}}(B\to D\tau^+\nu)^{[4.53,5.07]}$ & $6.1 \times 10^{-2}$ & $\left(9.1 \pm 5.4\right) \times 10^{-2}$ & \cite{Huschle:2015rga} & $0.6\sigma$\\
$\frac{\langle \text{BR} \rangle}{\text{BR}}(B\to D\tau^+\nu)^{[5.07,5.6]}$ & $7.5 \times 10^{-2}$ & $\left(7.2 \pm 5.0\right) \times 10^{-2}$ & \cite{Huschle:2015rga} & $0.1\sigma$\\
$\frac{\langle \text{BR} \rangle}{\text{BR}}(B\to D\tau^+\nu)^{[5.6,6.13]}$ & $8.6 \times 10^{-2}$ & $\left(9.3 \pm 5.1\right) \times 10^{-2}$ & \cite{Huschle:2015rga} & $0.1\sigma$\\
$\frac{\langle \text{BR} \rangle}{\text{BR}}(B\to D\tau^+\nu)^{[6.13,6.67]}$ & $9.4 \times 10^{-2}$ & $\left(5.3 \pm 5.2\right) \times 10^{-2}$ & \cite{Huschle:2015rga} & $0.8\sigma$\\
$\frac{\langle \text{BR} \rangle}{\text{BR}}(B\to D\tau^+\nu)^{[6.67,7.2]}$ & $0.095$ & $0.145 \pm 0.055$ & \cite{Huschle:2015rga} & $0.9\sigma$\\
$\frac{\langle \text{BR} \rangle}{\text{BR}}(B\to D\tau^+\nu)^{[7.2,7.73]}$ & $9.4 \times 10^{-2}$ & $\left(4.6 \pm 5.7\right) \times 10^{-2}$ & \cite{Huschle:2015rga} & $0.8\sigma$\\
$\frac{\langle \text{BR} \rangle}{\text{BR}}(B\to D\tau^+\nu)^{[7.73,8.27]}$ & $9.2 \times 10^{-2}$ & $\left(-1.0 \pm 5.4\right) \times 10^{-2}$ & \cite{Huschle:2015rga} & $1.9\sigma$\\
$\frac{\langle \text{BR} \rangle}{\text{BR}}(B\to D\tau^+\nu)^{[8.27,8.8]}$ & $8.4 \times 10^{-2}$ & $\left(5.3 \pm 5.4\right) \times 10^{-2}$ & \cite{Huschle:2015rga} & $0.6\sigma$\\
$\frac{\langle \text{BR} \rangle}{\text{BR}}(B\to D\tau^+\nu)^{[8.8,9.33]}$ & $0.076$ & $0.122 \pm 0.055$ & \cite{Huschle:2015rga} & $0.8\sigma$\\
$\frac{\langle \text{BR} \rangle}{\text{BR}}(B\to D\tau^+\nu)^{[9.33,9.86]}$ & $6.6 \times 10^{-2}$ & $\left(6.3 \pm 5.4\right) \times 10^{-2}$ & \cite{Huschle:2015rga} & $0.0\sigma$\\
$\frac{\langle \text{BR} \rangle}{\text{BR}}(B\to D\tau^+\nu)^{[9.86,10.4]}$ & $0.055$ & $0.121 \pm 0.055$ & \cite{Huschle:2015rga} & $1.2\sigma$\\
$\frac{\langle \text{BR} \rangle}{\text{BR}}(B\to D\tau^+\nu)^{[10.4,10.93]}$ & $4.0 \times 10^{-2}$ & $\left(-0.3 \pm 5.0\right) \times 10^{-2}$ & \cite{Huschle:2015rga} & $0.9\sigma$\\
$\frac{\langle \text{BR} \rangle}{\text{BR}}(B\to D\tau^+\nu)^{[10.93,11.47]}$ & $2.4 \times 10^{-2}$ & $\left(6.5 \pm 4.6\right) \times 10^{-2}$ & \cite{Huschle:2015rga} & $0.9\sigma$\\
$\frac{\langle \text{BR} \rangle}{\text{BR}}(B\to D\tau^+\nu)^{[11.47,12.0]}$ & $0.3 \times 10^{-2}$ & $\left(1.1 \pm 3.9\right) \times 10^{-2}$ & \cite{Huschle:2015rga} & $0.2\sigma$\\
\hline
\end{tabularx}

}
\caption{Charged-current LFU-testing observables except $R_{D^{(*)}}$ (part~1/2).}
\end{table}

\begin{table}[ht]
\resizebox{\textwidth}{!}{
\renewcommand{\arraystretch}{1.3}
\begin{tabularx}{1.36\textwidth}{lllll}
\hline
Observable ~~~~~~~~~~~~~~~~~~~~~~~~~~~~ & Prediction ~~~~~~~~~~~~~~~~~~~~~~~~~~~ & Measurement ~~~~~~~~~~~~~~~~~~~~~~~ & ~~~~~~~~~~~~~~~~~~ & Pull\\
\hline
$\frac{\langle \text{BR} \rangle}{\text{BR}}(B\to D\tau^+\nu)^{[4.0,4.5]}$ & $3.6 \times 10^{-2}$ & $\left(6.2 \pm 4.0\right) \times 10^{-2}$ & \cite{Lees:2013uzd} & $0.7\sigma$\\
$\frac{\langle \text{BR} \rangle}{\text{BR}}(B\to D\tau^+\nu)^{[4.5,5.0]}$ & $5.4 \times 10^{-2}$ & $\left(4.4 \pm 3.9\right) \times 10^{-2}$ & \cite{Lees:2013uzd} & $0.3\sigma$\\
$\frac{\langle \text{BR} \rangle}{\text{BR}}(B\to D\tau^+\nu)^{[5.0,5.5]}$ & $6.9 \times 10^{-2}$ & $\left(7.3 \pm 3.4\right) \times 10^{-2}$ & \cite{Lees:2013uzd} & $0.1\sigma$\\
$\frac{\langle \text{BR} \rangle}{\text{BR}}(B\to D\tau^+\nu)^{[5.5,6.0]}$ & $0.079$ & $0.118 \pm 0.043$ & \cite{Lees:2013uzd} & $0.9\sigma$\\
$\frac{\langle \text{BR} \rangle}{\text{BR}}(B\to D\tau^+\nu)^{[6.0,6.5]}$ & $0.086$ & $0.122 \pm 0.044$ & \cite{Lees:2013uzd} & $0.8\sigma$\\
$\frac{\langle \text{BR} \rangle}{\text{BR}}(B\to D\tau^+\nu)^{[6.5,7.0]}$ & $0.089$ & $0.104 \pm 0.044$ & \cite{Lees:2013uzd} & $0.3\sigma$\\
$\frac{\langle \text{BR} \rangle}{\text{BR}}(B\to D\tau^+\nu)^{[7.0,7.5]}$ & $8.9 \times 10^{-2}$ & $\left(8.3 \pm 4.1\right) \times 10^{-2}$ & \cite{Lees:2013uzd} & $0.2\sigma$\\
$\frac{\langle \text{BR} \rangle}{\text{BR}}(B\to D\tau^+\nu)^{[7.5,8.0]}$ & $0.087$ & $0.124 \pm 0.049$ & \cite{Lees:2013uzd} & $0.8\sigma$\\
$\frac{\langle \text{BR} \rangle}{\text{BR}}(B\to D\tau^+\nu)^{[8.0,8.5]}$ & $8.3 \times 10^{-2}$ & $\left(8.8 \pm 4.6\right) \times 10^{-2}$ & \cite{Lees:2013uzd} & $0.1\sigma$\\
$\frac{\langle \text{BR} \rangle}{\text{BR}}(B\to D\tau^+\nu)^{[8.5,9.0]}$ & $7.6 \times 10^{-2}$ & $\left(4.6 \pm 4.3\right) \times 10^{-2}$ & \cite{Lees:2013uzd} & $0.7\sigma$\\
$\frac{\langle \text{BR} \rangle}{\text{BR}}(B\to D\tau^+\nu)^{[9.0,9.5]}$ & $6.8 \times 10^{-2}$ & $\left(-0.2 \pm 4.3\right) \times 10^{-2}$ & \cite{Lees:2013uzd} & $1.6\sigma$\\
$\frac{\langle \text{BR} \rangle}{\text{BR}}(B\to D\tau^+\nu)^{[9.5,10.0]}$ & $5.9 \times 10^{-2}$ & $\left(1.8 \pm 4.7\right) \times 10^{-2}$ & \cite{Lees:2013uzd} & $0.9\sigma$\\
$\frac{\langle \text{BR} \rangle}{\text{BR}}(B\to D\tau^+\nu)^{[10.0,10.5]}$ & $4.8 \times 10^{-2}$ & $\left(9.2 \pm 5.3\right) \times 10^{-2}$ & \cite{Lees:2013uzd} & $0.8\sigma$\\
$\frac{\langle \text{BR} \rangle}{\text{BR}}(B\to D\tau^+\nu)^{[10.5,11.0]}$ & $3.6 \times 10^{-2}$ & $\left(0.7 \pm 4.0\right) \times 10^{-2}$ & \cite{Lees:2013uzd} & $0.7\sigma$\\
$\frac{\langle \text{BR} \rangle}{\text{BR}}(B\to D\tau^+\nu)^{[11.0,11.5]}$ & $2.1 \times 10^{-2}$ & $\left(0.4 \pm 3.7\right) \times 10^{-2}$ & \cite{Lees:2013uzd} & $0.5\sigma$\\
$\frac{\langle \text{BR} \rangle}{\text{BR}}(B\to D\tau^+\nu)^{[11.5,12.0]}$ & $0.2 \times 10^{-2}$ & $\left(1.7 \pm 2.9\right) \times 10^{-2}$ & \cite{Lees:2013uzd} & $0.5\sigma$\\
$\text{BR}(B_c\to \tau^+\nu_\tau)$ & $0.02$ & $< 0.1\ \text{@ 95\% CL}$ & \cite{Akeroyd:2017mhr} & $0.5\sigma$\\
$\text{BR}(\pi^+\to e^+\nu)$ & $1.234 \times 10^{-4}$ & $\left(1.233 \pm 0.002\right) \times 10^{-4}$ & \cite{Agashe:2014kda,Aguilar-Arevalo:2015cdf} & $0.5\sigma$\\
$F_L(B^0\to D^{\ast -}\tau^+\nu_\tau)$ & $0.442$ & $0.600 \pm 0.089$ & \cite{Adamczyk:2019wyt} & $1.8\sigma$\\
$R_{e\mu}(K^+\to \ell^+\nu)$ & $2.475 \times 10^{-5}$ & $\left(2.488 \pm 0.009\right) \times 10^{-5}$ & \cite{Agashe:2014kda} & $1.4\sigma$\\
\hline
\end{tabularx}

}
\caption{Charged-current LFU-testing observables except $R_{D^{(*)}}$ (part~2/2).}
\end{table}

\begin{table}[ht]
\resizebox{\textwidth}{!}{
\renewcommand{\arraystretch}{1.3}
\begin{tabularx}{1.36\textwidth}{lllll}
\hline
Observable ~~~~~~~~~~~~~~~~~~~~~~~~~~~~ & Prediction ~~~~~~~~~~~~~~~~~~~~~~~~~~~ & Measurement ~~~~~~~~~~~~~~~~~~~~~~~ & ~~~~~~~~~~~~~~~~~~ & Pull\\
\hline
$R_{\mu e}(B\to D^{\ast}\ell^+\nu)$ & $0.997$ & $0.982 \pm 0.027$ & \cite{Abdesselam:2017kjf,Abdesselam:2018nnh} & $0.6\sigma$\\
$R_{\tau \ell}(B\to D^{\ast}\ell^+\nu)$ & $0.255$ & $0.306 \pm 0.018$ & \cite{Hirose:2016wfn,Huschle:2015rga,Lees:2013uzd,Sato:2016svk} & $2.9\sigma$\\
$R_{\tau \ell}(B\to D\ell^+\nu)$ & $0.303$ & $0.406 \pm 0.050$ & \cite{Huschle:2015rga,Lees:2013uzd} & $2.1\sigma$\\
$R_{\tau \mu}(B\to D^{\ast}\ell^+\nu)$ & $0.255$ & $0.310 \pm 0.026$ & \cite{Aaij:2015yra,Aaij:2017uff} & $2.1\sigma$\\
\hline
\end{tabularx}

}
\caption{Charged-current LFU-testing observables $R_{D^{(*)}}$.}
\end{table}

\begin{table}[ht]
\resizebox{\textwidth}{!}{
\renewcommand{\arraystretch}{1.3}
\begin{tabularx}{1.36\textwidth}{lllll}
\hline
Observable ~~~~~~~~~~~~~~~~~~~~~~~~~~~~ & Prediction ~~~~~~~~~~~~~~~~~~~~~~~~~~~ & Measurement ~~~~~~~~~~~~~~~~~~~~~~~ & ~~~~~~~~~~~~~~~~~~ & Pull\\
\hline
$\langle R_{\mu e} \rangle(B^\pm\to K^\pm \ell^+\ell^-)^{[1.0,6.0]}$ & $1.001$ & $0.745 \pm 0.097$ & \cite{Aaij:2014ora} & $2.6\sigma$\\
$\langle R_{\mu e} \rangle(B^0\to K^{\ast 0}\ell^+\ell^-)^{[0.045,1.1]}$ & $0.93$ & $0.65 \pm 0.12$ & \cite{Aaij:2017vbb} & $2.4\sigma$\\
$\langle R_{\mu e} \rangle(B^0\to K^{\ast 0}\ell^+\ell^-)^{[1.1,6.0]}$ & $1.00$ & $0.68 \pm 0.12$ & \cite{Aaij:2017vbb} & $2.5\sigma$\\
$\text{BR}(B^\pm\to K^\pm \tau^+\tau^-)$ & $0.0002 \times 10^{-3}$ & $\left(1.31 \pm 0.77\right) \times 10^{-3}$ & \cite{TheBaBar:2016xwe} & $2.0\sigma$\\
$\text{BR}(B^0\to \tau^+\tau^-)$ & $0.00002 \times 10^{-3}$ & $< 1.8 \times 10^{-3}\ \text{@ 95\% CL}$ & \cite{Aaij:2017xqt,Aubert:2005qw} & $0.0\sigma$\\
$\overline{\text{BR}}(B_s\to \tau^+\tau^-)$ & $0.0008 \times 10^{-3}$ & $\left(-0.8 \pm 3.5\right) \times 10^{-3}$ & \cite{Aaij:2017xqt} & $0.3\sigma$\\
\hline
\end{tabularx}

}
\caption{Neutral-current LFU-testing observables.}
\end{table}

\begin{table}[ht]
\resizebox{\textwidth}{!}{
\renewcommand{\arraystretch}{1.3}
\begin{tabularx}{1.36\textwidth}{lllll}
\hline
Observable ~~~~~~~~~~~~~~~~~~~~~~~~~~~~ & Prediction ~~~~~~~~~~~~~~~~~~~~~~~~~~~ & Measurement ~~~~~~~~~~~~~~~~~~~~~~~ & ~~~~~~~~~~~~~~~~~~ & Pull\\
\hline
$\text{BR}(B^+\to K^{*+}\nu\bar\nu)$ & $1.0 \times 10^{-5}$ & $< 4.8 \times 10^{-5}\ \text{@ 95\% CL}$ & \cite{Grygier:2017tzo,Lees:2013kla,Lutz:2013ftz} & $1.1\sigma$\\
$\text{BR}(B^+\to K^+\nu\bar\nu)$ & $0.4 \times 10^{-5}$ & $< 1.7 \times 10^{-5}\ \text{@ 95\% CL}$ & \cite{Grygier:2017tzo,Lees:2013kla,Lutz:2013ftz,delAmoSanchez:2010bk} & $1.4\sigma$\\
$\text{BR}(B^+\to \pi^+\nu\bar\nu)$ & $0.01 \times 10^{-5}$ & $< 1.8 \times 10^{-5}\ \text{@ 95\% CL}$ & \cite{Grygier:2017tzo,Lutz:2013ftz} & $0.7\sigma$\\
$\text{BR}(B^+\to \rho^{+}\nu\bar\nu)$ & $0.04 \times 10^{-5}$ & $< 3.7 \times 10^{-5}\ \text{@ 95\% CL}$ & \cite{Grygier:2017tzo,Lutz:2013ftz} & $0.7\sigma$\\
$\text{BR}(B^0\to K^{*0}\nu\bar\nu)$ & $1.0 \times 10^{-5}$ & $< 2.0 \times 10^{-5}\ \text{@ 95\% CL}$ & \cite{Grygier:2017tzo,Lees:2013kla,Lutz:2013ftz} & $1.3\sigma$\\
$\text{BR}(B^0\to K^0\nu\bar\nu)$ & $0.4 \times 10^{-5}$ & $< 2.9 \times 10^{-5}\ \text{@ 95\% CL}$ & \cite{Grygier:2017tzo,Lees:2013kla,Lutz:2013ftz,delAmoSanchez:2010bk} & $0.5\sigma$\\
$\text{BR}(B^0\to \pi^0\nu\bar\nu)$ & $0.006 \times 10^{-5}$ & $< 1.3 \times 10^{-5}\ \text{@ 95\% CL}$ & \cite{Grygier:2017tzo,Lutz:2013ftz} & $0.6\sigma$\\
$\text{BR}(B^0\to \rho^{0}\nu\bar\nu)$ & $0.02 \times 10^{-5}$ & $< 4.6 \times 10^{-5}\ \text{@ 95\% CL}$ & \cite{Grygier:2017tzo,Lutz:2013ftz} & $1.3\sigma$\\
\hline
\end{tabularx}

}
\caption{$b\to q\nu\nu$ observables.}
\end{table}

\begin{table}[ht]
\resizebox{\textwidth}{!}{
\renewcommand{\arraystretch}{1.3}
\begin{tabularx}{1.36\textwidth}{lllll}
\hline
Observable ~~~~~~~~~~~~~~~~~~~~~~~~~~~~ & Prediction ~~~~~~~~~~~~~~~~~~~~~~~~~~~ & Measurement ~~~~~~~~~~~~~~~~~~~~~~~ & ~~~~~~~~~~~~~~~~~~ & Pull\\
\hline
$\langle A_7\rangle(B^0\to K^{\ast 0}\mu^+\mu^-)^{[1.1,6]}$ & $0.2 \times 10^{-2}$ & $\left(-4.5 \pm 5.0\right) \times 10^{-2}$ & \cite{Aaij:2015oid} & $0.9\sigma$\\
$\langle A_7\rangle(B^0\to K^{\ast 0}\mu^+\mu^-)^{[15,19]}$ & $0.010 \times 10^{-2}$ & $\left(-4.0 \pm 4.5\right) \times 10^{-2}$ & \cite{Aaij:2015oid} & $0.9\sigma$\\
$\langle A_8\rangle(B^0\to K^{\ast 0}\mu^+\mu^-)^{[1.1,6]}$ & $0.1 \times 10^{-2}$ & $\left(-4.7 \pm 5.8\right) \times 10^{-2}$ & \cite{Aaij:2015oid} & $0.8\sigma$\\
$\langle A_8\rangle(B^0\to K^{\ast 0}\mu^+\mu^-)^{[15,19]}$ & $0.007 \times 10^{-2}$ & $\left(2.5 \pm 4.8\right) \times 10^{-2}$ & \cite{Aaij:2015oid} & $0.5\sigma$\\
$\langle A_9\rangle(B^0\to K^{\ast 0}\mu^+\mu^-)^{[1.1,6]}$ & $0.01 \times 10^{-2}$ & $\left(-3.3 \pm 4.1\right) \times 10^{-2}$ & \cite{Aaij:2015oid} & $0.8\sigma$\\
$\langle A_9\rangle(B^0\to K^{\ast 0}\mu^+\mu^-)^{[15,19]}$ & $0.006 \times 10^{-2}$ & $\left(6.1 \pm 4.4\right) \times 10^{-2}$ & \cite{Aaij:2015oid} & $1.4\sigma$\\
\hline
\end{tabularx}

}
\caption{$B$ decays CPV observables.}
\end{table}

\begin{table}[ht]
\resizebox{\textwidth}{!}{
\renewcommand{\arraystretch}{1.3}
\begin{tabularx}{1.36\textwidth}{lllll}
\hline
Observable ~~~~~~~~~~~~~~~~~~~~~~~~~~~~ & Prediction ~~~~~~~~~~~~~~~~~~~~~~~~~~~ & Measurement ~~~~~~~~~~~~~~~~~~~~~~~ & ~~~~~~~~~~~~~~~~~~ & Pull\\
\hline
$A_ b$ & $0.935$ & $0.923 \pm 0.020$ & \cite{ALEPH:2005ab} & $0.6\sigma$\\
$A_ c$ & $0.668$ & $0.670 \pm 0.027$ & \cite{ALEPH:2005ab} & $0.1\sigma$\\
$A_ e$ & $0.1470$ & $0.1513 \pm 0.0019$ & \cite{ALEPH:2005ab} & $2.2\sigma$\\
$A_\mu$ & $0.147$ & $0.142 \pm 0.015$ & \cite{ALEPH:2005ab} & $0.3\sigma$\\
$A_\tau$ & $0.1470$ & $0.1433 \pm 0.0041$ & \cite{ALEPH:2005ab} & $0.9\sigma$\\
$A_\text{FB}^{0, b}$ & $10.31 \times 10^{-2}$ & $\left(9.92 \pm 0.16\right) \times 10^{-2}$ & \cite{ALEPH:2005ab} & $2.4\sigma$\\
$A_\text{FB}^{0, c}$ & $7.36 \times 10^{-2}$ & $\left(7.07 \pm 0.35\right) \times 10^{-2}$ & \cite{ALEPH:2005ab} & $0.8\sigma$\\
$A_\text{FB}^{0, e}$ & $1.62 \times 10^{-2}$ & $\left(1.45 \pm 0.25\right) \times 10^{-2}$ & \cite{ALEPH:2005ab} & $0.7\sigma$\\
$A_\text{FB}^{0,\mu}$ & $1.62 \times 10^{-2}$ & $\left(1.69 \pm 0.13\right) \times 10^{-2}$ & \cite{ALEPH:2005ab} & $0.5\sigma$\\
$A_\text{FB}^{0,\tau}$ & $1.62 \times 10^{-2}$ & $\left(1.88 \pm 0.17\right) \times 10^{-2}$ & \cite{ALEPH:2005ab} & $1.5\sigma$\\
$\text{BR}(W^\pm\to  e^\pm\nu)$ & $0.1084$ & $0.1071 \pm 0.0016$ & \cite{Schael:2013ita} & $0.8\sigma$\\
$\text{BR}(W^\pm\to \mu^\pm\nu)$ & $0.1084$ & $0.1063 \pm 0.0015$ & \cite{Schael:2013ita} & $1.4\sigma$\\
$\text{BR}(W^\pm\to \tau^\pm\nu)$ & $0.1084$ & $0.1138 \pm 0.0021$ & \cite{Schael:2013ita} & $2.6\sigma$\\
$\Gamma_W$ & $2.092$ & $2.085 \pm 0.042$ & \cite{Patrignani:2016xqp} & $0.2\sigma$\\
$\Gamma_Z$ & $2.494$ & $2.495 \pm 0.002$ & \cite{ALEPH:2005ab} & $0.5\sigma$\\
$R_ b^0$ & $0.2158$ & $0.2163 \pm 0.0007$ & \cite{ALEPH:2005ab} & $0.7\sigma$\\
$R_ c^0$ & $0.1722$ & $0.1721 \pm 0.0030$ & \cite{ALEPH:2005ab} & $0.0\sigma$\\
$R_ e^0$ & $20.73$ & $20.80 \pm 0.05$ & \cite{ALEPH:2005ab} & $1.4\sigma$\\
$R_\mu^0$ & $20.73$ & $20.78 \pm 0.03$ & \cite{ALEPH:2005ab} & $1.5\sigma$\\
$R_\tau^0$ & $20.78$ & $20.76 \pm 0.04$ & \cite{ALEPH:2005ab} & $0.4\sigma$\\
$m_W$ & $80.36$ & $80.38 \pm 0.01$ & \cite{Aaboud:2017svj,Aaltonen:2013iut} & $1.7\sigma$\\
$\sigma_\text{had}^0$ & $1.0654 \times 10^{-4}$ & $\left(1.0668 \pm 0.0009\right) \times 10^{-4}$ & \cite{ALEPH:2005ab} & $1.5\sigma$\\
\hline
\end{tabularx}

}
\caption{Electroweak precision observables.}
\end{table}

\begin{table}[ht]
\resizebox{\textwidth}{!}{
\renewcommand{\arraystretch}{1.3}
\begin{tabularx}{1.36\textwidth}{lllll}
\hline
Observable ~~~~~~~~~~~~~~~~~~~~~~~~~~~~ & Prediction ~~~~~~~~~~~~~~~~~~~~~~~~~~~ & Measurement ~~~~~~~~~~~~~~~~~~~~~~~ & ~~~~~~~~~~~~~~~~~~ & Pull\\
\hline
$\text{BR}(B^-\to K^{*-} e^+\mu^-)$ & $0$ & $\left(0.9 \pm 6.9\right) \times 10^{-7}$ & \cite{Aubert:2006vb} & $0.2\sigma$\\
$\text{BR}(B^-\to K^{*-} \mu^+e^-)$ & $0$ & $\left(-3.2 \pm 6.5\right) \times 10^{-7}$ & \cite{Aubert:2006vb} & $0.4\sigma$\\
$\text{BR}(B^-\to K^- e^+\mu^-)$ & $0$ & $\left(-1.21 \pm 0.78\right) \times 10^{-7}$ & \cite{Aubert:2006vb} & $1.5\sigma$\\
$\text{BR}(B^-\to K^- e^+\tau^-)$ & $0$ & $\left(0.2 \pm 2.1\right) \times 10^{-5}$ & \cite{Lees:2012zz} & $0.2\sigma$\\
$\text{BR}(B^-\to K^- \mu^+e^-)$ & $0$ & $\left(-2.9 \pm 7.7\right) \times 10^{-8}$ & \cite{Aubert:2006vb} & $0.3\sigma$\\
$\text{BR}(B^-\to K^- \mu^+\tau^-)$ & $0$ & $\left(0.8 \pm 1.9\right) \times 10^{-5}$ & \cite{Lees:2012zz} & $0.6\sigma$\\
$\text{BR}(B^-\to K^- \tau^+e^-)$ & $0$ & $\left(-1.3 \pm 1.8\right) \times 10^{-5}$ & \cite{Lees:2012zz} & $0.9\sigma$\\
$\text{BR}(B^-\to K^- \tau^+\mu^-)$ & $0$ & $\left(-0.4 \pm 1.4\right) \times 10^{-5}$ & \cite{Lees:2012zz} & $0.3\sigma$\\
$\text{BR}(B^-\to \pi^- e^\pm\mu^\mp)$ & $0$ & $< 2.0 \times 10^{-7}\ \text{@ 95\% CL}$ & \cite{Aubert:2007mm} & $0.0\sigma$\\
$\text{BR}(B^-\to \pi^- e^+\tau^-)$ & $0$ & $\left(2.8 \pm 2.4\right) \times 10^{-5}$ & \cite{Lees:2012zz} & $1.5\sigma$\\
$\text{BR}(B^-\to \pi^- \mu^+\tau^-)$ & $0$ & $\left(0.4 \pm 3.1\right) \times 10^{-5}$ & \cite{Lees:2012zz} & $0.2\sigma$\\
$\text{BR}(B^-\to \pi^- \tau^+e^-)$ & $0$ & $\left(-3.1 \pm 2.4\right) \times 10^{-5}$ & \cite{Lees:2012zz} & $1.3\sigma$\\
$\text{BR}(B^-\to \pi^- \tau^+\mu^-)$ & $0$ & $\left(0.0 \pm 2.6\right) \times 10^{-5}$ & \cite{Lees:2012zz} & $0.0\sigma$\\
$\text{BR}(\bar B^0\to \bar K^{*0} e^+\mu^-)$ & $0$ & $\left(0.7 \pm 2.4\right) \times 10^{-7}$ & \cite{Aubert:2006vb} & $0.5\sigma$\\
$\text{BR}(\bar B^0\to \bar K^{*0} \mu^+e^-)$ & $0$ & $\left(-0.7 \pm 2.3\right) \times 10^{-7}$ & \cite{Aubert:2006vb} & $0.2\sigma$\\
$\text{BR}(\bar B^0\to e^\pm \mu^\mp)$ & $0$ & $< 1.3 \times 10^{-9}\ \text{@ 95\% CL}$ & \cite{Aaij:2017cza} & $0.0\sigma$\\
$\text{BR}(\bar B^0\to e^\pm \tau^\mp)$ & $0$ & $\left(0.0 \pm 1.5\right) \times 10^{-5}$ & \cite{Aubert:2008cu} & $0.0\sigma$\\
$\text{BR}(\bar B^0\to \mu^\pm \tau^\mp)$ & $0$ & $\left(0.0 \pm 1.1\right) \times 10^{-5}$ & \cite{Aubert:2008cu} & $0.0\sigma$\\
$\text{BR}(\bar B^0\to \pi^0 e^\pm\mu^\mp)$ & $0$ & $< 1.7 \times 10^{-7}\ \text{@ 95\% CL}$ & \cite{Aubert:2007mm} & $0.0\sigma$\\
$\text{BR}(\bar B_s\to e^\pm \mu^\mp)$ & $0$ & $< 6.3 \times 10^{-9}\ \text{@ 95\% CL}$ & \cite{Aaij:2017cza} & $0.0\sigma$\\
$\text{BR}(K_L\to e^\pm\mu^\mp)$ & $0$ & $< 5.6 \times 10^{-12}\ \text{@ 95\% CL}$ & \cite{Tanabashi:2018oca} & $0.0\sigma$\\
$\text{BR}(\mu^-\to e^-e^+e^-)$ & $0$ & $< 1.2 \times 10^{-12}\ \text{@ 95\% CL}$ & \cite{Patrignani:2016xqp} & $0.0\sigma$\\
$\text{BR}(\mu\to e\gamma)$ & $0$ & $< 5.0 \times 10^{-13}\ \text{@ 95\% CL}$ & \cite{Patrignani:2016xqp} & $0.0\sigma$\\
$\text{BR}(\tau\to e\gamma)$ & $0$ & $< 3.9 \times 10^{-8}\ \text{@ 95\% CL}$ & \cite{Patrignani:2016xqp} & $0.0\sigma$\\
$\text{BR}(\tau^-\to e^-\mu^+e^-)$ & $0$ & $< 2.0 \times 10^{-8}\ \text{@ 95\% CL}$ & \cite{Hayasaka:2010np} & $0.6\sigma$\\
$\text{BR}(\tau^-\to e^-\mu^+\mu^-)$ & $0$ & $< 3.5 \times 10^{-8}\ \text{@ 95\% CL}$ & \cite{Hayasaka:2010np} & $0.0\sigma$\\
$\text{BR}(\tau^-\to \mu^-e^+e^-)$ & $0$ & $< 2.3 \times 10^{-8}\ \text{@ 95\% CL}$ & \cite{Hayasaka:2010np} & $0.5\sigma$\\
$\text{BR}(\tau^-\to \mu^-e^+\mu^-)$ & $0$ & $< 2.2 \times 10^{-8}\ \text{@ 95\% CL}$ & \cite{Hayasaka:2010np} & $0.5\sigma$\\
$\text{BR}(\tau\to \mu\gamma)$ & $0$ & $< 5.2 \times 10^{-8}\ \text{@ 95\% CL}$ & \cite{Patrignani:2016xqp} & $0.0\sigma$\\
$\text{BR}(\tau^-\to \mu^-\mu^+\mu^-)$ & $0$ & $< 2.7 \times 10^{-8}\ \text{@ 95\% CL}$ & \cite{Hayasaka:2010np} & $0.0\sigma$\\
$\text{BR}(\tau^+\to \phi e^+)$ & $0$ & $< 3.7 \times 10^{-8}\ \text{@ 95\% CL}$ & \cite{Patrignani:2016xqp} & $0.0\sigma$\\
$\text{BR}(\tau^+\to \phi\mu^+)$ & $0$ & $< 1.0 \times 10^{-7}\ \text{@ 95\% CL}$ & \cite{Patrignani:2016xqp} & $0.0\sigma$\\
$\text{BR}(\tau^+\to \rho^0 e^+)$ & $0$ & $< 2.1 \times 10^{-8}\ \text{@ 95\% CL}$ & \cite{Patrignani:2016xqp} & $0.0\sigma$\\
$\text{BR}(\tau^+\to \rho^0\mu^+)$ & $0$ & $< 1.4 \times 10^{-8}\ \text{@ 95\% CL}$ & \cite{Patrignani:2016xqp} & $0.0\sigma$\\
\hline
\end{tabularx}

}
\caption{LFV observables.}
\end{table}

\begin{table}[ht]
\resizebox{\textwidth}{!}{
\renewcommand{\arraystretch}{1.3}
\begin{tabularx}{1.36\textwidth}{lllll}
\hline
Observable ~~~~~~~~~~~~~~~~~~~~~~~~~~~~ & Prediction ~~~~~~~~~~~~~~~~~~~~~~~~~~~ & Measurement ~~~~~~~~~~~~~~~~~~~~~~~ & ~~~~~~~~~~~~~~~~~~ & Pull\\
\hline
$\text{BR}(Z^0\to  e^\pm\mu^\mp)$ & $0$ & $< 5.2 \times 10^{-7}\ \text{@ 95\% CL}$ & \cite{Aad:2014bca,Abreu:1996mj,Akers:1995gz} & $0.0\sigma$\\
$\text{BR}(Z^0\to  e^\pm\tau^\mp)$ & $0$ & $< 6.8 \times 10^{-6}\ \text{@ 95\% CL}$ & \cite{Abreu:1996mj,Akers:1995gz} & $0.0\sigma$\\
$\text{BR}(Z^0\to \mu^\pm\tau^\mp)$ & $0$ & $< 7.0 \times 10^{-6}\ \text{@ 95\% CL}$ & \cite{Abreu:1996mj,Akers:1995gz} & $0.0\sigma$\\
\hline
\end{tabularx}

}
\caption{$Z$ decay LFV observables.}
\end{table}

\FloatBarrier

\bibliographystyle{JHEP}
\bibliography{bibliography}

\end{document}